# Reconsidering Global Comparisons of Scientist Trustworthiness

Yuke Yang[a], Zhihao Ma[a]

[a] *Communication Collaboratory, School of Journalism and Communication, Nanjing University*

*Corresponding Author: Zhihao Ma

**Address:**

Computational Communication Collaboratory

School of Journalism and Communication

Nanjing University

Room 345, Zijin Building, Nanjing University (Xianlin Campus), 163 Xianlin Road, Qixia District, Jiangsu Nanjing, 210023 China

Email: redclass@163.com

**Competing Interests:** None.

# Reconsidering Global Comparisons of Scientist Trustworthiness

**Abstract:** In a reanalysis of perceived scientist trustworthiness across 68 countries/regions, measurement-aligned indicated limitations in the cross-national comparability of the original measurement structure. Some country/region rankings and predictor associations varied across alternative scoring approaches, suggesting the value of considering measurement invariance when interpreting cross-national findings. Future research could use diverse methods to strengthen cross-cultural measurement validity.

Large-scale cross-national and big-team science studies integrate data from diverse contexts, increasing the possibility that common measures function differently across countries and regions. If such variation is not adequately considered, observed rankings and associations may partly reflect measurement non-invariance rather than substantive differences. Given the policy and public relevance of such findings, greater attention to measurement invariance (MI) is warranted.

Whether MI is a necessary condition for meaningful cross-national comparison remains debated. One view holds that concerns about non-invariance may be overstated and that valid comparisons can remain possible under some forms of imperfect equivalence (Welzel et al., 2023). A contrasting view emphasizes that non-invariance can affect latent means, rankings, associations, and nomological validity in ways that are difficult to distinguish from substantive group differences (Meuleman et al., 2023). This debate has recently extended to cross-cultural psychology, with Kusano et al. (2025) questioning conventional invariance requirements and Sterner and Goretzko (2026) underscoring the consequences of non-invariance for comparability. The present study examines the sensitivity of cross-national findings to how measurement non-invariance is handled through a reanalysis of Cologna et al. (2025), who assessed perceived scientist trustworthiness among 69,534 respondents in 68 countries/regions.

We reanalyzed the original dataset in three steps (Supplementary Fig. S1 and Table S1). First, we evaluated MI for the 12-item measure of perceived scientist trustworthiness using two strategies: a pooled confirmatory factor analysis (CFA) including all countries/regions, followed by multigroup MI testing across the full sample; and country/region specific CFAs used as a screening step, after which only countries/regions with acceptable CFA fit were included in the subsequent multigroup MI testing. Second, where scalar invariance was not supported, we applied measurement alignment (MA) to accommodate partial non-invariance and compared the resulting estimates with those reported in the original article based on weighted observed scores. Third, all analyses were repeated with and without survey weights.

For the four-factor perceived scientist trustworthiness model, configural and metric invariance were supported, whereas scalar and strict invariance were not (Supplementary Table S4). A partial-invariance model did not converge. We therefore implemented four MA specifications following Han (2024). The main text reports the specification most closely corresponding to the original CFA model (Path B1, Step 5 in Fig. S1); the remaining results are provided in Supplementary Tables S5–S10.

Admissible aligned solutions were obtained for benevolence and openness, which were retained for subsequent ranking and correlational analyses. To assess the impact of factor reduction, we compared rankings and regression results across three scores based on the same sample: the original four-factor score, a six-item unaligned score from the two retained factors, and the aligned two-factor score (see Supplementary Comparison section).

The results showed that alignment preserved the broad ordering while altering the specific rank positions of many countries. The overall rank ordering was consistent (Spearman's $\rho$ = .971; Kendall's $\tau$ = .863), but 62 of 68 countries/regions (91.2%) still changed their exact rank position after alignment, with a median absolute displacement of 3 positions (IQR = 1.75–5; range = 0–15). The unaligned two-factor score and the aligned two-factor score showed greater overall similarity; however, 36 countries still changed their rank positions (Table 1; Supplementary Table S15).

(insert Table 1 here)

The significant negative association with science-related populist attitudes in the original article (Fig.2, p. 716) was not replicated using aligned scores (Supplementary Table S25, Block 3). Direct comparison based on the same sample confirmed a significant difference between the original and aligned scores ($\Delta\beta$ = −0.033, 95% CI [−0.040, −0.026], BH-adjusted $p$ < .001; Supplementary Table S29). Pairwise comparisons revealed significant differences in regression coefficients between the original score and both the unaligned and aligned two-factor scores, but not between the two-factor scores themselves (Supplementary TableS30). Moreover, some predictors, including science-related populist attitudes, were not tested for MI, which may also affect cross-national regression estimates (Yan & Ma, 2026).

Taken together, these findings highlight two considerations for large-scale cross-national research. First, uncertainty regarding measurement comparability warrants caution when interpreting country/region rankings and substantive associations. In our study, measurement alignment did not yield an acceptable solution for all factors: two of the four original factors could not be successfully aligned, suggesting possible limitations in their cross-national comparability. Our findings indicate that, in such situations, different analytical choices may lead to different results and interpretations. In the present study, we reported country/region rankings and regression results under three different analytical strategies. These findings may suggest that examining measurement invariance can help clarify the structure of the measure and provide a basis for a more cautious interpretation of subsequent analyses. Second, the measurement framework used in the original article was developed using U.S. online samples (Besley et al., 2026), raising questions about its conceptual equivalence across 68 culturally, politically, and institutionally diverse countries/regions. Future cross-cultural measure development may integrate qualitative and mixed-method evidence to help clarify how the focal construct is understood across contexts and which cues inform respondents' judgments. Such evidence may guide item refinement, factor-structure evaluation, and subsequent psychometric validation, thereby strengthening cross-cultural validity and the interpretation of cross-context comparisons.

In summary, the credibility of large-scale comparative findings depends not only on sample size, geographic coverage, and analytic sophistication, but also on the conceptual and psychometric comparability of the constructs examined. Measurement comparability merits consideration when interpreting global comparative findings.

Yan, X., & Ma, Z. (2026). Revisiting the SciPop Scale: Measurement alignment, scoring rules, and cross-context comparability. *Political Psychology*, *47*(5), e70191. https://doi.org/10.1111/pops.70191

**Acknowledgments**

We thank Mede, Cologna, and their colleagues for their insightful and valuable work, which provided the data and opportunity for us to further examine the question of measurement comparability. We are also grateful for our continued correspondence with Mede during the preparation of the manuscript, which helped deepen our understanding of the issues addressed in this study. We are sincerely grateful to our two anonymous reviewers for their thoughtful and constructive comments. Many of their practical and valuable suggestions directly informed our revisions and prompted us to reflect more deeply on the research questions addressed in this study.

**Data and Code Availability**

The dataset employed in our study is accessible via OSF at https://osf.io/5c3qd/. The R analysis code used in this study are available via OSF at https://osf.io/9stzn/.

**Disclosure of Conflict of Interest**

The authors declare no competing interests.

**Ethical Approval Statement**

The present study used publicly available, de-identified data. Ethical approval for data collection was obtained by the original study team. No additional ethical approval was required for this secondary data analysis.

**Table 1. Rank-Order Agreement and Country-Level Rank Displacement Across Three Scoring Approaches**

| Comparison | Spearman's $\rho$ | Kendall's $\tau$ | Countries changed rank, *n* (%) | Median absolute rank change | IQR | Range |
|---|---|---|---|---|---|---|
| Raw4 vs Raw2 | 0.971 | 0.863 | 62 (91.176%) | 3 | 2–5 | 0–15 |
| Raw4 vs Aligned | 0.971 | 0.863 | 62 (91.176%) | 3 | 1.75–5 | 0–15 |
| Raw2 vs Aligned | 0.995 | 0.963 | 36 (52.941%) | 1 | 0–1 | 0–11 |

*Note.* "Raw4" refers to country rankings based on the scoring approach used in the original article, combining all 12 items across the four dimensions of perceived scientist trustworthiness. "Raw2" refers to rankings based on the six original, unaligned items from the two retained dimensions, benevolence and openness. "Aligned" refers to rankings based on the aligned factor scores for benevolence and openness. Spearman's $\rho$ and Kendall's $\tau$ indicate overall rank-order agreement between each pair of scoring approaches. Interquartile range (IQR) denotes the interquartile range of these absolute changes, and Range denotes the minimum to maximum absolute rank change.

## Supplementary Information

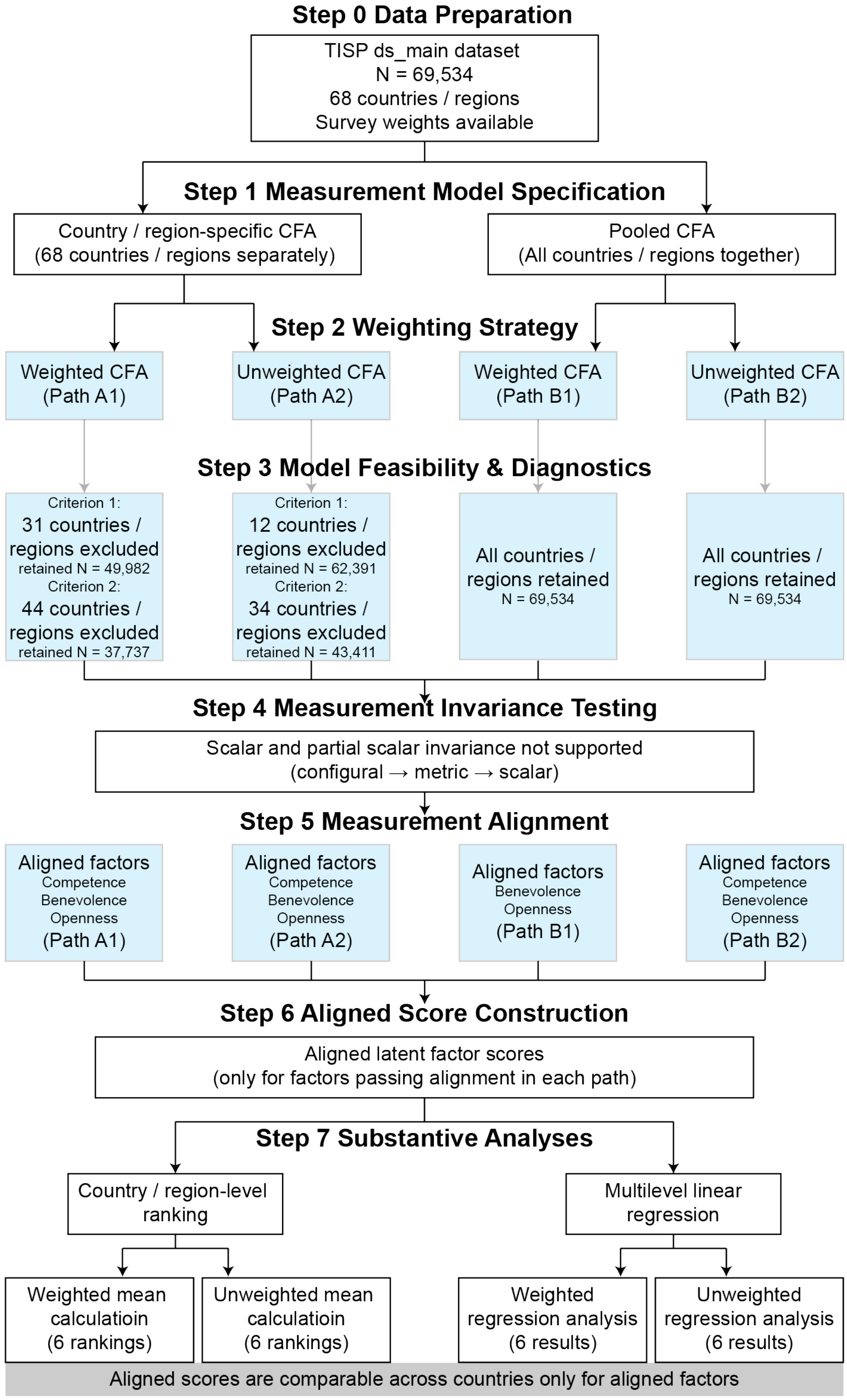

**Figure S1. Analytic Workflow of Parallel CFA–Alignment Pipelines**

*Note.* Path A1: country/region-specific CFA with weights; Path A2 country/region-specific CFA without weights; Path B1: pooled CFA with weights; Path B2: pooled CFA without weights. In the country/region-specific CFA approach, the four-factor CFA was first fitted separately within each country/region, and only countries/regions meeting the specified fit criteria were retained for subsequent multigroup MI testing. In the pooled CFA approach, all

countries/regions were included from the outset and proceeded directly to multigroup MI testing. Criterion 1: CFI and TLI ≥ 0.90, RMSEA ≤ 0.08, and SRMR ≤ 0.10; Criterion 2: CFI and TLI ≥ 0.95, RMSEA ≤ 0.08, and SRMR ≤ 0.10.

**Table S1. Roadmap for analysis and results**

| | Path A1-C1 | Path A1-C2 | Path A2-C1 | Path A2-C2 | Path B1 | Path B2 |
|---|---|---|---|---|---|---|
| CFA | Table S2 | Table S2 | Table S3 | Table S3 | Table S4 | Table S4 |
| Country set | 37 | 24 | 56 | 34 | 68 | 68 |
| MI | Table S4 | Table S4 | Table S4 | Table S4 | Table S4 | Table S4 |
| MA | Table S5 | Table S6 | Table S7 | Table S8 | Table S9 | Table S10 |
| Retained factor | Competence, Benevolence, Openness | Competence, Benevolence, Openness | Competence, Benevolence, Openness | Competence, Benevolence, Openness | Benevolence, Openness | Competence, Benevolence, Openness |
| Weighted rankings | Figure S2 | Figure S4 | Figure S6 | Figure S8 | Figure S10 | Figure S12 |
| Unweighted rankings | Figure S3 | Figure S5 | Figure S7 | Figure S9 | Figure S11 | Figure S13 |
| Weighted ranking change | Table S11 | Table S12 | Table S13 | Table S14 | Table S15 | Table S16 |
| Unweighted ranking change | Table S11 | Table S12 | Table S13 | Table S14 | Table S15 | Table S16 |
| Weighted regression | Table S17 | Table S19 | Table S21 | Table S23 | Table S25 | Table S27 |
| Unweighted rankings | Table S18 | Table S20 | Table S22 | Table S24 | Table S26 | Table S28 |

*Note*. Hyperlinks enable direct navigation to the relevant content upon clicking. Path A1-C1 refers to weighted country/region-specific CFA with Criterion 1; Path A1-C2 refers to weighted country/region-specific CFA with Criterion 2; Path A2-C1 refers to unweighted country/region-specific CFA with Criterion 1; and Path A2-C2 refers to unweighted country/region-specific CFA with Criterion 2. Path B1 refers to weighted pooled CFA, and Path B2 refers to unweighted pooled CFA. In Paths A1 and A2, four-factor CFA models were fitted separately for each country/region, and only those meeting the specified fit criteria were retained for subsequent multigroup measurement invariance (MI) testing. In Paths B1 and B2, all countries/regions were included without country/region-specific CFA screening and proceeded directly to multigroup MI testing. Criterion 1 required CFI and TLI $\geq 0.90$, RMSEA $\leq 0.08$, and SRMR $\leq 0.10$; Criterion 2 required CFI and TLI $\geq 0.95$, RMSEA $\leq 0.08$, and SRMR $\leq 0.10$.

## CFA and Measurement invariance

In group comparisons, researchers frequently presume that the instruments share an equivalent measurement structure. The soundness of this presumption is essential for interpreting group-related difference (Vandenberg & Lance, 2000). Without measurement invariance, one cannot confidently maintain that the construct has the same meaning across groups (Little, 1997). Accordingly, comparisons of means or structural parameters should be based on measurement models that function similarly across groups (Ployhart & Oswald, 2004).

We noted that current studies differ in how they operationalize cross-group measurement invariance testing. On the one hand, because configural invariance is typically regarded as the foundation for subsequent, higher levels of measurement invariance (Byrne, 2013; Putnick & Bornstein, 2016), some researchers diagnose the measurement model within each group before fitting a multi-group model, in order to identify and address potential problems that may lead to estimation failure or unacceptable solutions. In extreme cases, when certain groups show serious data-quality issues, or when the model is not identified / yields unacceptable solutions that cannot be resolved through reasonable modifications, researchers may exclude these groups from cross-group comparability inferences (Cosma et al., 2022; Sawicki et al., 2022) . On the other hand, others have argued that when a large number of groups are involved (e.g., cross-national or large-scale surveys), requiring every group to achieve a traditionally defined “good fit” is often unrealistic (Rutkowski & Svetina, 2014). Accordingly, some researchers do not screen groups via separate single-group CFAs prior to invariance testing; instead, they fit a multigroup CFA (MG-CFA) including all groups simultaneously, beginning with the configural (baseline) model and then imposing equality constraints to evaluate measurement invariance (Hirschfeld & Brachel, 2014).

Considering that Cologna et al. (2025) conducted CFA simultaneously across all countries when testing measurement invariance, we decided to adopt two different CFA strategies (see Fig. S1). Specifically, we evaluated the four-factor model of perceived scientist trustworthiness using two CFA strategies: (a) a pooled, multi-country CFA in which all countries were analyzed simultaneously, and (b) separate CFAs conducted within each of the 68 countries/regions.

For the country-specific approach, the four-factor CFA was first fitted separately within each country, and only countries with acceptable model fit were retained for subsequent multigroup measurement invariance testing. Following established recommendations (Bentler & Bonett, 1980; Brown, 2015; Hu & Bentler, 1999; Schermelleh-Engel et al., 2003), model fit was evaluated using multiple indices, including the comparative fit index (CFI), Tucker–Lewis index (TLI), root mean square error of approximation (RMSEA), and standardized root mean square residual (SRMR). Two sets of thresholds were applied: Criterion 1 (CFI ≥ 0.90, TLI ≥ 0.90, RMSEA ≤ 0.08, SRMR ≤ 0.10) and Criterion 2 (CFI ≥ 0.95, TLI ≥ 0.95, RMSEA ≤ 0.08, SRMR ≤ 0.10).

We noted that Cologna et al. (2025) provided and applied post-stratification weights in their data analysis strategy to align statistical estimates more closely with the target population distributions on key demographic variables such as sex, age, and education. Post-stratification/calibration weights are primarily used to reduce bias arising from sample selection and nonresponse by matching the sample to known marginal or joint distributions of auxiliary variables (e.g., sex, age, education), thereby improving the population representativeness of point estimates (e.g., means, regression coefficients). However, obtaining appropriate variances and standard errors depends on using suitable design-based variance estimation methods and on assumptions such as approximate randomness within post-strata (Battaglia et al., 2009). At the same time, incorporating weights in structural equation modeling and multi-group CFA can affect parameter estimates, standard errors, and fit statistics. This concern is particularly salient when weights are primarily derived from post-stratification calibration rather than from a clearly specified probability sampling design, in which case the appropriateness of the weighting scheme and its impact on substantive conclusions should be evaluated via sensitivity analyses (Asparouhov, 2005). Accordingly, to assess the robustness of measurement-model fit and measurement-invariance conclusions to weighting decisions, we report results from both weighted and unweighted analyses in our primary models: the weighted analyses used the country-level (“WEIGHT_CNTRY”) or the pooled-sample (“WEIGHT_GLOBL”).

Under the country-specific CFA approach, when sampling weights were applied, 37 countries met Criterion 1 and 24 countries met Criterion 2; when sampling weights were not applied, 56 countries met Criterion 1 and 34 countries met Criterion 2 (see Fig. S1, Step 3).

The corresponding CFA fit statistics and measurement invariance results are reported below.

**Table S2. CFA model fit of the perceived scientist trustworthiness scale based on the four-factor model across 68 countries/regions (Fig. S1, Path A1: country-specific CFA with weights)**

| Country | *N* | $\chi^2$ | *df* | RMSEA | CFI | TLI | SRMR | Criterion 1 | Criterion 2 |
|---|---|---|---|---|---|---|---|---|---|
| Albania | 343 | 241.868 | 48 | 0.236 | 0.591 | 0.438 | 0.167 | | |
| Argentina | 495 | 80.895 | 48 | 0.054 | 0.963 | 0.949 | 0.047 | ✓ | |
| Australia | 3,523 | 311.634 | 48 | 0.058 | 0.976 | 0.967 | 0.030 | ✓ | ✓ |
| Austria | 1,035 | 242.522 | 48 | 0.072 | 0.965 | 0.951 | 0.032 | ✓ | ✓ |
| Bangladesh | 484 | 66.414 | 48 | 0.068 | 0.951 | 0.933 | 0.063 | ✓ | |
| Belgium | 2,035 | 281.396 | 48 | 0.066 | 0.964 | 0.950 | 0.031 | ✓ | ✓ |
| Bolivia | 493 | 109.562 | 48 | 0.083 | 0.948 | 0.929 | 0.045 | | |
| Botswana | 401 | 78.231 | 48 | 0.091 | 0.936 | 0.912 | 0.061 | | |
| Brazil | 1,214 | 112.795 | 48 | 0.061 | 0.975 | 0.966 | 0.027 | ✓ | ✓ |
| Bulgaria | 497 | 125.702 | 48 | 0.090 | 0.950 | 0.932 | 0.042 | | |
| Cameroon | 473 | 81.661 | 48 | 0.094 | 0.916 | 0.884 | 0.065 | | |
| Canada | 2,507 | 288.777 | 48 | 0.059 | 0.977 | 0.969 | 0.025 | ✓ | ✓ |
| Chile | 1,003 | 103.662 | 48 | 0.059 | 0.972 | 0.961 | 0.032 | ✓ | ✓ |
| China | 497 | 108.642 | 48 | 0.129 | 0.922 | 0.892 | 0.052 | | |
| Colombia | 499 | 113.367 | 48 | 0.082 | 0.945 | 0.925 | 0.048 | | |
| Congo DR | 389 | 84.318 | 48 | 0.129 | 0.798 | 0.722 | 0.104 | | |
| Costa Rica | 539 | 120.679 | 48 | 0.074 | 0.959 | 0.943 | 0.038 | ✓ | |
| Côte d'Ivoire | 466 | 61.804 | 48 | 0.066 | 0.951 | 0.933 | 0.063 | ✓ | |
| Cyprus | 502 | 104.305 | 48 | 0.067 | 0.967 | 0.955 | 0.036 | ✓ | ✓ |
| Czech Republic | 495 | 81.396 | 48 | 0.043 | 0.983 | 0.976 | 0.029 | ✓ | ✓ |
| Denmark | 1,208 | 127.788 | 48 | 0.056 | 0.969 | 0.958 | 0.031 | ✓ | ✓ |

| Country | *N* | $\chi^2$ | *df* | RMSEA | CFI | TLI | SRMR | Criterion 1 | Criterion 2 |
|---|---|---|---|---|---|---|---|---|---|
| Egypt | 420 | 56.030 | 48 | 0.046 | 0.983 | 0.977 | 0.042 | ✓ | ✓ |
| Ethiopia | 363 | 127.505 | 48 | 0.163 | 0.777 | 0.693 | 0.104 | | |
| Finland | 998 | 141.264 | 48 | 0.058 | 0.975 | 0.966 | 0.029 | ✓ | ✓ |
| France | 2,007 | 195.506 | 48 | 0.055 | 0.978 | 0.970 | 0.025 | ✓ | ✓ |
| Georgia | 493 | 121.789 | 48 | 0.078 | 0.962 | 0.948 | 0.034 | ✓ | |
| Germany | 8,014 | 1266.664 | 48 | 0.069 | 0.964 | 0.951 | 0.031 | ✓ | ✓ |
| Ghana | 474 | 154.910 | 48 | 0.156 | 0.761 | 0.671 | 0.091 | | |
| Greece | 1,436 | 283.146 | 48 | 0.075 | 0.959 | 0.944 | 0.035 | ✓ | |
| Hong Kong | 594 | 162.317 | 48 | 0.087 | 0.937 | 0.913 | 0.048 | | |
| Hungary | 498 | 103.585 | 48 | 0.058 | 0.975 | 0.966 | 0.029 | ✓ | ✓ |
| India | 473 | 93.378 | 48 | 0.131 | 0.878 | 0.832 | 0.073 | | |
| Indonesia | 2,056 | 150.948 | 48 | 0.079 | 0.959 | 0.944 | 0.035 | ✓ | |
| Ireland | 498 | 90.126 | 48 | 0.054 | 0.972 | 0.962 | 0.036 | ✓ | ✓ |
| Israel | 1,010 | 158.823 | 48 | 0.067 | 0.958 | 0.942 | 0.037 | ✓ | |
| Italy | 1,505 | 261.239 | 48 | 0.068 | 0.969 | 0.957 | 0.027 | ✓ | ✓ |
| Japan | 1,000 | 164.913 | 48 | 0.082 | 0.935 | 0.910 | 0.044 | | |
| Kazakhstan | 513 | 113.940 | 48 | 0.073 | 0.955 | 0.938 | 0.040 | ✓ | |
| Kenya | 456 | 133.309 | 48 | 0.143 | 0.819 | 0.751 | 0.090 | | |
| Malaysia | 985 | 101.982 | 48 | 0.072 | 0.963 | 0.949 | 0.036 | ✓ | |
| Mexico | 498 | 110.423 | 48 | 0.084 | 0.941 | 0.918 | 0.049 | | |
| Morocco | 375 | 80.820 | 48 | 0.072 | 0.953 | 0.935 | 0.050 | ✓ | |
| Netherlands | 1,410 | 219.415 | 48 | 0.066 | 0.959 | 0.944 | 0.038 | ✓ | |
| New Zealand | 2,009 | 169.108 | 48 | 0.058 | 0.977 | 0.968 | 0.029 | ✓ | ✓ |

| Country | $N$ | $\chi^2$ | $df$ | RMSEA | CFI | TLI | SRMR | Criterion 1 | Criterion 2 |
|---|---|---|---|---|---|---|---|---|---|
| Nicaragua | 399 | 109.851 | 48 | 0.106 | 0.931 | 0.905 | 0.051 | | |
| Nigeria | 988 | 117.452 | 48 | 0.096 | 0.924 | 0.896 | 0.056 | | |
| Norway | 494 | 94.404 | 48 | 0.058 | 0.975 | 0.966 | 0.034 | ✓ | ✓ |
| Peru | 496 | 103.195 | 48 | 0.086 | 0.938 | 0.914 | 0.048 | | |
| Philippines | 545 | 91.053 | 48 | 0.084 | 0.940 | 0.917 | 0.058 | | |
| Poland | 3,002 | 629.975 | 48 | 0.082 | 0.955 | 0.938 | 0.034 | | |
| Portugal | 499 | 208.229 | 48 | 0.106 | 0.925 | 0.896 | 0.052 | | |
| Romania | 432 | 91.762 | 48 | 0.113 | 0.861 | 0.809 | 0.079 | | |
| Russia | 1,503 | 446.021 | 48 | 0.083 | 0.950 | 0.931 | 0.032 | | |
| Serbia | 499 | 136.728 | 48 | 0.092 | 0.947 | 0.927 | 0.041 | | |
| Slovakia | 530 | 148.912 | 48 | 0.082 | 0.964 | 0.950 | 0.036 | | |
| Slovenia | 501 | 151.597 | 48 | 0.085 | 0.959 | 0.944 | 0.032 | | |
| South Africa | 1,000 | 66.105 | 48 | 0.040 | 0.988 | 0.983 | 0.030 | ✓ | ✓ |
| South Korea | 500 | 128.890 | 48 | 0.086 | 0.928 | 0.900 | 0.060 | | |
| Spain | 1,009 | 247.599 | 48 | 0.084 | 0.957 | 0.941 | 0.035 | | |
| Sweden | 1,004 | 182.875 | 48 | 0.067 | 0.963 | 0.950 | 0.033 | ✓ | ✓ |
| Switzerland | 997 | 202.454 | 48 | 0.065 | 0.967 | 0.955 | 0.030 | ✓ | ✓ |
| Taiwan | 1,204 | 108.428 | 48 | 0.054 | 0.980 | 0.973 | 0.026 | ✓ | ✓ |
| Türkiye | 500 | 98.293 | 48 | 0.087 | 0.946 | 0.926 | 0.052 | | |
| Uganda | 387 | 95.536 | 48 | 0.132 | 0.866 | 0.816 | 0.094 | | |
| Ukraine | 1,008 | 145.114 | 48 | 0.066 | 0.963 | 0.950 | 0.034 | ✓ | ✓ |
| United Kingdom | 1,983 | 326.361 | 48 | 0.076 | 0.962 | 0.948 | 0.032 | ✓ | |
| United States | 2,559 | 239.407 | 48 | 0.054 | 0.984 | 0.978 | 0.021 | ✓ | ✓ |

| Country | *N* | $\chi^2$ | *df* | RMSEA | CFI | TLI | SRMR | Criterion 1 | Criterion 2 |
|---|---|---|---|---|---|---|---|---|---|
| Uruguay | 312 | 90.982 | 48 | 0.086 | 0.953 | 0.936 | 0.046 | | |

*Note.* Path A1 represents the country-specific CFA with weights. Under this approach, the four-factor CFA was first fitted separately within each country, and only countries with acceptable model fit were retained for subsequent multigroup measurement invariance testing. Criterion 1 required CFI and TLI ≥ 0.90, RMSEA ≤ 0.08, and SRMR ≤ 0.10; 37 countries met this criterion. Criterion 2 required CFI and TLI ≥ 0.95, RMSEA ≤ 0.08, and SRMR ≤ 0.10; 24 countries met this criterion.

**Table S3. CFA model fit of the perceived scientist trustworthiness scale based on the four-factor model across 68 countries/regions (Fig. S1, Path A2: country-specific CFA without weights)**

| Country | *N* | $\chi^2$ | *df* | RMSEA | CFI | TLI | SRMR | Criterion 1 | Criterion 2 |
|---|---|---|---|---|---|---|---|---|---|
| Albania | 343 | 192.943 | 48 | 0.107 | 0.898 | 0.859 | 0.063 | | |
| Argentina | 495 | 127.037 | 48 | 0.065 | 0.954 | 0.936 | 0.041 | ✓ | |
| Australia | 3,523 | 457.581 | 48 | 0.057 | 0.975 | 0.966 | 0.031 | ✓ | ✓ |
| Austria | 1,035 | 248.662 | 48 | 0.073 | 0.963 | 0.949 | 0.033 | ✓ | |
| Bangladesh | 484 | 116.375 | 48 | 0.064 | 0.969 | 0.957 | 0.033 | ✓ | ✓ |
| Belgium | 2,035 | 379.233 | 48 | 0.068 | 0.963 | 0.949 | 0.032 | ✓ | |
| Bolivia | 493 | 139.487 | 48 | 0.072 | 0.960 | 0.945 | 0.036 | ✓ | |
| Botswana | 401 | 118.102 | 48 | 0.071 | 0.959 | 0.943 | 0.040 | ✓ | |
| Brazil | 1,214 | 239.855 | 48 | 0.070 | 0.966 | 0.953 | 0.028 | ✓ | ✓ |
| Bulgaria | 497 | 132.533 | 48 | 0.072 | 0.967 | 0.954 | 0.033 | ✓ | ✓ |
| Cameroon | 473 | 135.169 | 48 | 0.073 | 0.952 | 0.934 | 0.037 | ✓ | |
| Canada | 2,507 | 340.455 | 48 | 0.059 | 0.978 | 0.969 | 0.025 | ✓ | ✓ |
| Chile | 1,003 | 208.513 | 48 | 0.067 | 0.965 | 0.951 | 0.031 | ✓ | ✓ |
| China | 497 | 104.452 | 48 | 0.058 | 0.981 | 0.974 | 0.030 | ✓ | ✓ |
| Colombia | 499 | 200.990 | 48 | 0.093 | 0.930 | 0.904 | 0.046 | | |
| Congo DR | 389 | 154.321 | 48 | 0.088 | 0.922 | 0.893 | 0.058 | | |
| Costa Rica | 539 | 178.560 | 48 | 0.080 | 0.951 | 0.932 | 0.036 | ✓ | |
| Côte d'Ivoire | 466 | 87.005 | 48 | 0.049 | 0.976 | 0.966 | 0.032 | ✓ | ✓ |
| Cyprus | 502 | 148.044 | 48 | 0.076 | 0.955 | 0.938 | 0.037 | ✓ | |
| Czech Republic | 495 | 79.270 | 48 | 0.042 | 0.983 | 0.976 | 0.030 | ✓ | ✓ |
| Denmark | 1,208 | 196.409 | 48 | 0.059 | 0.967 | 0.955 | 0.030 | ✓ | ✓ |

| Country | $N$ | $\chi^2$ | $df$ | RMSEA | CFI | TLI | SRMR | Criterion 1 | Criterion 2 |
|---|---|---|---|---|---|---|---|---|---|
| Egypt | 420 | 82.462 | 48 | 0.052 | 0.979 | 0.972 | 0.031 | ✓ | ✓ |
| Ethiopia | 363 | 93.341 | 48 | 0.060 | 0.966 | 0.953 | 0.037 | ✓ | ✓ |
| Finland | 998 | 175.201 | 48 | 0.061 | 0.972 | 0.961 | 0.032 | ✓ | ✓ |
| France | 2,007 | 234.774 | 48 | 0.053 | 0.978 | 0.970 | 0.024 | ✓ | ✓ |
| Georgia | 493 | 161.355 | 48 | 0.081 | 0.958 | 0.942 | 0.033 | | |
| Germany | 8,014 | 1325.609 | 48 | 0.068 | 0.965 | 0.952 | 0.031 | ✓ | ✓ |
| Ghana | 474 | 106.804 | 48 | 0.061 | 0.969 | 0.958 | 0.031 | ✓ | ✓ |
| Greece | 1,436 | 329.701 | 48 | 0.075 | 0.958 | 0.942 | 0.034 | ✓ | |
| Hong Kong | 594 | 214.792 | 48 | 0.088 | 0.938 | 0.915 | 0.045 | | |
| Hungary | 498 | 118.250 | 48 | 0.063 | 0.971 | 0.960 | 0.031 | ✓ | ✓ |
| India | 473 | 76.362 | 48 | 0.046 | 0.982 | 0.976 | 0.033 | ✓ | ✓ |
| Indonesia | 2,056 | 353.495 | 48 | 0.070 | 0.967 | 0.955 | 0.028 | ✓ | ✓ |
| Ireland | 498 | 108.711 | 48 | 0.059 | 0.966 | 0.953 | 0.036 | ✓ | ✓ |
| Israel | 1,010 | 221.686 | 48 | 0.068 | 0.955 | 0.939 | 0.035 | ✓ | |
| Italy | 1,505 | 283.590 | 48 | 0.063 | 0.972 | 0.961 | 0.025 | ✓ | ✓ |
| Japan | 1,000 | 213.048 | 48 | 0.069 | 0.956 | 0.940 | 0.035 | ✓ | |
| Kazakhstan | 513 | 170.510 | 48 | 0.078 | 0.952 | 0.934 | 0.037 | ✓ | |
| Kenya | 456 | 128.709 | 48 | 0.070 | 0.956 | 0.940 | 0.039 | ✓ | |
| Malaysia | 985 | 212.824 | 48 | 0.070 | 0.968 | 0.957 | 0.029 | ✓ | ✓ |
| Mexico | 498 | 180.277 | 48 | 0.087 | 0.944 | 0.923 | 0.039 | | |
| Morocco | 375 | 121.587 | 48 | 0.072 | 0.956 | 0.939 | 0.037 | ✓ | |
| Netherlands | 1,410 | 249.841 | 48 | 0.065 | 0.961 | 0.947 | 0.038 | ✓ | |
| New Zealand | 2,009 | 258.454 | 48 | 0.055 | 0.977 | 0.969 | 0.029 | ✓ | ✓ |

| Country | $N$ | $\chi^2$ | $df$ | RMSEA | CFI | TLI | SRMR | Criterion 1 | Criterion 2 |
|---|---|---|---|---|---|---|---|---|---|
| Nicaragua | 399 | 191.603 | 48 | 0.099 | 0.938 | 0.915 | 0.042 | | |
| Nigeria | 988 | 181.339 | 48 | 0.068 | 0.962 | 0.948 | 0.033 | ✓ | |
| Norway | 494 | 124.830 | 48 | 0.065 | 0.968 | 0.956 | 0.034 | ✓ | ✓ |
| Peru | 496 | 160.096 | 48 | 0.077 | 0.953 | 0.935 | 0.036 | ✓ | |
| Philippines | 545 | 105.496 | 48 | 0.057 | 0.975 | 0.965 | 0.030 | ✓ | ✓ |
| Poland | 3,002 | 647.662 | 48 | 0.080 | 0.954 | 0.937 | 0.034 | ✓ | |
| Portugal | 499 | 232.203 | 48 | 0.100 | 0.928 | 0.901 | 0.052 | | |
| Romania | 432 | 112.968 | 48 | 0.064 | 0.966 | 0.953 | 0.034 | ✓ | ✓ |
| Russia | 1,503 | 437.731 | 48 | 0.082 | 0.951 | 0.933 | 0.032 | | |
| Serbia | 499 | 126.980 | 48 | 0.068 | 0.970 | 0.959 | 0.031 | ✓ | ✓ |
| Slovakia | 530 | 189.861 | 48 | 0.085 | 0.960 | 0.945 | 0.036 | | |
| Slovenia | 501 | 147.437 | 48 | 0.078 | 0.963 | 0.949 | 0.031 | ✓ | |
| South Africa | 1,000 | 181.801 | 48 | 0.063 | 0.972 | 0.962 | 0.029 | ✓ | ✓ |
| South Korea | 500 | 146.728 | 48 | 0.070 | 0.952 | 0.934 | 0.046 | ✓ | |
| Spain | 1,009 | 269.369 | 48 | 0.081 | 0.959 | 0.943 | 0.035 | | |
| Sweden | 1,004 | 194.990 | 48 | 0.065 | 0.965 | 0.952 | 0.033 | ✓ | ✓ |
| Switzerland | 997 | 203.333 | 48 | 0.064 | 0.968 | 0.956 | 0.029 | ✓ | ✓ |
| Taiwan | 1,204 | 200.172 | 48 | 0.061 | 0.974 | 0.964 | 0.027 | ✓ | ✓ |
| Türkiye | 500 | 125.676 | 48 | 0.069 | 0.971 | 0.960 | 0.032 | ✓ | ✓ |
| Uganda | 387 | 147.205 | 48 | 0.085 | 0.933 | 0.908 | 0.046 | | |
| Ukraine | 1,008 | 228.471 | 48 | 0.070 | 0.958 | 0.942 | 0.033 | ✓ | |
| United Kingdom | 1,983 | 394.987 | 48 | 0.071 | 0.965 | 0.952 | 0.031 | ✓ | ✓ |
| United States | 2,559 | 261.015 | 48 | 0.056 | 0.984 | 0.977 | 0.022 | ✓ | ✓ |

| Country | $N$ | $\chi^2$ | $df$ | RMSEA | CFI | TLI | SRMR | Criterion 1 | Criterion 2 |
|---|---|---|---|---|---|---|---|---|---|
| Uruguay | 312 | 121.886 | 48 | 0.075 | 0.962 | 0.948 | 0.034 | ✓ | |

*Note.* Path A2 represents the country-specific CFA without weights. Under this approach, the four-factor CFA was first fitted separately within each country, and only countries with acceptable model fit were retained for subsequent multigroup measurement invariance testing. Criterion 1 required CFI and TLI ≥ 0.90, RMSEA ≤ 0.08, and SRMR ≤ 0.10; 56 countries met this criterion. Criterion 2 required CFI and TLI ≥ 0.95, RMSEA ≤ 0.08, and SRMR ≤ 0.10; 34 countries met this criterion.

**Table S4. Measurement invariance of the perceived scientist trustworthiness scale based on the four-factor model**

| | $\chi^2$ | *df* | RMSEA | CFI | TLI | SRMR | ΔRMSEA | ΔCFI | ΔTLI | ΔSRMR | Accept |
|---|---|---|---|---|---|---|---|---|---|---|---|
| **Pooled CFA weighted** | | | | | | | | | | | |
| Configural model | 9,563.712 | 3,264 | 0.077 | 0.956 | 0.940 | 0.035 | | | | | |
| Metric model | 10,881.563 | 3,800 | 0.077 | 0.950 | 0.941 | 0.056 | 0.000 | -0.006 | 0.001 | 0.021 | ✓ |
| Scalar model | 15,401.763 | 4,336 | 0.090 | 0.921 | 0.918 | 0.068 | 0.013 | -0.029 | -0.023 | 0.012 | |
| Strict model | 22,184.416 | 5,140 | 0.106 | 0.870 | 0.887 | 0.074 | 0.016 | -0.051 | -0.031 | 0.006 | |
| **Pooled CFA unweighted** | | | | | | | | | | | |
| Configural model | 14,577.975 | 3,264 | 0.068 | 0.965 | 0.952 | 0.030 | | | | | |
| Metric model | 16,954.764 | 3,800 | 0.068 | 0.960 | 0.953 | 0.046 | 0.000 | -0.005 | 0.001 | 0.016 | ✓ |
| Scalar model | 27,600.484 | 4,336 | 0.083 | 0.932 | 0.930 | 0.058 | 0.015 | -0.028 | -0.023 | 0.012 | |
| Strict model | 39,762.100 | 5,140 | 0.095 | 0.895 | 0.908 | 0.063 | 0.012 | -0.037 | -0.022 | 0.005 | |
| **Weighted country-specific CFA (Criterion 1, *N* = 37)** | | | | | | | | | | | |
| Configural model | 5,619.107 | 1,776 | 0.064 | 0.970 | 0.958 | 0.029 | | | | | |
| Metric model | 6,587.223 | 2,064 | 0.064 | 0.965 | 0.958 | 0.048 | 0.000 | -0.005 | 0.000 | 0.019 | ✓ |
| Scalar model | 10,293.567 | 2,352 | 0.079 | 0.939 | 0.937 | 0.059 | 0.015 | -0.026 | -0.021 | 0.011 | |
| Strict model | 14,189.495 | 2,784 | 0.089 | 0.908 | 0.919 | 0.062 | 0.010 | -0.031 | -0.018 | 0.003 | |
| **Weighted country-specific CFA (Criterion 2, *N* = 24)** | | | | | | | | | | | |
| Configural model | 4,209.177 | 1,152 | 0.061 | 0.973 | 0.963 | 0.027 | | | | | |
| Metric model | 4,960.580 | 1,336 | 0.062 | 0.968 | 0.962 | 0.045 | 0.001 | -0.005 | -0.001 | 0.018 | ✓ |
| Scalar model | 7,665.523 | 1,520 | 0.074 | 0.948 | 0.945 | 0.055 | 0.012 | -0.020 | -0.017 | 0.010 | |
| Strict model | 10,258.105 | 1,796 | 0.082 | 0.924 | 0.933 | 0.057 | 0.008 | -0.024 | -0.012 | 0.002 | |

| | $\chi^2$ | *df* | RMSEA | CFI | TLI | SRMR | ΔRMSEA | ΔCFI | ΔTLI | ΔSRMR | Accept |
|---|---|---|---|---|---|---|---|---|---|---|---|
| **Unweighted country-specific (Criterion 1, *N* = 56)** | | | | | | | | | | | |
| Configural model | 12,018.100 | 2,688 | 0.066 | 0.968 | 0.956 | 0.029 | | | | | |
| Metric model | 14,047.054 | 3,128 | 0.065 | 0.963 | 0.956 | 0.045 | -0.001 | -0.005 | 0.000 | 0.017 | ✓ |
| Scalar model | 22,898.976 | 3,568 | 0.080 | 0.937 | 0.934 | 0.056 | 0.015 | -0.026 | -0.022 | 0.011 | |
| Strict model | 32,423.234 | 4,228 | 0.091 | 0.904 | 0.916 | 0.061 | 0.011 | -0.033 | -0.018 | 0.005 | |
| **Unweighted country-specific CFA (Criterion 2, *N* = 34)** | | | | | | | | | | | |
| Configural model | 7,381.569 | 1,632 | 0.063 | 0.972 | 0.961 | 0.027 | | | | | |
| Metric model | 8,659.705 | 1,896 | 0.062 | 0.967 | 0.961 | 0.043 | -0.001 | -0.005 | 0.000 | 0.016 | ✓ |
| Scalar model | 14,265.906 | 2,160 | 0.077 | 0.944 | 0.941 | 0.054 | 0.015 | -0.023 | -0.020 | 0.011 | |
| Strict model | 19,576.478 | 2,556 | 0.086 | 0.917 | 0.927 | 0.057 | 0.009 | -0.027 | -0.014 | 0.003 | |

*Note.* Two CFA strategies were used. In the pooled strategy, all countries were included simultaneously and measurement invariance was evaluated across the full set of countries using multi-group CFA. In the country-specific strategy, the four-factor CFA was first estimated separately within each country, and only countries meeting the prespecified CFA criteria were retained for subsequent multi-group invariance testing. The reported configural, metric, scalar, and strict fit statistics for the country-specific strategy were therefore obtained from the multi-group CFA models fitted to the retained countries. Based on the existing literature (Chen, 2007; Rutkowski & Svetina, 2014), we applied the following criteria: for metric invariance, the thresholds were |ΔCFI| < 0.02, |ΔRMSEA| < 0.03, and ΔSRMR < 0.03; for scalar or strict invariance, the thresholds were |ΔCFI| < 0.01, |ΔRMSEA| < 0.015, and ΔSRMR < 0.015.

## Measurement Alignment

Given that scalar and strict measurement invariance of the scale was not supported under any of the four CFA specifications, we first attempted to establish partial measurement invariance. However, the partial invariance model did not converge and failed to yield an acceptable solution. Therefore, to enable cross-group comparisons of the perceived scientist trustworthiness scale in a multi-country/region sample (12 items; four-factor structure: competence, integrity, benevolence, and openness), following Han's approach (Han, 2024), we further adopted measurement alignment to address cross-group non-equivalence in measurement parameters.

Measurement alignment was implemented based on six CFA configurational models:

(1) Pooled CFA with weights (see Fig. S1, Path B1);

(2) Pooled CFA without weights (see Fig. S1, Path B2);

(3) Country -specific CFA with weights (see Fig. S1, Path A1) retaining countries/regions that met Criterion 1 (37 countries/regions);

(4) Country -specific CFA with weights (see Fig. S1, Path A1) retaining countries/regions that met Criterion 2 (24 countries/regions);

(5) Country -specific CFA with weights (see Fig. S1, Path A2) retaining countries/regions that met Criterion 1 (56 countries/regions);

(6) Country -specific CFA with weights (see Fig. S1, Path A2) retaining countries/regions that met (34 countries/regions).

For each configurational model, we extracted item loadings (λ) and item intercepts (ν; corresponding to the "~1" term for continuous indicators) for each country/region from the *lavaan* parameter estimates, and constructed "group × item" parameter matrices separately for each factor (3 items per factor). We then conducted alignment optimization for each factor separately using the invariance. alignment function in the R package *sirt*, estimating aligned measurement parameters and comparable latent factor levels while allowing a subset of parameters to vary across groups.

Measurement alignment was evaluated using two complementary diagnostics: the alignment $R^2$ values for factor loadings and intercepts, and the proportion of parameters identified as noninvariant. Following Asparouhov and Muthén (2023), the proportion of noninvariant parameters was required to remain below 20%. For the alignment $R^2$ values, we adopted the criterion of $R^2 > .90$ proposed by Cao (2026); values below .80, particularly for intercepts, were considered indicative of poor alignment performance. An alignment solution was classified as acceptable only when both criteria were satisfied. Dimensions that failed either criterion were considered unsuccessfully aligned and were excluded from subsequent analyses. The loading and intercept tolerances used to identify noninvariant parameters, together with the corresponding $R^2$ values, are reported in Supplementary Tables S4–S9.

## Rationale for the alignment power specification

Because the success of alignment was evaluated primarily using the alignment indices $R^2$, we set `align.pow = c(2, 2)` for the loading and intercept loss functions. This choice followed implementation-specific guidance for the *sirt* alignment procedure, which—drawing on comments from Alexander Robitzsch—recommends the quadratic loss function when $R^2$ is used as the principal diagnostic of alignment quality. The default setting, `align.pow = c(.25, .25)`, applies a more robust loss function but may produce biased $R^2$ values under some conditions. Thus, `c(2, 2)` was selected to align the estimation criterion with the diagnostic used to evaluate alignment, rather than because it is uniformly preferable to the default specification.

Details of the measurement alignment procedure and results are reported below.

**Table S5. Measurement alignment for the perceived scientist trustworthiness scale: four-factor CFA model (aligned based on country-specific weighted CFA; Fig. S1, Path A1; 37 countries/regions)**

| Factor | Item | Mean | Joint item | N non-invariance | $R^2$ | Percentage of non-invariance (%) | Noninvariant Groups | Alignment status |
|---|---|---|---|---|---|---|---|---|
| **Loadings** | | | | | | | | |
| Competence | | | | | 0.993 | 0.00 | | Success |
| | TRUST _ SCI _ expert | 0.586 | 0.518 | 0 | | | | |
| | TRUST _ SCI _ intellig | 0.555 | 0.493 | 0 | | | | |
| | TRUST _ SCI _ qualified | 0.626 | 0.546 | 0 | | | | |
| Integrity | | | | | -62019.020 | 0.90 | | Failed |
| | TRUST _ SCI _ honest | 0.648 | 0.330 | 0 | | | | |
| | TRUST _ SCI _ ethical | 0.600 | 0.313 | 0 | | | | |
| | TRUST _ SCI _ sincere | 0.666 | 0.348 | 1 | | | Argentina | |
| Benevolence | | | | | 0.992 | 0.00 | | Success |
| | TRUST _ SCI _ concerned | 0.687 | 0.573 | 0 | | | | |
| | TRUST _ SCI _ improve | 0.682 | 0.574 | 0 | | | | |
| | TRUST _ SCI _ otherint | 0.689 | 0.580 | 0 | | | | |
| Openness | | | | | 0.993 | 0.00 | | Success |
| | TRUST _SCI _ open | 0.680 | 0.500 | 0 | | | | |
| | TRUST _SCI _ trans | 0.744 | 0.530 | 0 | | | | |
| | TRUST _SCI _ otherviews | 0.706 | 0.501 | 0 | | | | |
| **Intercepts** | | | | | | | | |
| Competence | | | | | 0.999 | 8.10 | | Success |

| Factor | Item | Mean | Joint item | N non-invariance | $R^2$ | Percentage of non-invariance (%) | Noninvariant Groups | Alignment status |
|---|---|---|---|---|---|---|---|---|
| | TRUST _ SCI _ expert | 3.881 | 4.044 | 4 | | | Côte d'Ivoire, Finland, Kazakhstan, Ukraine | |
| | TRUST _ SCI _ intellig | 4.205 | 4.346 | 3 | | | Finland, Georgia, Italy | |
| | TRUST _ SCI _ qualified | 4.038 | 4.228 | 2 | | | Italy, Taiwan | |
| Integrity | | | | | -0.496 | 5.40 | | Failed |
| | TRUST _ SCI _ honest | 3.582 | 3.524 | 3 | | | Côte d'Ivoire, Czech Republic, Egypt | |
| | TRUST _ SCI _ ethical | 3.582 | 3.467 | 2 | | | Argentina, Egypt | |
| | TRUST _ SCI _ sincere | 3.640 | 3.535 | 1 | | | Côte d'Ivoire | |
| Benevolence | | | | | 0.999 | 4.50 | | Success |
| | TRUST _ SCI _ concerned | 3.527 | 3.697 | 3 | | | Egypt, Hungary, Morocco | |
| | TRUST _ SCI _ improve | 3.724 | 3.863 | 0 | | | | |
| | TRUST _ SCI _ otherint | 3.445 | 3.574 | 2 | | | Egypt, Morocco, | |
| Openness | | | | | 0.999 | 7.20 | | Success |
| | TRUST _SCI _ open | 3.320 | 3.579 | 3 | | | Bangladesh, Côte d'Ivoire, Morocco | |
| | TRUST _SCI _ trans | 3.404 | 3.685 | 2 | | | Bangladesh, Taiwan | |
| | TRUST _SCI _ otherviews | 3.285 | 3.496 | 3 | | | Argentina, Côte d'Ivoire, Indonesia | |

*Note.* For integrity, the alignment solution produced a negative $R^2$ value of both factor-loading and intercept, indicating an unsatisfactory alignment solution and rendering the $R^2$ fit index difficult to interpret substantively. Cao (2026) noted that an intercept $R^2$ of .80 already suggests that approximate measurement invariance is very unlikely to hold and that a loading $R^2$ around .40 is a strong signal that loading invariance does not hold. Given the negative $R^2$ observed here, the alignment solution for integrity was classified as failed, and this factor was not used to construct aligned composite score.

**Table S6. Measurement alignment for the perceived scientist trustworthiness scale: four-factor CFA model (aligned based on country-specific weighted CFA; Fig. S1, Path A1; 24 countries/regions)**

| Factor | Item | Mean | Joint item | N non-invariance | $R^2$ | Percentage of non-invariance (%) | Noninvariant Groups | Alignment status |
|---|---|---|---|---|---|---|---|---|
| **Loadings** | | | | | | | | |
| Competence | | | | | 0.994 | 0.00 | | Success |
| | TRUST _ SCI _ expert | 0.591 | 0.558 | 0 | | | | |
| | TRUST _ SCI _ intellig | 0.564 | 0.529 | 0 | | | | |
| | TRUST _ SCI _ qualified | 0.618 | 0.582 | 0 | | | | |
| Integrity | | | | | -<br>103478<br>0.923 | 1.40 | | Failed |
| | TRUST _ SCI _ honest | 0.660 | 0.343 | 0 | | | | |
| | TRUST _ SCI _ ethical | 0.591 | 0.319 | 1 | | | Australia | |
| | TRUST _ SCI _ sincere | 0.648 | 0.353 | 0 | | | | |
| Benevolence | | | | | 0.996 | 0.00 | | Success |
| | TRUST _ SCI _ concerned | 0.689 | 0.622 | 0 | | | | |
| | TRUST _ SCI _ improve | 0.682 | 0.621 | 0 | | | | |
| | TRUST _ SCI _ otherint | 0.693 | 0.631 | 0 | | | | |
| Openness | | | | | 0.995 | 0.00 | | Success |
| | TRUST _SCI _ open | 0.696 | 0.706 | 0 | | | | |
| | TRUST _SCI _ trans | 0.737 | 0.739 | 0 | | | | |
| | TRUST _SCI _ otherviews | 0.705 | 0.689 | 0 | | | | |
| **Intercepts** | | | | | | | | |
| Competence | | | | | 0.999 | 8.30 | | Success |

| Factor | Item | Mean | Joint item | N non-invariance | $R^2$ | Percentage of non-invariance (%) | Noninvariant Groups | Alignment status |
|---|---|---|---|---|---|---|---|---|
| | TRUST _ SCI _ expert | 3.923 | 3.993 | 2 | | | Finland, Ukraine | |
| | TRUST _ SCI _ intellig | 4.221 | 4.290 | 2 | | | Finland, Italy | |
| | TRUST _ SCI _ qualified | 4.071 | 4.177 | 2 | | | Italy, Taiwan | |
| Integrity | | | | | -544.1635 | 8.30 | | Failed |
| | TRUST _ SCI _ honest | 3.618 | 3.631 | 3 | | | Australia, Czech Republic, Egypt | |
| | TRUST _ SCI _ ethical | 3.585 | 3.551 | 2 | | | Australia, Egypt | |
| | TRUST _ SCI _ sincere | 3.659 | 3.643 | 1 | | | Australia | |
| Benevolence | | | | | 1.000 | 6.90 | | Success |
| | TRUST _ SCI _ concerned | 3.543 | 3.822 | 4 | | | Czech Republic, Egypt, Hungary, Norway, | |
| | TRUST _ SCI _ improve | 3.750 | 3.989 | 0 | | | | |
| | TRUST _ SCI _ otherint | 3.443 | 3.691 | 1 | | | Egypt | |
| Openness | | | | | 1.000 | 1.40 | | Success |
| | TRUST _SCI _ open | 3.348 | 3.637 | 0 | | | | |
| | TRUST _SCI _ trans | 3.403 | 3.745 | 1 | | | Taiwan | |
| | TRUST _SCI _ otherviews | 3.258 | 3.541 | 0 | | | | |

*Note.* For integrity, the alignment solution produced a negative $R^2$ value of both factor-loading and intercept, indicating an unsatisfactory alignment solution and rendering the $R^2$ fit index difficult to interpret substantively. Cao (2026) noted that an intercept $R^2$ of .80 already suggests that approximate measurement invariance is very unlikely to hold and that a loading $R^2$ around .40 is a strong signal that loading invariance does not hold. Given the negative $R^2$ observed here, the alignment solution for integrity was classified as failed, and this factor was not used to construct aligned composite score.

**Table S7. Measurement alignment for the perceived scientist trustworthiness scale: four-factor CFA model (aligned based on country-specific unweighted CFA; Fig. S1, Path A2; 56 countries/regions)**

| Factor | Item | Mean | Joint item | N non-invariance | $R^2$ | Percentage of non-invariance (%) | Noninvariant Groups | Alignment status |
|---|---|---|---|---|---|---|---|---|
| **Loadings** | | | | | | | | |
| Competence | | | | | 0.992 | 0.00 | | Success |
| | TRUST _ SCI _ expert | 0.589 | 0.497 | 0 | | | | |
| | TRUST _ SCI _ intellig | 0.550 | 0.463 | 0 | | | | |
| | TRUST _ SCI _ qualified | 0.643 | 0.527 | 0 | | | | |
| Integrity | | | | | -45778.037 | 0.60 | | Failed |
| | TRUST _ SCI _ honest | 0.651 | 0.338 | 0 | | | | |
| | TRUST _ SCI _ ethical | 0.633 | 0.338 | 0 | | | | |
| | TRUST _ SCI _ sincere | 0.686 | 0.364 | 1 | | | Argentina | |
| Benevolence | | | | | 0.994 | 0.00 | | Success |
| | TRUST _ SCI _ concerned | 0.708 | 0.584 | 0 | | | | |
| | TRUST _ SCI _ improve | 0.704 | 0.581 | 0 | | | | |
| | TRUST _ SCI _ otherint | 0.712 | 0.587 | 0 | | | | |
| Openness | | | | | 0.996 | 0.00 | | Success |
| | TRUST _SCI _ open | 0.715 | 0.530 | 0 | | | | |
| | TRUST _SCI _ trans | 0.755 | 0.555 | 0 | | | | |
| | TRUST _SCI _ otherviews | 0.717 | 0.522 | 0 | | | | |
| **Intercepts** | | | | | | | | |
| Competence | | | | | 0.999 | 6.50 | | Success |

| Factor | Item | Mean | Joint item | N non-invariance | $R^2$ | Percentage of non-invariance (%) | Noninvariant Groups | Alignment status |
|---|---|---|---|---|---|---|---|---|
| | TRUST _ SCI _ expert | 3.867 | 4.042 | 6 | | | Bolivia, Côte d'Ivoire, Finland, Japan, Sweden, Ukraine | |
| | TRUST _ SCI _ intellig | 4.229 | 4.397 | 2 | | | Finland, Italy | |
| | TRUST _ SCI _ qualified | 4.029 | 4.246 | 3 | | | Italy, Japan, Taiwan | |
| Integrity | | | | | -39.825 | 3.00 | | Failed |
| | TRUST _ SCI _ honest | 3.586 | 3.451 | 3 | | | Argentina, Bolivia, Czech Republic | |
| | TRUST _ SCI _ ethical | 3.583 | 3.420 | 1 | | | Argentina | |
| | TRUST _ SCI _ sincere | 3.641 | 3.470 | 1 | | | Argentina | |
| Benevolence | | | | | 0.999 | 3.00 | | Success |
| | TRUST _ SCI _ concerned | 3.524 | 3.670 | 4 | | | Egypt, Hungary, Morocco, Türkiye | |
| | TRUST _ SCI _ improve | 3.708 | 3.837 | 0 | | | | |
| | TRUST _ SCI _ otherint | 3.447 | 3.558 | 1 | | | Morocco | |
| Openness | | | | | 0.999 | 2.40 | | Success |
| | TRUST _SCI _ open | 3.325 | 3.534 | 2 | | | Bangladesh, Morocco | |
| | TRUST _SCI _ trans | 3.422 | 3.638 | 0 | | | | |
| | TRUST _SCI _ otherviews | 3.294 | 3.491 | 2 | | | Argentina, India | |

*Note.* For integrity, the alignment solution produced a negative $R^2$ value of both factor-loading and intercept, indicating an unsatisfactory alignment solution and rendering the $R^2$ fit index difficult to interpret substantively. Cao (2026) noted that an intercept $R^2$ of .80 already suggests that approximate measurement invariance is very unlikely to hold and that a loading $R^2$ around .40 is a strong signal that loading invariance does not hold. Given the negative $R^2$ observed here, the alignment solution for integrity was classified as failed, and this factor was not used to construct aligned composite score.

**Table S8. Measurement alignment for the perceived scientist trustworthiness scale: four-factor CFA model (aligned based on country-specific unweighted CFA; Fig. S1, Path A2; 34 countries/regions)**

| Factor | Item | Mean | Joint item | N non-invariance | $R^2$ | Percentage of non-invariance (%) | Noninvariant Groups | Alignment status |
|---|---|---|---|---|---|---|---|---|
| **Loadings** | | | | | | | | |
| Competence | | | | | 0.992 | 0.00 | | Success |
| | TRUST _ SCI _ expert | 0.592 | 0.477 | 0 | | | | |
| | TRUST _ SCI _ intellig | 0.548 | 0.444 | 0 | | | | |
| | TRUST _ SCI _ qualified | 0.637 | 0.505 | 0 | | 0.00 | | |
| Integrity | | | | | -145073.500 | | | Failed |
| | TRUST _ SCI _ honest | 0.656 | 0.321 | 0 | | | | |
| | TRUST _ SCI _ ethical | 0.629 | 0.324 | 0 | | | | |
| | TRUST _ SCI _ sincere | 0.675 | 0.346 | 0 | | | | |
| Benevolence | | | | | 0.995 | 0.00 | | Success |
| | TRUST _ SCI _ concerned | 0.695 | 0.558 | 0 | | | | |
| | TRUST _ SCI _ improve | 0.691 | 0.552 | 0 | | | | |
| | TRUST _ SCI _ otherint | 0.721 | 0.575 | 0 | | | | |
| Openness | | | | | 0.997 | 0.00 | | Success |
| | TRUST _SCI _ open | 0.731 | 0.688 | 0 | | | | |
| | TRUST _SCI _ trans | 0.766 | 0.724 | 0 | | | | |
| | TRUST _SCI _ otherviews | 0.737 | 0.679 | 0 | | | | |
| **Intercepts** | | | | | | | | |
| Competence | | | | | 0.999 | 6.90 | | Success |

| Factor | Item | Mean | Joint item | N non-invariance | $R^2$ | Percentage of non-invariance (%) | Noninvariant Groups | Alignment status |
|---|---|---|---|---|---|---|---|---|
| | TRUST _ SCI _ expert | 3.926 | 4.010 | 3 | | | Côte d'Ivoire, Finland, Sweden | |
| | TRUST _ SCI _ intellig | 4.267 | 4.355 | 2 | | | Finland, Italy | |
| | TRUST _ SCI _ qualified | 4.082 | 4.215 | 2 | | | Italy, Taiwan | |
| Integrity | | | | | 0.993 | 3.90 | | Failed |
| | TRUST _ SCI _ honest | 3.670 | 3.581 | 2 | | | Australia, Czech Republic | |
| | TRUST _ SCI _ ethical | 3.651 | 3.542 | 1 | | | Australia | |
| | TRUST _ SCI _ sincere | 3.724 | 3.607 | 1 | | | Australia | |
| Benevolence | | | | | 0.999 | 2.90 | | Success |
| | TRUST _ SCI _ concerned | 3.603 | 3.806 | 3 | | | Egypt, Norway, Türkiye | |
| | TRUST _ SCI _ improve | 3.786 | 3.978 | 0 | | | | |
| | TRUST _ SCI _ otherint | 3.539 | 3.712 | 0 | | | | |
| Openness | | | | | 0.999 | 2.90 | | Success |
| | TRUST _SCI _ open | 3.404 | 3.658 | 2 | | | Bangladesh, Hungary | |
| | TRUST _SCI _ trans | 3.503 | 3.775 | 0 | | | | |
| | TRUST _SCI _ otherviews | 3.386 | 3.642 | 1 | | | Bulgaria | |

*Note.* For integrity, the alignment solution produced a negative factor-loading $R^2$ value, indicating an unsatisfactory alignment solution and rendering the $R^2$ fit index difficult to interpret substantively. Cao (2026) noted that an intercept $R^2$ of .80 already suggests that approximate measurement invariance is very unlikely to hold and that a loading $R^2$ around .40 is a strong signal that loading invariance does not hold. Given the negative factor-loading $R^2$ observed here, the alignment solution for integrity was classified as failed, and this factor was not used to construct aligned composite score.

**Table S9. Measurement alignment for the perceived scientist trustworthiness scale: four-factor CFA model (aligned based on pooled weighted CFA; Fig. S1, Path B1; 68 countries/regions)**

| Factor | Item | Mean | Joint item | N non-invariance | $R^2$ | Percentage of non-invariance (%) | Noninvariant Groups | Alignment status |
|---|---|---|---|---|---|---|---|---|
| **Loadings** | | | | | | | | |
| Competence | | | | | -1314550000000 | 1.50 | | Fail |
| | TRUST _ SCI _ expert | 0.597 | 0.000 | 1 | | | Albania | |
| | TRUST _ SCI _ intellig | 0.541 | 0.000 | 1 | | | Albania | |
| | TRUST _ SCI _ qualified | 0.635 | 0.000 | 1 | | | Albania | |
| Integrity | | | | | -9220812294 | 1.00 | | Fail |
| | TRUST _ SCI _ honest | 0.651 | 0.000 | 0 | | | | |
| | TRUST _ SCI _ ethical | 0.621 | 0.000 | 1 | | | Albania | |
| | TRUST _ SCI _ sincere | 0.669 | 0.000 | 1 | | | Albania | |
| Benevolence | | | | | 0.981 | 1.00 | | Success |
| | TRUST _ SCI _ concerned | 0.699 | 0.295 | 0 | | | | |
| | TRUST _ SCI _ improve | 0.698 | 0.296 | 1 | | | Albania | |
| | TRUST _ SCI _ otherint | 0.722 | 0.294 | 1 | | | Albania | |
| Openness | | | | | 0.986 | 1.00 | | Success |
| | TRUST _SCI _ open | 0.696 | 0.273 | 1 | | | Albania | |
| | TRUST _SCI _ trans | 0.754 | 0.288 | 0 | | | | |

| Factor | Item | Mean | Joint item | N non-invariance | $R^2$ | Percentage of non-invariance (%) | Noninvariant Groups | Alignment status |
|---|---|---|---|---|---|---|---|---|
| | TRUST _SCI _ otherviews | 0.725 | 0.273 | 1 | | | Albania | |
| **Intercepts** | | | | | | | | |
| Competence | | | | | - 114169 40000 | 9.80 | | Failed |
| | TRUST _ SCI _ expert | 3.817 | 3.491 | 9 | | | Albania, Bolivia, Côte d'Ivoire, Finland, Ghana, Japan, Nicaragua, Portugal, Sweden | |
| | TRUST _ SCI _ intellig | 4.214 | 3.883 | 7 | | | Bolivia, China, Congo DR, Finland, Georgia, Ghana, Türkiye | |
| | TRUST _ SCI _ qualified | 4.010 | 3.677 | 4 | | | Albania, China, Italy, Taiwan, | |
| Integrity | | | | | - 149357 757 | 11.30 | | Failed |
| | TRUST _ SCI _ honest | 3.532 | 3.147 | 9 | | | Albania, Côte d'Ivoire, Congo DR, Czech Republic, Egypt, Ghana, Nicaragua, Philippines, Portugal | |
| | TRUST _ SCI _ ethical | 3.561 | 3.128 | 5 | | | Egypt, India, Japan, Morocco, Nicaragua | |
| | TRUST _ SCI _ sincere | 3.603 | 3.136 | 9 | | | Albania, Côte d'Ivoire, Egypt, Ghana, India, Kenya, South Korea, Philippines, Portugal | |
| Benevolence | | | | | 0.999 | 9.80 | | Success |
| | TRUST _ SCI _ concerned | 3.509 | 3.244 | 6 | | | Albania, Egypt, India, Morocco, Portugal, Türkiye | |
| | TRUST _ SCI _ improve | 3.709 | 3.426 | 6 | | | Albania, Ethiopia, India, Kenya, Portugal, Uganda, | |
| | TRUST _ SCI _ otherint | 3.414 | 3.132 | 8 | | | Albania, Egypt, Ethiopia, Ghana, India, Kenya, Morocco, Uganda | |

| Factor | Item | Mean | Joint item | N non-invariance | $R^2$ | Percentage of non-invariance (%) | Noninvariant Groups | Alignment status |
|---|---|---|---|---|---|---|---|---|
| Openness | | | | | 0.999 | 7.40 | | Success |
| | TRUST _SCI _ open | 3.307 | 3.031 | 5 | | | Bangladesh, Cyprus, Ethiopia, Greece, Portugal | |
| | TRUST _SCI _ trans | 3.410 | 3.119 | 6 | | | Argentina, Bangladesh, Congo DR, Egypt, Taiwan, Uganda, | |
| | TRUST _SCI _ otherviews | 3.270 | 2.974 | 4 | | | Albania, Argentina, China, Poland | |

*Note.* For Competence and Integrity, the alignment solution produced a negative $R^2$ value of both factor-loading and intercept, indicating an unsatisfactory alignment solution and rendering the $R^2$ fit index difficult to interpret substantively. Cao (2026) noted that an intercept $R^2$ of .80 already suggests that approximate measurement invariance is very unlikely to hold and that a loading $R^2$ around .40 is a strong signal that loading invariance does not hold. Given the negative $R^2$ observed here, the alignment solutions for integrity and competence were classified as failed, and these two factors were not used to construct aligned composite score.

**Table S10. Measurement alignment for the perceived scientist trustworthiness scale: four-factor CFA model (aligned based on unweighted CFA; Fig. S1, Path B2; 68 countries/regions)**

| Factor | Item | Mean | Joint item | N non-invariance | $R^2$ | Percentage of non-invariance (%) | Noninvariant Groups | Alignment status |
|---|---|---|---|---|---|---|---|---|
| **Loadings** | | | | | | | | |
| Competence | | | | | 0.992 | 0.00 | | Success |
| | TRUST _ SCI _ expert | 0.590 | 0.411 | 0 | | | | |
| | TRUST _ SCI _ intellig | 0.552 | 0.385 | 0 | | | | |
| | TRUST _ SCI _ qualified | 0.648 | 0.443 | 0 | | | | |
| Integrity | | | | | -62.099 | 0.00 | | Failed |
| | TRUST _ SCI _ honest | 0.659 | 0.275 | 0 | | | | |
| | TRUST _ SCI _ ethical | 0.644 | 0.276 | 0 | | | | |
| | TRUST _ SCI _ sincere | 0.697 | 0.297 | 0 | | | | |
| Benevolence | | | | | 0.994 | 0.00 | | Success |
| | TRUST _ SCI _ concerned | 0.718 | 0.556 | 0 | | | | |
| | TRUST _ SCI _ improve | 0.716 | 0.552 | 0 | | | | |
| | TRUST _ SCI _ otherint | 0.719 | 0.554 | 0 | | | | |
| Openness | | | | | 0.995 | 0.00 | | Success |
| | TRUST _SCI _ open | 0.711 | 0.534 | 0 | | | | |
| | TRUST _SCI _ trans | 0.761 | 0.563 | 0 | | | | |
| | TRUST _SCI _ otherviews | 0.721 | 0.528 | 0 | | | | |
| **Intercepts** | | | | | | | | |
| Competence | | | | | 0.999 | 6.40 | | Success |

| Factor | Item | Mean | Joint item | N non-invariance | $R^2$ | Percentage of non-invariance (%) | Noninvariant Groups | Alignment status |
|---|---|---|---|---|---|---|---|---|
| | TRUST _ SCI _ expert | 3.839 | 3.482 | 7 | | | Albania, Bolivia, Finland, Japan, Nicaragua, Sweden, Portugal | |
| | TRUST _ SCI _ intellig | 4.217 | 3.883 | 4 | | | China, Finland, Georgia, Türkiye, | |
| | TRUST _ SCI _ qualified | 4.011 | 3.632 | 2 | | | Italy, Uganda | |
| Integrity | | | | | 0.162 | 3.40 | | Failed |
| | TRUST _ SCI _ honest | 3.560 | 3.113 | 4 | | | Albania, Czech Republic, Nicaragua, Portugal, | |
| | TRUST _ SCI _ ethical | 3.573 | 3.098 | 1 | | | Albania, | |
| | TRUST _ SCI _ sincere | 3.610 | 3.112 | 2 | | | South Korea, Portugal, | |
| Benevolence | | | | | 0.999 | 2.50 | | Success |
| | TRUST _ SCI _ concerned | 3.510 | 3.093 | 4 | | | Albania, Egypt, Morocco, Portugal | |
| | TRUST _ SCI _ improve | 3.688 | 3.252 | 0 | | | | |
| | TRUST _ SCI _ otherint | 3.427 | 2.987 | 1 | | | Morocco | |
| Openness | | | | | 0.999 | 4.90 | | Success |
| | TRUST _SCI _ open | 3.310 | 2.982 | 5 | | | Congo DR, Greece, Morocco, Portugal, Cyprus, | |
| | TRUST _SCI _ trans | 3.408 | 3.065 | 2 | | | Bangladesh, Egypt, | |
| | TRUST _SCI _ otherviews | 3.282 | 2.931 | 3 | | | Argentina, Indonesia, India | |

*Note.* For integrity, the alignment solution produced a negative factor-loading $R^2$ value, indicating an unsatisfactory alignment solution and rendering the $R^2$ fit index difficult to interpret substantively. Cao (2026) noted that an intercept $R^2$ of .80 already suggests that approximate measurement invariance is very unlikely to hold and that a loading $R^2$ around .40 is a strong signal that loading invariance does not hold. Given the negative factor-loading $R^2$ observed here, the alignment solution for integrity was classified as failed, and this factor was not used to construct aligned composite score.

## Cross-Country/Region Differences in the Ranking of Perceived Scientist Trustworthiness

Based on the aligned scores from factors that could be successfully aligned, we averaged factor scores to create an overall perceived scientist trustworthiness score (i.e., overall trustworthiness of scientists). At the individual level, the overall score was computed as the mean of that respondent's aligned factor scores. At the country/region level, scores were obtained by averaging individuals' overall scores within each country/region, and both weighted and unweighted country-level means were reported below.

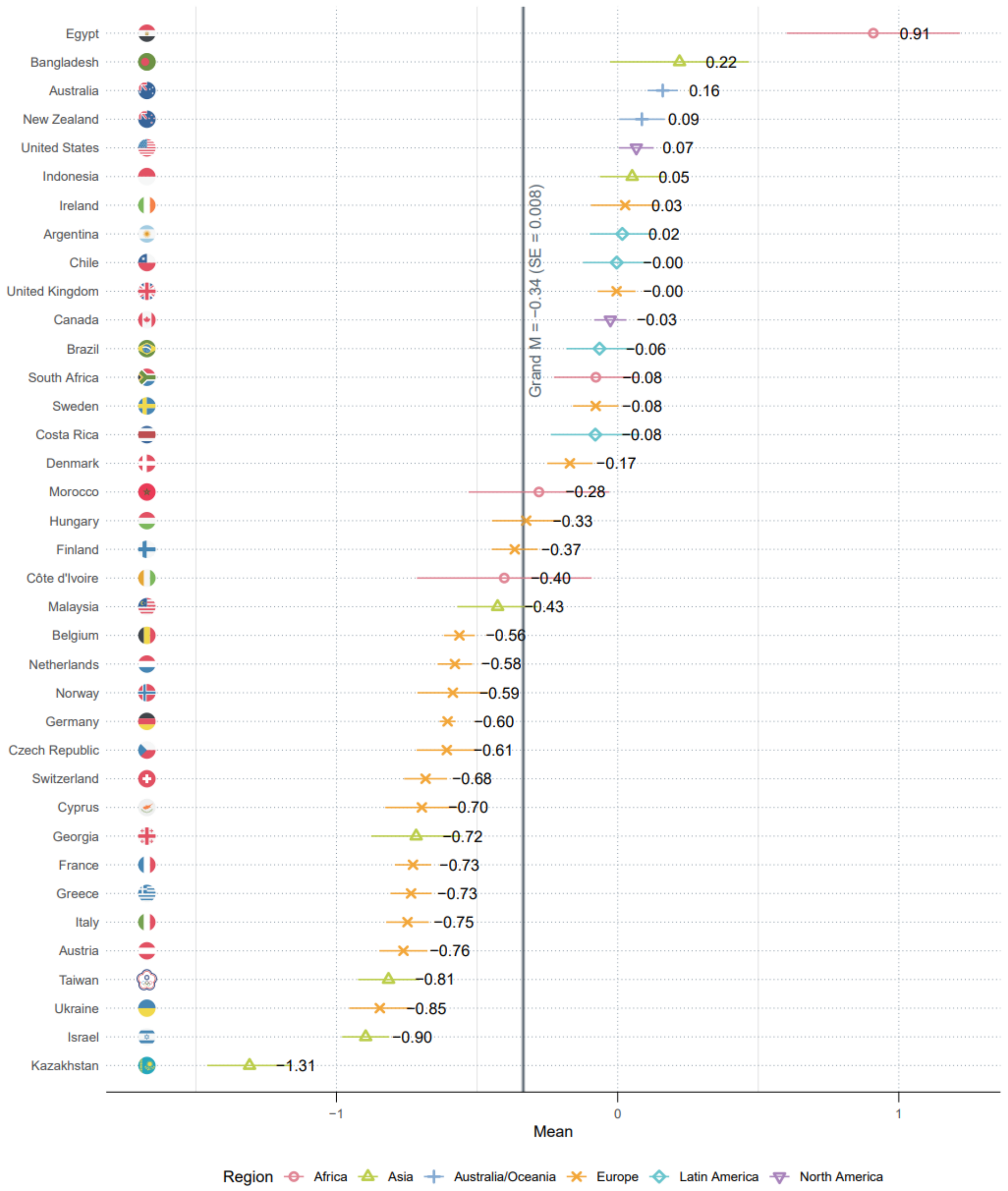

**Figure S2. Weighted means for perceived scientist trustworthiness across 37 countries/regions (aligned scores based on Fig. S1, Path A1)**

*Note.* Total $n$ = 49,959. Country ns range between 375 and 8,009. All statistics are computed using the aligned scores based on Path A1: country-specific weighted CFA (see Fig.1 Path A1). The vertical line denotes the weighted grand mean (based on aligned scores). The horizontal lines indicate means ± standard errors. Country-level standard errors range between 0.077 and 0.157.

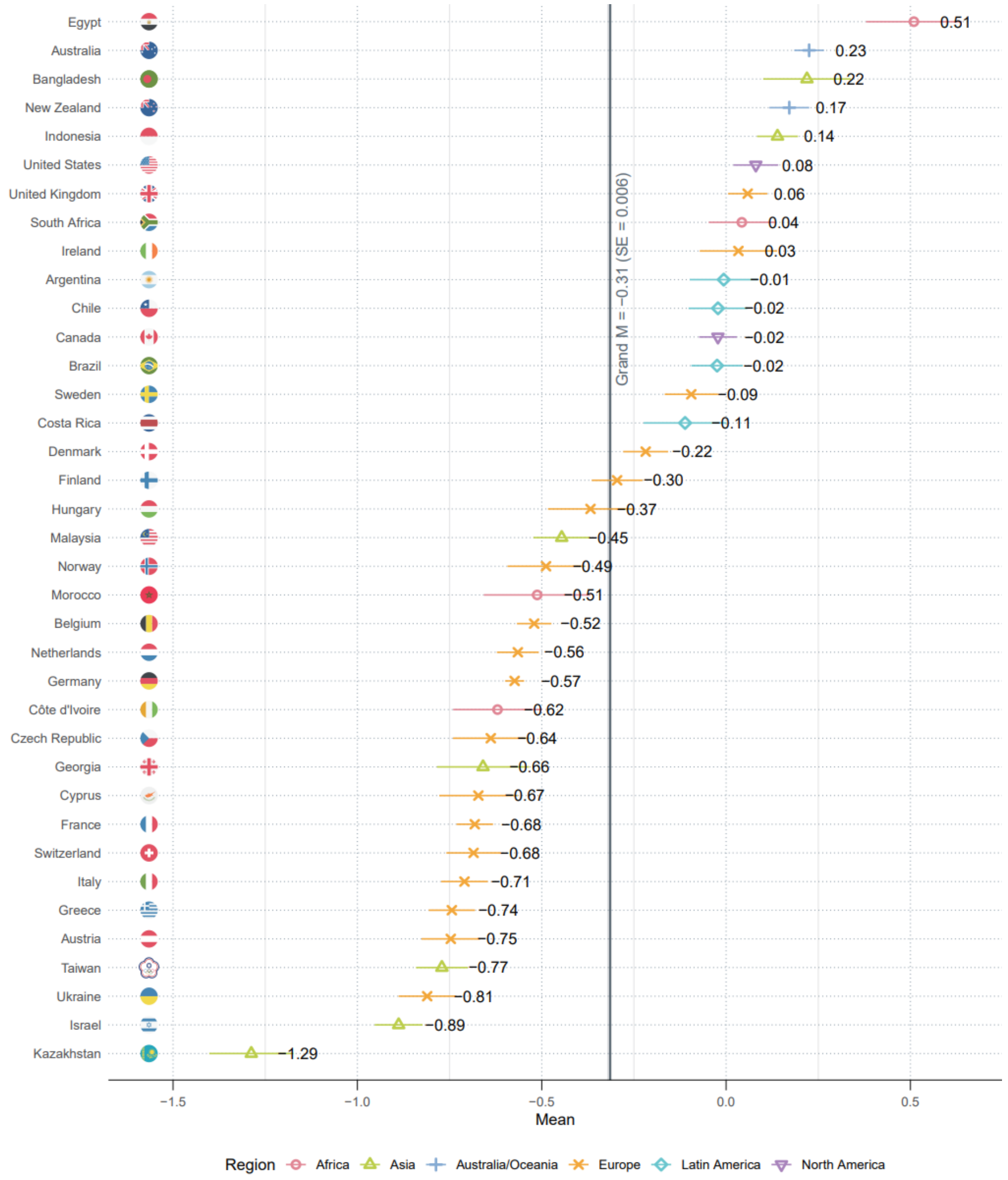

**Figure S3. Unweighted means for perceived scientist trustworthiness across 37 countries/regions (aligned scores based on Fig. S1, Path A1)**

*Note.* Total *n* = 49,959. Country ns range between 375 and 8,009. All statistics are computed using the aligned scores based on Path A1: country-specific weighted CFA (see Fig.1 Path A1). The vertical line denotes the weighted grand mean (based on aligned scores). The horizontal lines indicate means ± standard errors. Country-level standard errors range between 0.013 and 0.074.

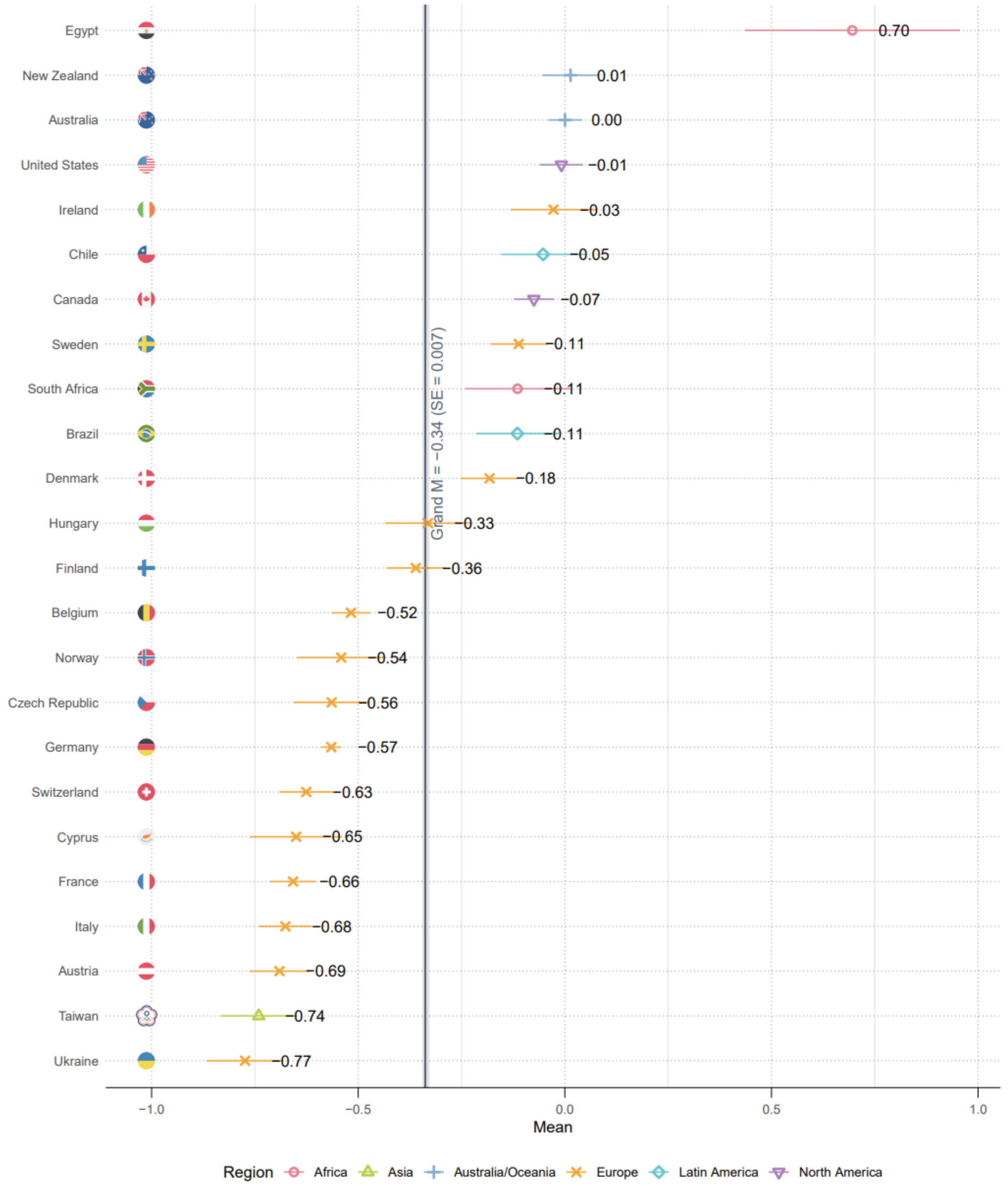

**Figure S4. Weighted means for perceived scientist trustworthiness across 24 countries/regions (aligned scores based on Fig. S1, Path A1)**

*Note.* Total $n$ = 37,715. Country ns range between 420 and 8,009. All statistics are computed using the aligned scores based on Path A1: country-specific weighted CFA (see Fig.1 Path A1). The vertical line denotes the weighted grand mean (based on aligned scores). The horizontal lines indicate means ± standard errors. Country-level standard errors range between 0.047 and 0.133.

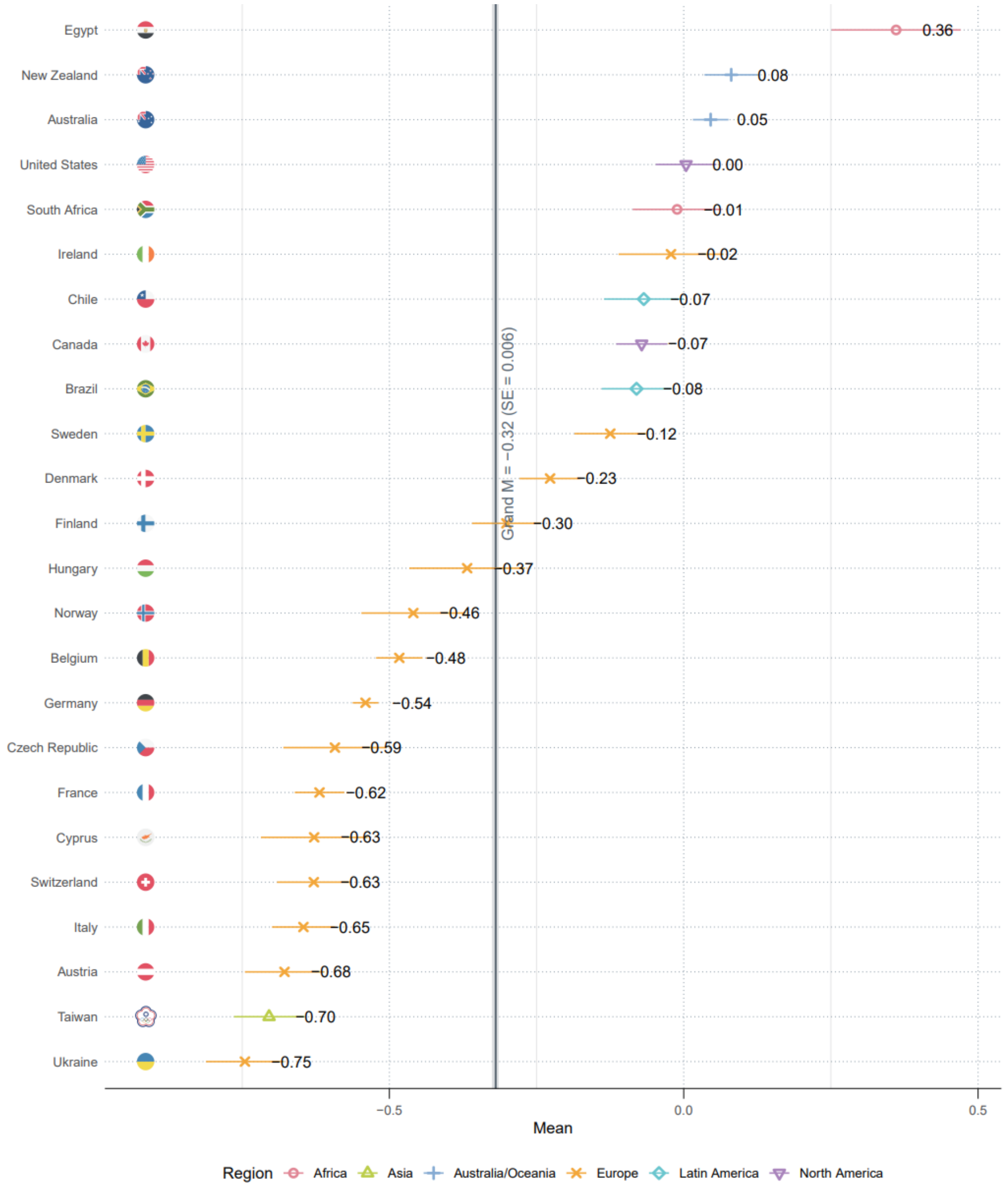

**Figure S5. Unweighted means for perceived scientist trustworthiness across 24 countries/regions (aligned scores based on Fig. S1, Path A1)**

*Note.* Total $n$ = 37,715. Country ns range between 420 and 8,009. All statistics are computed using the aligned scores based on Path A1: country-specific weighted CFA (see Fig.1 Path A1). The vertical line denotes the weighted grand mean (based on aligned scores). The horizontal lines indicate means ± standard errors. Country-level standard errors range between 0.011 and 0.056.

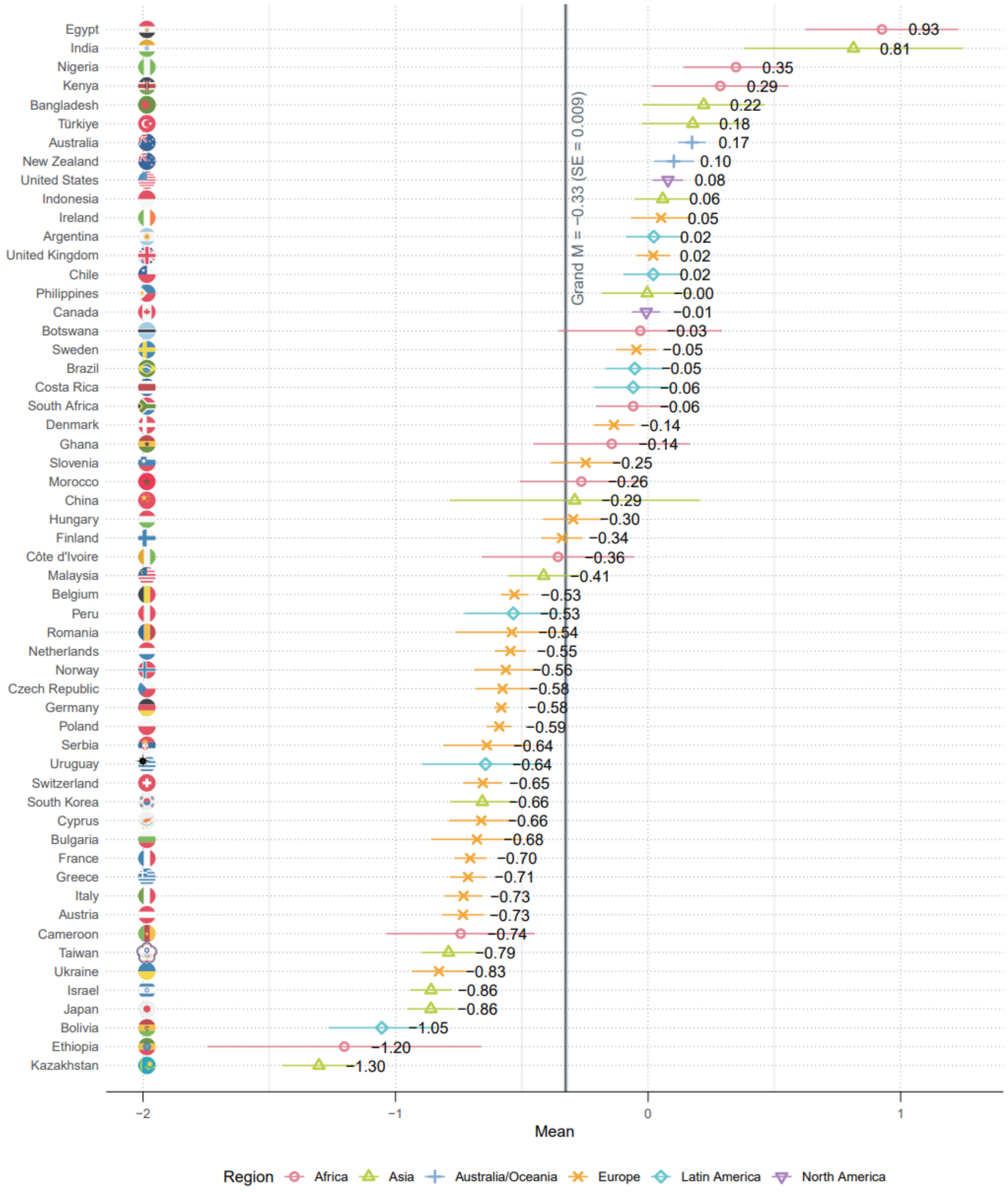

**Figure S6. Weighted means for perceived scientist trustworthiness across 56 countries/regions (aligned scores based on Fig. S1, Path A2)**

*Note.* N = 62,362. Country Ns range between 311 and 8,009. All statistics are computed using the aligned scores based on Path A2: country-specific unweighted CFA (see Fig.1 Path A2). The vertical line denotes the weighted grand mean (based on aligned scores). The horizontal lines indicate means ± standard errors. Country-level standard errors range between 0.014 and 0.277.

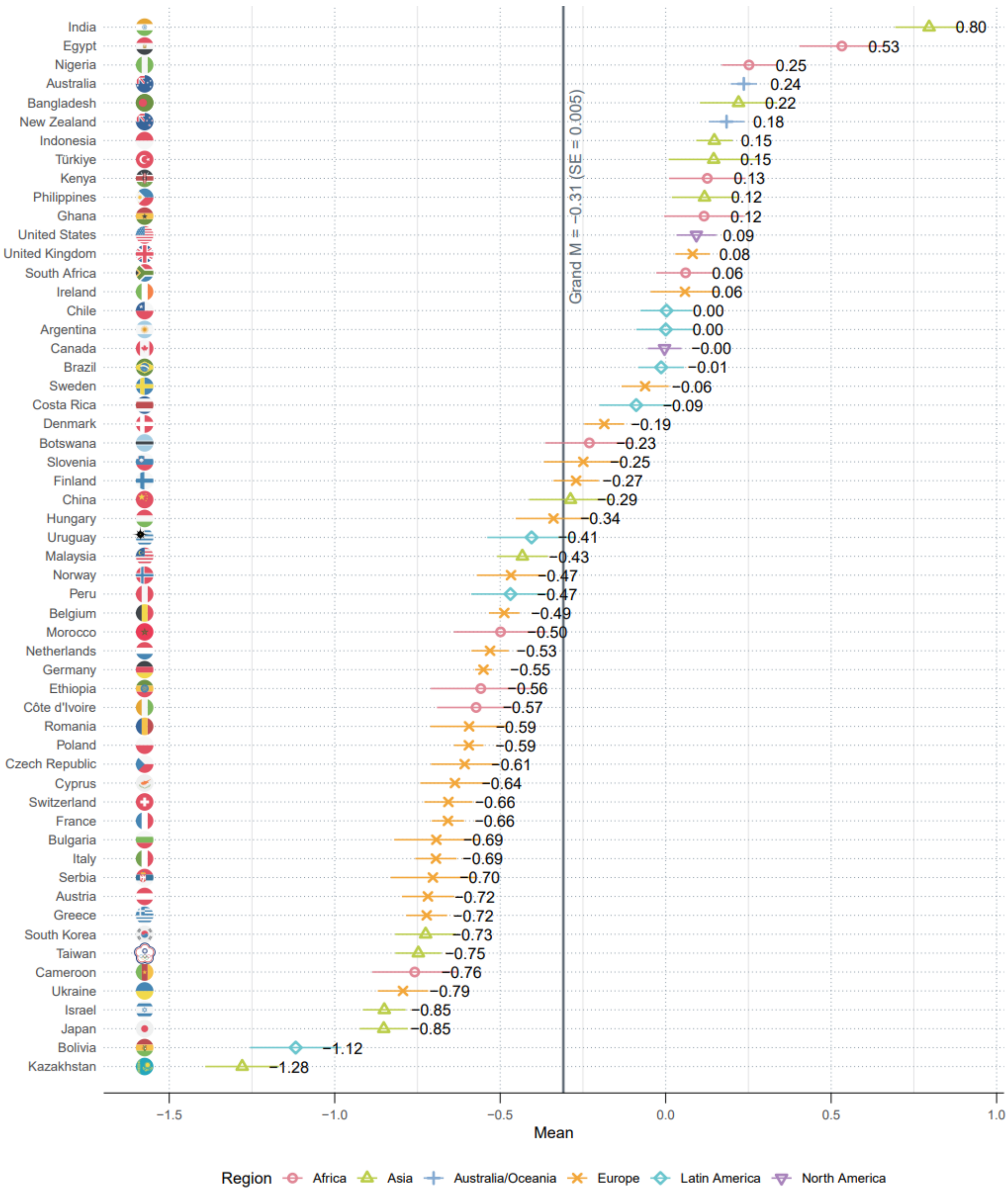

**Figure S7. Unweighted means for perceived scientist trustworthiness across 56 countries/regions (aligned scores based on Fig. S1, Path A2)**

*Note.* N = 62,362. Country Ns range between 311 and 8,009. All statistics are computed using the aligned scores based on Path A2: country-specific unweighted CFA (see Fig.1 Path A2). The vertical line denotes the weighted grand mean (based on aligned scores). The horizontal lines indicate means ± standard errors. Country-level standard errors range between 0.013 and 0.077.

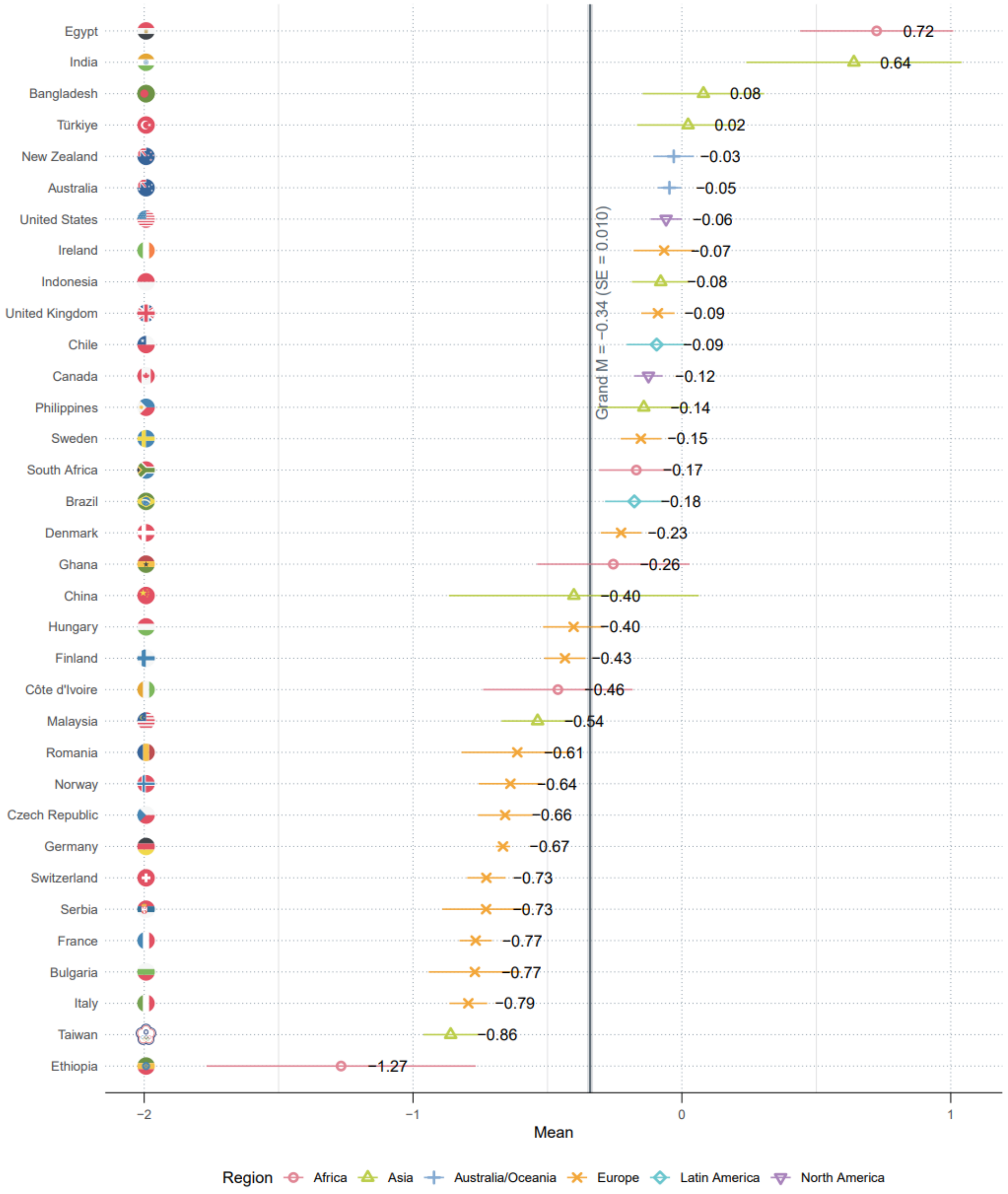

**Figure S8. Weighted means for perceived scientist trustworthiness across 34 countries/regions (aligned scores based on Fig. S1, Path A2)**

*Note.* N = 43,393. Country Ns range between 363 and 8,009. All statistics are computed using the aligned scores based on Path A2: country-specific unweighted CFA (see Fig.1 Path A2). The vertical line denotes the weighted grand mean (based on aligned scores). The horizontal lines indicate means ± standard errors. Country-level standard errors range between 0.014 and 0.255.

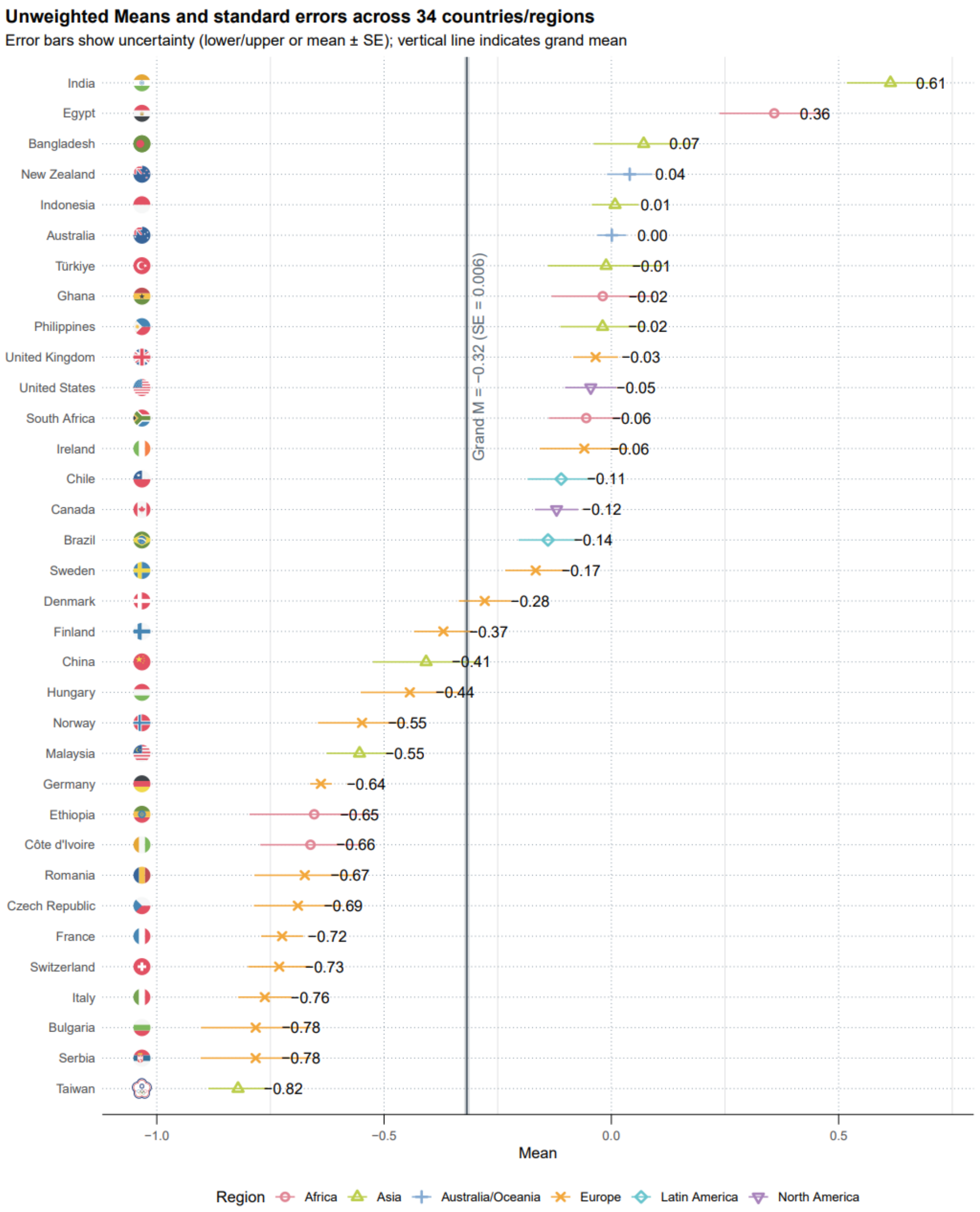

**Figure S9. Unweighted means for perceived scientist trustworthiness across 34 countries/regions (aligned scores based on Fig. S1, Path A2)**

*Note.* N = 43,393. Country Ns range between 363 and 8,009. All statistics are computed using the aligned scores based on Path A2: country-specific unweighted CFA (see Fig.1 Path A2). The vertical line denotes the weighted grand mean (based on aligned scores). The horizontal lines indicate means ± standard errors. Country-level standard errors range between 0.012 and 0.073.

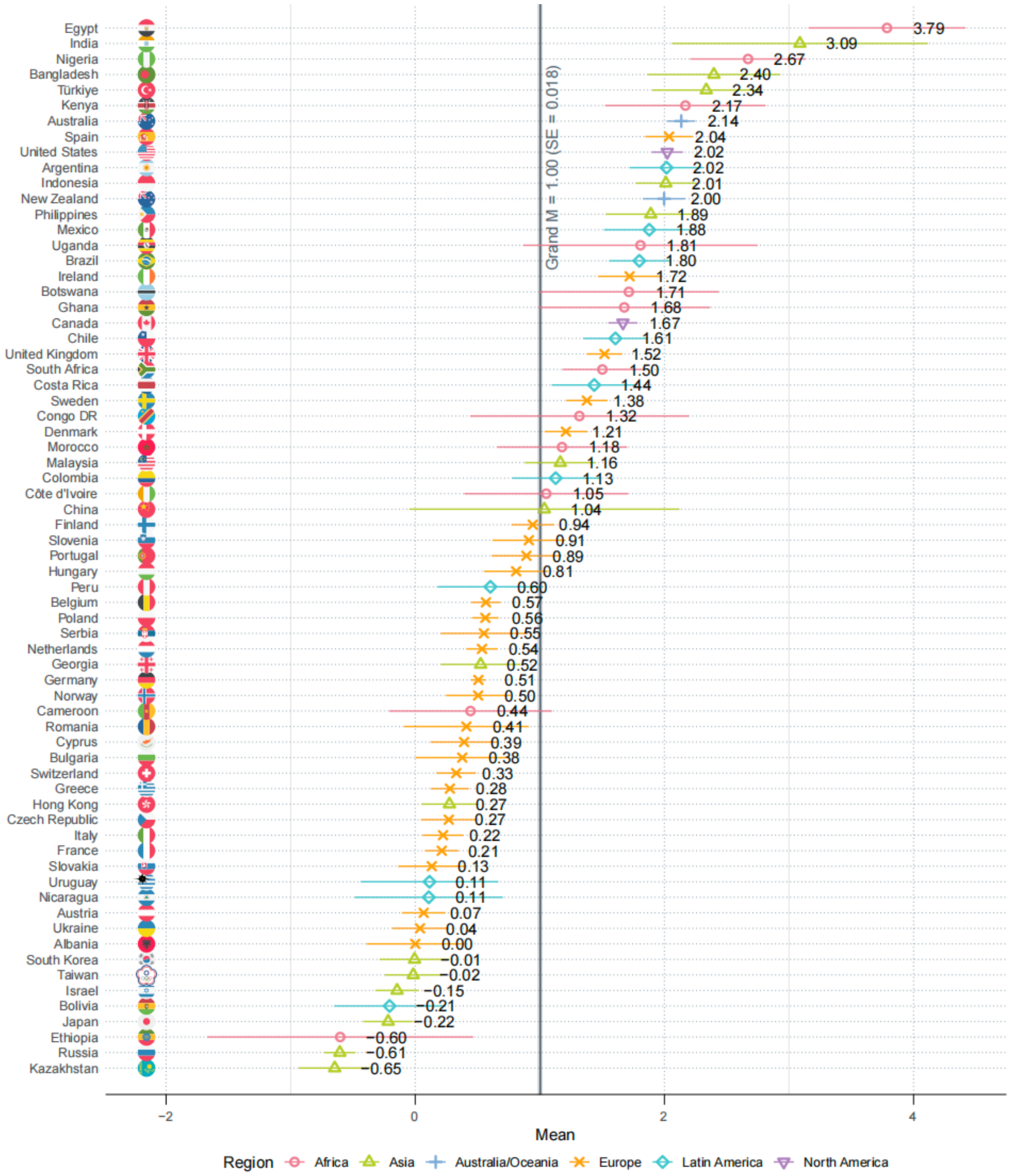

**Figure S10. Weighted means for perceived scientist trustworthiness across countries and regions (aligned scores derived via workflow Path B1)**

*Note.* Total $n$ = 69,495. Country $n$s range between 310 and 8,009. All statistics are computed using the aligned scores based on Path B1: pooled weighted CFA (see Fig.1 Path B1). The vertical line denotes the weighted grand mean (based on aligned scores). The horizontal lines indicate means ± standard errors. Country-level standard errors range between 0.027 and 0.169.

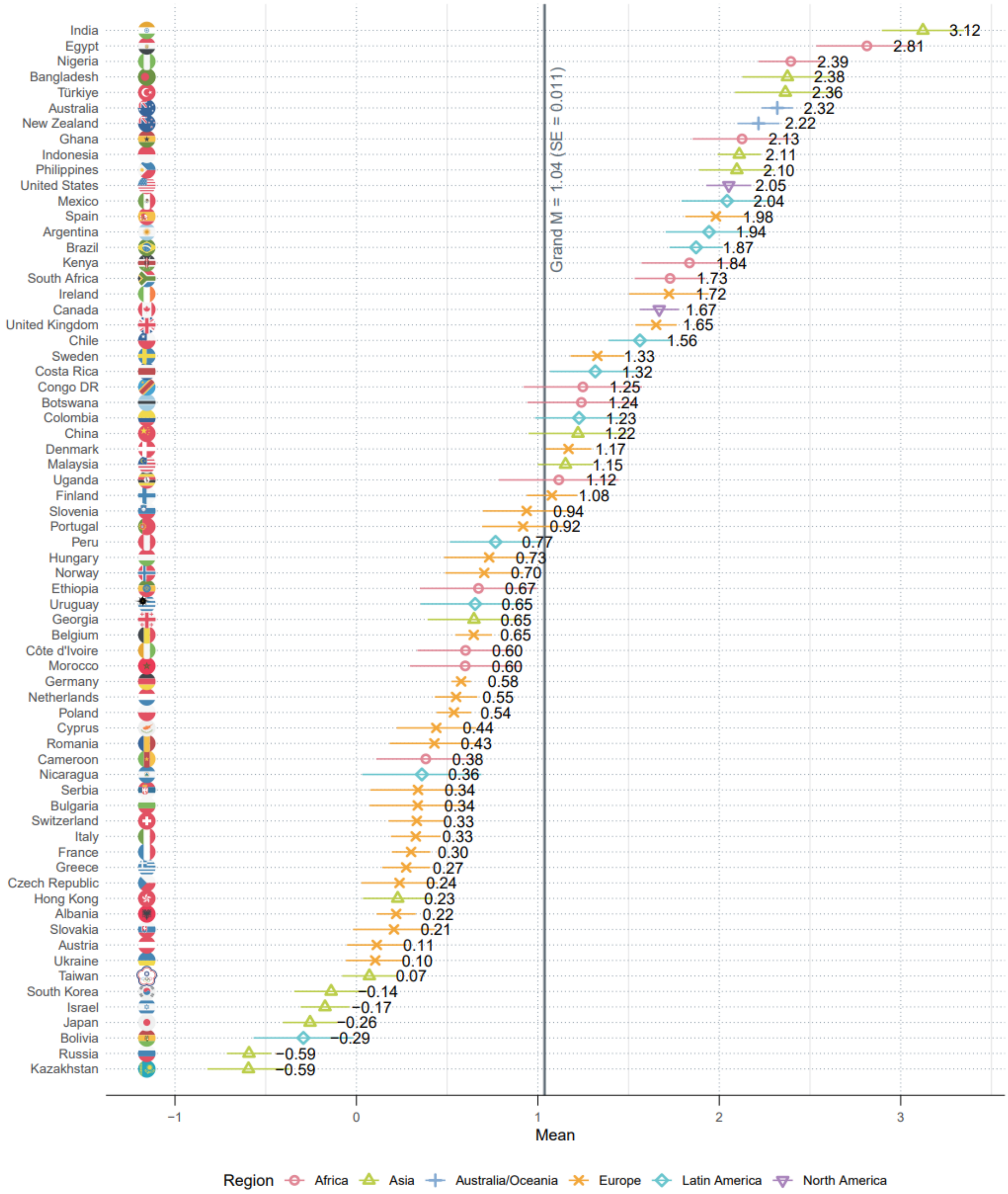

**Figure S11. Unweighted means for perceived scientist trustworthiness across 68 countries /regions (aligned scores based on Fig. S1, Path B1)**

*Note.* Total $n$ = 69,495. Country ns range between 310 and 8,009. All statistics are computed using the aligned scores based on Path B1: pooled weighted CFA (see Fig.1 Path B1). The vertical line denotes the weighted grand mean (based on aligned scores). The horizontal lines indicate means ± standard errors. Country-level standard errors range between 0.027 and 0.169.

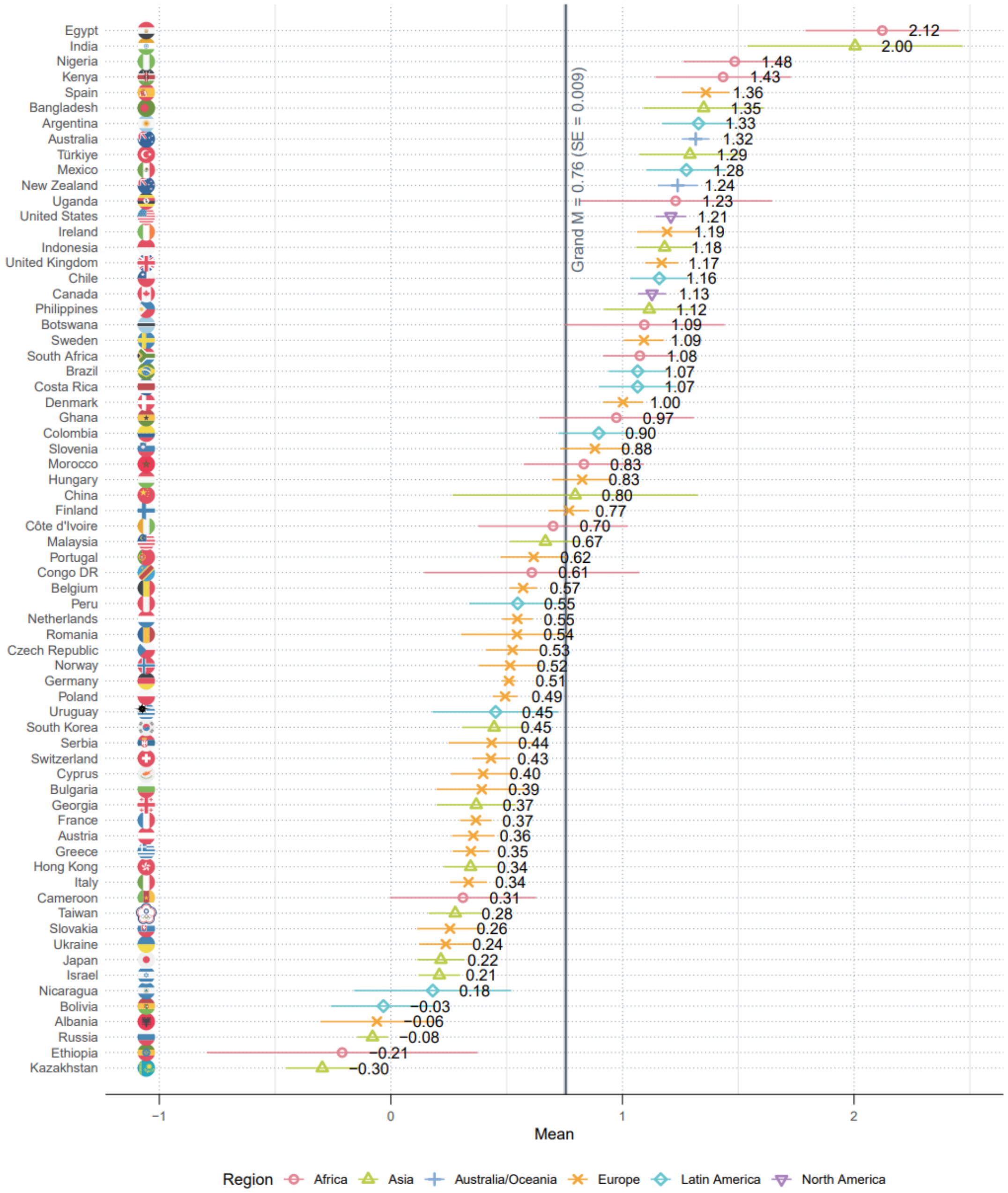

**Figure S12. Weighted means for perceived scientist trustworthiness across 68 countries/regions (aligned scores based on Fig. S1, Path B2)**

*Note.* Total $n$ = 69,503. Country ns range between 310 and 8,009. All statistics are computed using the aligned scores based on Path B2: pooled unweighted CFA (see Fig.1 Path B2). The vertical line denotes the weighted grand mean (based on aligned scores). The horizontal lines indicate means ± standard errors. Country-level standard errors range between 0.016 and 0.298.

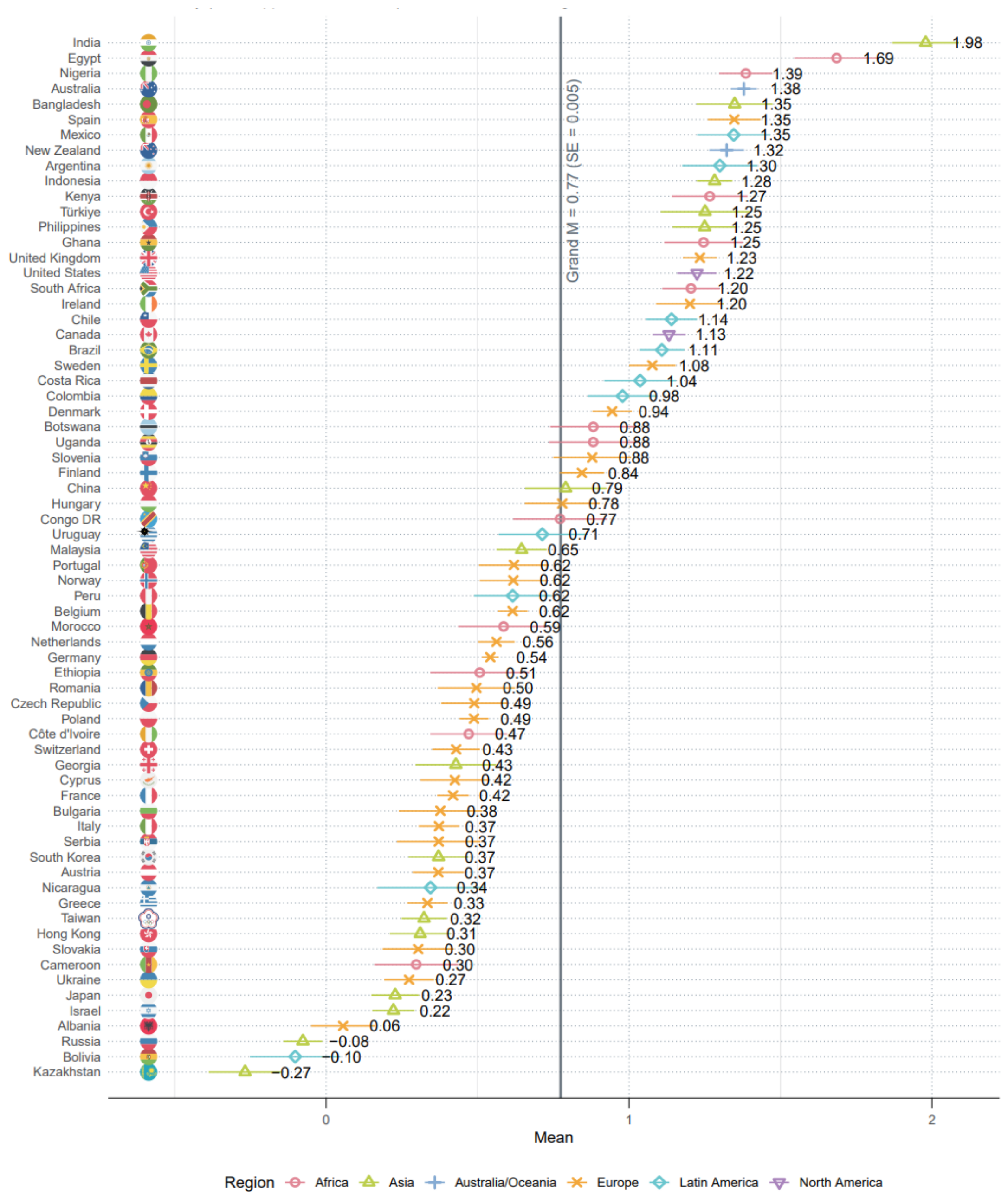

**Figure S13. Unweighted means for perceived scientist trustworthiness across 68 countries/regions (aligned scores based on Fig. S1, Path B2)**

*Note.* Total $n$ = 69,503. Country ns range between 310 and 8,009. All statistics are computed using the aligned scores based on Path B2: pooled unweighted CFA (see Fig.1 Path B2). The vertical line denotes the weighted grand mean (based on aligned scores). The horizontal lines indicate means ± standard errors. Country-level standard errors range between 0.014 and 0.089.

**Table S11. Changes in 37 country/region rankings of perceived scientist trustworthiness after measurement alignment (aligned scores based on Fig. S1, Path A1)**

| Country/Region | Rank | Alignment rank (Weight mean) | Alignment rank (Unweight mean) | Rank change (Weight mean) | Rank change (Unweight mean) |
|---|---|---|---|---|---|
| Egypt | 1 | 1 | 1 | 0 | 0 |
| Australia | 2 | 3 | 2 | -1 | 0 |
| Bangladesh | 3 | 2 | 3 | 1 | 0 |
| New Zealand | 4 | 4 | 4 | 0 | 0 |
| Argentina | 5 | 8 | 10 | -3 | -5 |
| United States | 6 | 5 | 6 | 1 | 0 |
| Indonesia | 7 | 6 | 5 | 1 | 2 |
| Ireland | 8 | 7 | 9 | 1 | -1 |
| United Kingdom | 9 | 10 | 7 | -1 | 2 |
| Canada | 10 | 11 | 12 | -1 | -2 |
| Chile | 11 | 9 | 11 | 2 | 0 |
| Sweden | 12 | 14 | 14 | -2 | -2 |
| Brazil | 13 | 12 | 13 | 1 | 0 |
| Denmark | 14 | 16 | 16 | -2 | -2 |
| South Africa | 15 | 13 | 8 | 2 | 7 |
| Costa Rica | 16 | 15 | 15 | 1 | 1 |
| Morocco | 17 | 17 | 21 | 0 | -4 |
| Finland | 18 | 19 | 17 | -1 | 1 |
| Hungary | 19 | 18 | 18 | 1 | 1 |
| Côte d'Ivoire | 20 | 20 | 25 | 0 | -5 |
| Malaysia | 21 | 21 | 19 | 0 | 2 |
| Belgium | 22 | 22 | 22 | 0 | 0 |
| Norway | 23 | 24 | 20 | -1 | 3 |
| Netherlands | 24 | 23 | 23 | 1 | 1 |
| Germany | 25 | 25 | 24 | 0 | 1 |
| Georgia | 26 | 29 | 27 | -3 | -1 |
| Switzerland | 27 | 27 | 30 | 0 | -3 |
| Czech Republic | 28 | 26 | 26 | 2 | 2 |
| France | 29 | 30 | 29 | -1 | 0 |
| Cyprus | 30 | 28 | 28 | 2 | 2 |
| Austria | 31 | 33 | 33 | -2 | -2 |
| Greece | 32 | 31 | 32 | 1 | 0 |
| Italy | 33 | 32 | 31 | 1 | 2 |
| Ukraine | 34 | 35 | 35 | -1 | -1 |
| Taiwan | 35 | 34 | 34 | 1 | 1 |
| Israel | 36 | 36 | 36 | 0 | 0 |
| Kazakhstan | 37 | 37 | 37 | 0 | 0 |

*Note.* The aligned country rankings were computed from the aligned scores based on Path A1: country-specific weighted CFA (see Fig. S1, Path A1). The ranking change score was computed as the original ranking reported in the main text minus the aligned ranking; positive values indicate an improvement in rank, whereas negative values indicate a decline in rank. The rank refers to these 37 countries' relative ranking as reported in the original article.

**Table S12. Changes in 24 country/region rankings of perceived scientist trustworthiness after measurement alignment (aligned scores based on Fig. S1, Path A1)**

| Country/Region | Rank | Alignment rank (Weight mean) | Alignment rank (Unweight mean) | Rank change (Weight mean) | Rank change (Unweight mean) |
|---|---|---|---|---|---|
| Egypt | 1 | 1 | 1 | 0 | 0 |
| Australia | 2 | 3 | 3 | -1 | -1 |
| New Zealand | 3 | 2 | 2 | 1 | 1 |
| United States | 4 | 4 | 4 | 0 | 0 |
| Ireland | 5 | 5 | 6 | 0 | -1 |
| Canada | 6 | 7 | 8 | -1 | -2 |
| Chile | 7 | 6 | 7 | 1 | 0 |
| Sweden | 8 | 8 | 10 | 0 | -2 |
| Brazil | 9 | 10 | 9 | -1 | 0 |
| Denmark | 10 | 11 | 11 | -1 | -1 |
| South Africa | 11 | 9 | 5 | 2 | 6 |
| Finland | 12 | 13 | 12 | -1 | 0 |
| Hungary | 13 | 12 | 13 | 1 | 0 |
| Belgium | 14 | 14 | 15 | 0 | -1 |
| Norway | 15 | 15 | 14 | 0 | 1 |
| Germany | 16 | 17 | 16 | -1 | 0 |
| Switzerland | 17 | 18 | 20 | -1 | -3 |
| Czech Republic | 18 | 16 | 17 | 2 | 1 |
| France | 19 | 20 | 18 | -1 | 1 |
| Cyprus | 20 | 19 | 19 | 1 | 1 |
| Austria | 21 | 22 | 22 | -1 | -1 |
| Italy | 22 | 21 | 21 | 1 | 1 |
| Ukraine | 23 | 24 | 24 | -1 | -1 |
| Taiwan | 24 | 23 | 23 | 1 | 1 |

*Note.* The aligned country rankings were computed from the aligned scores based on Path A1: country-specific weighted CFA (see Fig. S1, Path A1). The ranking change score was computed as the original ranking reported in the main text minus the aligned ranking; positive values indicate an improvement in rank, whereas negative values indicate a decline in rank. The rank refers to these 24 countries' relative ranking as reported in the original article.

**Table S13. Changes in 56 country/region rankings of perceived scientist trustworthiness after measurement alignment (aligned scores based on Fig. S1, Path A2)**

| Country/Region | Rank | Alignment rank (Weight mean) | Alignment rank (Unweight mean) | Rank change (Weight mean) | Rank change (Unweight mean) |
|---|---|---|---|---|---|
| Egypt | 1 | 1 | 2 | 0 | -1 |
| India | 2 | 2 | 1 | 0 | 1 |
| Nigeria | 3 | 3 | 3 | 0 | 0 |
| Kenya | 4 | 4 | 9 | 0 | -5 |
| Australia | 5 | 7 | 4 | -2 | 1 |
| Bangladesh | 6 | 5 | 5 | 1 | 1 |
| Türkiye | 7 | 6 | 8 | 1 | -1 |
| New Zealand | 8 | 8 | 6 | 0 | 2 |
| Argentina | 9 | 12 | 17 | -3 | -8 |
| United States | 10 | 9 | 12 | 1 | -2 |
| Indonesia | 11 | 10 | 7 | 1 | 4 |
| Ireland | 12 | 11 | 15 | 1 | -3 |
| United Kingdom | 13 | 13 | 13 | 0 | 0 |
| Canada | 14 | 16 | 18 | -2 | -4 |
| Philippines | 15 | 15 | 10 | 0 | 5 |
| Chile | 16 | 14 | 16 | 2 | 0 |
| Sweden | 17 | 18 | 20 | -1 | -3 |
| Brazil | 18 | 19 | 19 | -1 | -1 |
| Denmark | 19 | 22 | 22 | -3 | -3 |
| Botswana | 20 | 17 | 23 | 3 | -3 |
| Ghana | 21 | 23 | 11 | -2 | 10 |
| South Africa | 22 | 21 | 14 | 1 | 8 |
| Costa Rica | 23 | 20 | 21 | 3 | 2 |
| Morocco | 24 | 25 | 33 | -1 | -9 |
| Finland | 25 | 28 | 25 | -3 | 0 |
| China | 26 | 26 | 26 | 0 | 0 |
| Slovenia | 27 | 24 | 24 | 3 | 3 |
| Hungary | 28 | 27 | 27 | 1 | 1 |
| Côte d'Ivoire | 29 | 29 | 37 | 0 | -8 |
| Malaysia | 30 | 30 | 29 | 0 | 1 |
| Romania | 31 | 33 | 38 | -2 | -7 |
| Belgium | 32 | 31 | 32 | 1 | 0 |
| Norway | 33 | 35 | 30 | -2 | 3 |
| Netherlands | 34 | 34 | 34 | 0 | 0 |
| Uruguay | 35 | 40 | 28 | -5 | 7 |
| Poland | 36 | 38 | 39 | -2 | -3 |
| Peru | 37 | 32 | 31 | 5 | 6 |
| Germany | 38 | 37 | 35 | 1 | 3 |
| Serbia | 39 | 39 | 46 | 0 | -7 |
| Switzerland | 40 | 41 | 42 | -1 | -2 |
| Bulgaria | 41 | 44 | 44 | -3 | -3 |
| Czech Republic | 42 | 36 | 40 | 6 | 2 |
| South Korea | 43 | 42 | 49 | 1 | -6 |
| France | 44 | 45 | 43 | -1 | 1 |
| Cyprus | 45 | 43 | 41 | 2 | 4 |
| Austria | 46 | 48 | 47 | -2 | -1 |
| Cameroon | 47 | 49 | 51 | -2 | -4 |
| Greece | 48 | 46 | 48 | 2 | 0 |
| Italy | 49 | 47 | 45 | 2 | 4 |
| Ukraine | 50 | 51 | 52 | -1 | -2 |
| Japan | 51 | 53 | 54 | -2 | -3 |
| Taiwan | 52 | 50 | 50 | 2 | 2 |
| Israel | 53 | 52 | 53 | 1 | 0 |
| Ethiopia | 54 | 55 | 36 | -1 | 18 |
| Bolivia | 55 | 54 | 55 | 1 | 0 |
| Kazakhstan | 56 | 56 | 56 | 0 | 0 |

*Note.* The aligned country rankings were computed from the aligned scores based on Path A2: country-specific unweighted CFA (see Fig. S1, Path A2). The ranking change score was computed as the original ranking reported in the main text minus the aligned ranking; positive values indicate an improvement in rank, whereas negative values indicate a decline in rank. The rank refers to these 56 countries' relative ranking as reported in the original article.

**Table S14. Changes in 34 country/region rankings of perceived scientist trustworthiness after measurement alignment (aligned scores based on Fig. S1, Path A2)**

| Country/Region | Rank | Alignment rank (Weight mean) | Alignment rank (Unweight mean) | Rank change (Weight mean) | Rank change (Unweight mean) |
|---|---|---|---|---|---|
| Egypt | 1 | 1 | 2 | 0 | -1 |
| India | 2 | 2 | 1 | 0 | 1 |
| Australia | 3 | 6 | 6 | -3 | -3 |
| Bangladesh | 4 | 3 | 3 | 1 | 1 |
| Türkiye | 5 | 4 | 7 | 1 | -2 |
| New Zealand | 6 | 5 | 4 | 1 | 2 |
| United States | 7 | 7 | 11 | 0 | -4 |
| Indonesia | 8 | 9 | 5 | -1 | 3 |
| Ireland | 9 | 8 | 13 | 1 | -4 |
| United Kingdom | 10 | 10 | 10 | 0 | 0 |
| Canada | 11 | 12 | 15 | -1 | -4 |
| Philippines | 12 | 13 | 9 | -1 | 3 |
| Chile | 13 | 11 | 14 | 2 | -1 |
| Sweden | 14 | 14 | 17 | 0 | -3 |
| Brazil | 15 | 16 | 16 | -1 | -1 |
| Denmark | 16 | 17 | 18 | -1 | -2 |
| Ghana | 17 | 18 | 8 | -1 | 9 |
| South Africa | 18 | 15 | 12 | 3 | 6 |
| Finland | 19 | 21 | 19 | -2 | 0 |
| China | 20 | 19 | 20 | 1 | 0 |
| Hungary | 21 | 20 | 21 | 1 | 0 |
| Côte d'Ivoire | 22 | 22 | 26 | 0 | -4 |
| Malaysia | 23 | 23 | 23 | 0 | 0 |
| Romania | 24 | 24 | 27 | 0 | -3 |
| Norway | 25 | 25 | 22 | 0 | 3 |
| Germany | 26 | 27 | 24 | -1 | 2 |
| Serbia | 27 | 29 | 33 | -2 | -6 |
| Switzerland | 28 | 28 | 30 | 0 | -2 |
| Bulgaria | 29 | 31 | 32 | -2 | -3 |
| Czech Republic | 30 | 26 | 28 | 4 | 2 |
| France | 31 | 30 | 29 | 1 | 2 |
| Italy | 32 | 32 | 31 | 0 | 1 |
| Taiwan | 33 | 33 | 34 | 0 | -1 |
| Ethiopia | 34 | 34 | 25 | 0 | 9 |

*Note.* The aligned country rankings were computed from the aligned scores based on Path A2: country-specific unweighted CFA (see Fig. S1, Path A2). The ranking change score was computed as the original ranking reported in the main text minus the aligned ranking;; positive

values indicate an improvement in rank, whereas negative values indicate a decline in rank. The rank refers to these 34 countries' relative ranking as reported in the original article.

**Table S15. Changes in 68 country/region rankings of perceived scientist trustworthiness after measurement alignment (aligned scores based on Fig. S1, Path B1)**

| Country/Region | Rank | Alignment rank (Weight mean) | Alignment rank (Unweight mean) | Rank change (Weight mean) | Rank change (Unweight mean) |
|---|---|---|---|---|---|
| Egypt | 1 | 1 | 2 | 0 | -1 |
| India | 2 | 2 | 1 | 0 | 1 |
| Nigeria | 3 | 3 | 3 | 0 | 0 |
| Kenya | 4 | 6 | 16 | -2 | -12 |
| Australia | 5 | 7 | 6 | -2 | -1 |
| Bangladesh | 6 | 4 | 4 | 2 | 2 |
| Spain | 7 | 8 | 13 | -1 | -6 |
| Türkiye | 8 | 5 | 5 | 3 | 3 |
| New Zealand | 9 | 12 | 7 | -3 | 2 |
| Argentina | 10 | 10 | 14 | 0 | -4 |
| Mexico | 11 | 14 | 12 | -3 | -1 |
| United States | 12 | 9 | 11 | 3 | 1 |
| Indonesia | 13 | 11 | 9 | 2 | 4 |
| Ireland | 14 | 17 | 18 | -3 | -4 |
| United Kingdom | 15 | 22 | 20 | -7 | -5 |
| Uganda | 16 | 15 | 30 | 1 | -14 |
| Canada | 17 | 20 | 19 | -3 | -2 |
| Philippines | 18 | 13 | 10 | 5 | 8 |
| Chile | 19 | 21 | 21 | -2 | -2 |
| Sweden | 20 | 25 | 22 | -5 | -2 |
| Brazil | 21 | 16 | 15 | 5 | 6 |
| Denmark | 22 | 27 | 28 | -5 | -6 |
| Botswana | 23 | 18 | 25 | 5 | -2 |
| Ghana | 24 | 19 | 8 | 5 | 16 |
| South Africa | 25 | 23 | 17 | 2 | 8 |
| Costa Rica | 26 | 24 | 23 | 2 | 3 |
| Morocco | 27 | 28 | 42 | -1 | -15 |
| Finland | 28 | 33 | 31 | -5 | -3 |
| China | 29 | 32 | 27 | -3 | 2 |
| Slovenia | 30 | 34 | 32 | -4 | -2 |
| Colombia | 31 | 30 | 26 | 1 | 5 |
| Hungary | 32 | 36 | 35 | -4 | -3 |
| Côte d'Ivoire | 33 | 31 | 41 | 2 | -8 |
| Malaysia | 34 | 29 | 29 | 5 | 5 |
| Portugal | 35 | 35 | 33 | 0 | 2 |
| Romania | 36 | 46 | 47 | -10 | -11 |
| Belgium | 37 | 38 | 40 | -1 | -3 |
| Congo DR | 38 | 26 | 24 | 12 | 14 |

| Country/Region | Rank | Alignment rank (Weight mean) | Alignment rank (Unweight mean) | Rank change (Weight mean) | Rank change (Unweight mean) |
|---|---|---|---|---|---|
| Norway | 39 | 44 | 36 | -5 | 3 |
| Netherlands | 40 | 41 | 44 | -1 | -4 |
| Uruguay | 41 | 56 | 38 | -15 | 3 |
| Poland | 42 | 39 | 45 | 3 | -3 |
| Peru | 43 | 37 | 34 | 6 | 9 |
| Germany | 44 | 43 | 43 | 1 | 1 |
| Georgia | 45 | 42 | 39 | 3 | 6 |
| Serbia | 46 | 40 | 50 | 6 | -4 |
| Switzerland | 47 | 49 | 52 | -2 | -5 |
| Bulgaria | 48 | 48 | 51 | 0 | -3 |
| Czech Republic | 49 | 52 | 56 | -3 | -7 |
| South Korea | 50 | 61 | 63 | -11 | -13 |
| France | 51 | 54 | 54 | -3 | -3 |
| Cyprus | 52 | 47 | 46 | 5 | 6 |
| Austria | 53 | 58 | 60 | -5 | -7 |
| Hong Kong | 54 | 51 | 57 | 3 | -3 |
| Cameroon | 55 | 45 | 48 | 10 | 7 |
| Greece | 56 | 50 | 55 | 6 | 1 |
| Italy | 57 | 53 | 53 | 4 | 4 |
| Ukraine | 58 | 59 | 61 | -1 | -3 |
| Japan | 59 | 65 | 65 | -6 | -6 |
| Slovakia | 60 | 55 | 59 | 5 | 1 |
| Taiwan | 61 | 62 | 62 | -1 | -1 |
| Israel | 62 | 63 | 64 | -1 | -2 |
| Nicaragua | 63 | 57 | 49 | 6 | 14 |
| Ethiopia | 64 | 66 | 37 | -2 | 27 |
| Russia | 65 | 67 | 67 | -2 | -2 |
| Bolivia | 66 | 64 | 66 | 2 | 0 |
| Kazakhstan | 67 | 68 | 68 | -1 | -1 |
| Albania | 68 | 60 | 58 | 8 | 10 |

*Note.* The aligned country rankings were computed from the aligned scores based on Path B1: pooled weighted CFA (see Fig. S1, Path B1). The ranking change score was computed as the original ranking reported in the main text minus the aligned ranking;; positive values indicate an improvement in rank, whereas negative values indicate a decline in rank.

**Table S16. Changes in 68 country/region rankings of perceived scientist trustworthiness after measurement alignment (aligned scores based on Fig. S1, Path B2)**

| Country/Region | Rank | Alignment rank (Weight mean) | Alignment rank (Unweight mean) | Rank change (weight mean) | Rank change (Unweight mean) |
|---|---|---|---|---|---|
| Egypt | 1 | 1 | 2 | 0 | -1 |
| India | 2 | 2 | 1 | 0 | 1 |
| Nigeria | 3 | 3 | 3 | 0 | 0 |
| Kenya | 4 | 4 | 11 | 0 | -7 |
| Australia | 5 | 8 | 4 | -3 | 1 |
| Bangladesh | 6 | 6 | 5 | 0 | 1 |
| Spain | 7 | 5 | 6 | 2 | 1 |
| Türkiye | 8 | 9 | 12 | -1 | -4 |
| New Zealand | 9 | 11 | 8 | -2 | 1 |
| Argentina | 10 | 7 | 9 | 3 | 1 |
| Mexico | 11 | 10 | 7 | 1 | 4 |
| United States | 12 | 13 | 16 | -1 | -4 |
| Indonesia | 13 | 15 | 10 | -2 | 3 |
| Ireland | 14 | 14 | 18 | 0 | -4 |
| United Kingdom | 15 | 16 | 15 | -1 | 0 |
| Uganda | 16 | 12 | 27 | 4 | -11 |
| Canada | 17 | 18 | 20 | -1 | -3 |
| Philippines | 18 | 19 | 13 | -1 | 5 |
| Chile | 19 | 17 | 19 | 2 | 0 |
| Sweden | 20 | 21 | 22 | -1 | -2 |
| Brazil | 21 | 23 | 21 | -2 | 0 |
| Denmark | 22 | 25 | 25 | -3 | -3 |
| Botswana | 23 | 20 | 26 | 3 | -3 |
| Ghana | 24 | 26 | 14 | -2 | 10 |
| South Africa | 25 | 22 | 17 | 3 | 8 |
| Costa Rica | 26 | 24 | 23 | 2 | 3 |
| Morocco | 27 | 29 | 39 | -2 | -12 |
| Finland | 28 | 32 | 29 | -4 | -1 |
| China | 29 | 31 | 30 | -2 | -1 |
| Slovenia | 30 | 28 | 28 | 2 | 2 |
| Colombia | 31 | 27 | 24 | 4 | 7 |
| Hungary | 32 | 30 | 31 | 2 | 1 |
| Côte d'Ivoire | 33 | 33 | 46 | 0 | -13 |
| Malaysia | 34 | 34 | 34 | 0 | 0 |
| Portugal | 35 | 35 | 35 | 0 | 0 |
| Romania | 36 | 40 | 43 | -4 | -7 |
| Belgium | 37 | 37 | 38 | 0 | -1 |
| Congo DR | 38 | 36 | 32 | 2 | 6 |

| Country/Region | Rank | Alignment rank (Weight mean) | Alignment rank (Unweight mean) | Rank change (weight mean) | Rank change (Unweight mean) |
|---|---|---|---|---|---|
| Norway | 39 | 42 | 36 | -3 | 3 |
| Netherlands | 40 | 39 | 40 | 1 | 0 |
| Uruguay | 41 | 45 | 33 | -4 | 8 |
| Poland | 42 | 44 | 45 | -2 | -3 |
| Peru | 43 | 38 | 37 | 5 | 6 |
| Germany | 44 | 43 | 41 | 1 | 3 |
| Georgia | 45 | 51 | 48 | -6 | -3 |
| Serbia | 46 | 47 | 53 | -1 | -7 |
| Switzerland | 47 | 48 | 47 | -1 | 0 |
| Bulgaria | 48 | 50 | 51 | -2 | -3 |
| Czech Republic | 49 | 41 | 44 | 8 | 5 |
| South Korea | 50 | 46 | 54 | 4 | -4 |
| France | 51 | 52 | 50 | -1 | 1 |
| Cyprus | 52 | 49 | 49 | 3 | 3 |
| Austria | 53 | 53 | 55 | 0 | -2 |
| Hong Kong | 54 | 55 | 59 | -1 | -5 |
| Cameroon | 55 | 57 | 61 | -2 | -6 |
| Greece | 56 | 54 | 57 | 2 | -1 |
| Italy | 57 | 56 | 52 | 1 | 5 |
| Ukraine | 58 | 60 | 62 | -2 | -4 |
| Japan | 59 | 61 | 63 | -2 | -4 |
| Slovakia | 60 | 59 | 60 | 1 | 0 |
| Taiwan | 61 | 58 | 58 | 3 | 3 |
| Israel | 62 | 62 | 64 | 0 | -2 |
| Nicaragua | 63 | 63 | 56 | 0 | 7 |
| Ethiopia | 64 | 67 | 42 | -3 | 22 |
| Russia | 65 | 66 | 66 | -1 | -1 |
| Bolivia | 66 | 64 | 67 | 2 | -1 |
| Kazakhstan | 67 | 68 | 68 | -1 | -1 |
| Albania | 68 | 65 | 65 | 3 | 3 |

*Note.* The aligned country rankings were computed from the aligned scores based on Path B2: pooled unweighted CFA (see Fig. S1, Path B2). The ranking change score was computed as the original ranking reported in the main text minus the aligned ranking;; positive values indicate an improvement in rank, whereas negative values indicate a decline in rank

## Linear Multilevel Regression Testing

After completing the CFA and measurement alignment procedures, we obtained a total of 12 sets of aligned scores. These aligned scores were then entered into the regression analyses as the dependent variables (DVs), namely perceived scientist trustworthiness. Following the approach of Cologna et al., we used multilevel linear regression to examine the associations between perceived scientist trustworthiness and perceptions of demographic, ideological, attitudinal, and country-level factors. In the regression stage, we estimated models both with and without sampling weights.

The weights were based on WEIGHT_MLVLM, which contains rescaled post-stratification weights for weighted multilevel analyses using R's lme4 package.

**Table S17. Weighted linear multilevel regression testing the association of perceived scientist trustworthiness with demographic characteristics, attitudes toward science, and country-level indicators in 37 countries/regions (random intercepts across countries/regions, aligned scores from weighted country-specific CFA; Fig. S1, Path A1)**

| | Block 1: Demographic characteristics | | | | | | Block 2: Ideological views | | | | | |
|---|---|---|---|---|---|---|---|---|---|---|---|---|
| *Predictors* | *Beta* | *SE* | *CI* | *t* | *p* | *df* | *Beta* | *SE* | *CI* | *t* | *p* | *df* |
| Intercept | -0.270 | 0.072 | -0.417 – -0.123 | -3.736 | **0.001** | 35.020 | -0.201 | 0.076 | -0.357 – -0.046 | -2.637 | **0.013** | 33.549 |
| Sex (male) | -0.032 | 0.006 | -0.043 – -0.020 | -5.442 | **<0.001** | 45671.630 | -0.016 | 0.007 | -0.029 – -0.003 | -2.377 | **0.017** | 34198.413 |
| Age | 0.037 | 0.005 | 0.026 – 0.047 | 6.787 | **<0.001** | 45683.690 | 0.014 | 0.006 | 0.002 – 0.026 | 2.313 | **0.021** | 34206.782 |
| Education (tertiary) | 0.079 | 0.006 | 0.067 – 0.092 | 12.800 | **<0.001** | 45648.782 | 0.070 | 0.007 | 0.057 – 0.084 | 10.278 | **<0.001** | 34199.730 |
| Income | 0.076 | 0.006 | 0.065 – 0.088 | 12.909 | **<0.001** | 45673.135 | 0.054 | 0.007 | 0.041 – 0.068 | 7.989 | **<0.001** | 34197.569 |
| Residence place (urban) | 0.066 | 0.006 | 0.054 – 0.077 | 11.430 | **<0.001** | 45671.818 | 0.067 | 0.007 | 0.054 – 0.080 | 10.277 | **<0.001** | 34197.596 |
| Political orientation (right) | | | | | | | 0.007 | 0.008 | -0.008 – 0.023 | 0.923 | 0.356 | 34196.867 |
| Political orientation (conservative) | | | | | | | -0.073 | 0.008 | -0.089 – -0.058 | -9.290 | **<0.001** | 34200.784 |
| Religiosity | | | | | | | 0.097 | 0.007 | 0.084 – 0.110 | 14.238 | **<0.001** | 34198.492 |
| Social dominance orientation | | | | | | | -0.193 | 0.007 | -0.206 – -0.179 | -28.129 | **<0.001** | 34198.917 |
| Science-related populist attitudes | | | | | | | | | | | | |
| Perceived benefit of science | | | | | | | | | | | | |
| Willingness to be vulnerable to science | | | | | | | | | | | | |
| Trust in the scientific method | | | | | | | | | | | | |
| GDP per capita | | | | | | | | | | | | |
| Govt expenditure on education (% of GDP) | | | | | | | | | | | | |
| Gini index | | | | | | | | | | | | |
| Science literacy (PISA) | | | | | | | | | | | | |
| Academic freedom | | | | | | | | | | | | |
| Degree of populism in politics | | | | | | | | | | | | |
| **Random Effects** | | | | | | | | | | | | |
| $\sigma^2$ | 1.01 | | | | | | 0.95 | | | | | |
| $\tau_{00}$ | 0.19 COUNTRY_NAME | | | | | | 0.20 COUNTRY_NAME | | | | | |
| ICC | 0.16 | | | | | | 0.18 | | | | | |
| N | 37 COUNTRY_NAME | | | | | | 36 COUNTRY_NAME | | | | | |
| Observations | 45713 | | | | | | 34241 | | | | | |
| Marginal $R^2$ / Conditional $R^2$ | 0.017 / 0.173 | | | | | | 0.054 / 0.221 | | | | | |
| AIC | 168343.623 | | | | | | 123126.503 | | | | | |

| | Block 3: Attitudes to science | | | | | | Block 4: Country indicators | | | | | |
|---|---|---|---|---|---|---|---|---|---|---|---|---|
| *Predictors* | *Beta* | *SE* | *CI* | *t* | *p* | *df* | *Beta* | *SE* | *CI* | *t* | *p* | *df* |
| Intercept | -0.301 | 0.076 | -0.455 – -0.147 | -3.977 | **<0.001** | 34.224 | -0.289 | 0.084 | -0.462 – -0.116 | -3.434 | **0.002** | 25.090 |
| Sex (male) | -0.045 | 0.005 | -0.055 – -0.035 | -8.564 | **<0.001** | 33634.789 | -0.044 | 0.005 | -0.054 – -0.033 | -8.200 | **<0.001** | 32048.858 |
| Age | 0.004 | 0.005 | -0.005 – 0.014 | 0.864 | 0.387 | 33639.964 | 0.003 | 0.005 | -0.006 – 0.013 | 0.697 | 0.486 | 32051.731 |
| Education (tertiary) | -0.022 | 0.005 | -0.033 – -0.012 | -4.109 | **<0.001** | 33665.810 | -0.021 | 0.006 | -0.032 – -0.010 | -3.849 | **<0.001** | 32070.455 |
| Income | 0.009 | 0.005 | -0.002 – 0.019 | 1.613 | 0.107 | 33634.018 | 0.007 | 0.005 | -0.004 – 0.018 | 1.307 | 0.191 | 32048.432 |
| Residence place (urban) | 0.030 | 0.005 | 0.020 – 0.040 | 5.819 | **<0.001** | 33634.095 | 0.030 | 0.005 | 0.020 – 0.041 | 5.758 | **<0.001** | 32048.016 |
| Political orientation (right) | 0.020 | 0.006 | 0.008 – 0.033 | 3.255 | **0.001** | 33633.607 | 0.021 | 0.006 | 0.008 – 0.033 | 3.216 | **0.001** | 32048.052 |
| Political orientation (conservative) | -0.027 | 0.006 | -0.039 – -0.015 | -4.351 | **<0.001** | 33636.537 | -0.027 | 0.006 | -0.040 – -0.014 | -4.223 | **<0.001** | 32049.114 |
| Religiosity | 0.093 | 0.005 | 0.082 – 0.103 | 17.108 | **<0.001** | 33634.309 | 0.091 | 0.006 | 0.080 – 0.102 | 16.484 | **<0.001** | 32048.283 |
| Social dominance orientation | -0.033 | 0.006 | -0.044 – -0.022 | -6.001 | **<0.001** | 33634.553 | -0.035 | 0.006 | -0.046 – -0.024 | -6.176 | **<0.001** | 32048.697 |
| Science-related populist attitudes | -0.023 | 0.006 | -0.034 – -0.012 | -4.108 | **<0.001** | 33634.948 | -0.023 | 0.006 | -0.034 – -0.012 | -4.025 | **<0.001** | 32048.541 |
| Perceived benefit of science | 0.272 | 0.006 | 0.260 – 0.284 | 43.858 | **<0.001** | 33635.313 | 0.269 | 0.006 | 0.257 – 0.281 | 42.361 | **<0.001** | 32048.371 |
| Willingness to be vulnerable to science | 0.376 | 0.006 | 0.364 – 0.389 | 59.773 | **<0.001** | 33633.669 | 0.377 | 0.006 | 0.365 – 0.390 | 58.568 | **<0.001** | 32047.991 |
| Trust in the scientific method | 0.361 | 0.006 | 0.349 – 0.373 | 57.381 | **<0.001** | 33633.445 | 0.362 | 0.006 | 0.349 – 0.374 | 56.124 | **<0.001** | 32047.959 |
| GDP per capita | | | | | | | 0.052 | 0.087 | -0.128 – 0.232 | 0.599 | 0.554 | 25.174 |
| Govt expenditure on education (% of GDP) | | | | | | | -0.003 | 0.075 | -0.157 – 0.150 | -0.046 | 0.963 | 25.238 |
| Gini index | | | | | | | 0.290 | 0.084 | 0.117 – 0.463 | 3.453 | **0.002** | 25.304 |
| Science literacy (PISA) | | | | | | | 0.087 | 0.097 | -0.112 – 0.286 | 0.895 | 0.379 | 26.031 |
| Academic freedom | | | | | | | -0.054 | 0.094 | -0.248 – 0.140 | -0.574 | 0.571 | 25.432 |
| Degree of populism in politics | | | | | | | -0.090 | 0.073 | -0.241 – 0.060 | -1.233 | 0.229 | 26.267 |
| **Random Effects** | | | | | | | | | | | | |
| $\sigma^2$ | 0.58 | | | | | | 0.59 | | | | | |
| $\tau_{00}$ | 0.20 COUNTRY_NAME | | | | | | 0.12 COUNTRY_NAME | | | | | |
| ICC | 0.26 | | | | | | 0.16 | | | | | |
| N | 36COUNTRY_NAME | | | | | | 32 COUNTRY_NAME | | | | | |
| Observations | 33682 | | | | | | 32092 | | | | | |
| Marginal $R^2$ / Conditional $R^2$ | 0.437 / 0.584 | | | | | | 0.477 / 0.563 | | | | | |
| AIC | 104472.563 | | | | | | 96682.859 | | | | | |

*Note.* Significant testing based on two-sided t tests. AIC = Akaike information criterion, ICC = Intraclass Correlation Coefficient. $\sigma^2$ = within country (residual) variance. $\tau_{00}$ = between-country variance (variation between individual intercepts and average intercept).

**Table S18. Unweighted linear multilevel regression testing the association of perceived scientist trustworthiness with demographic characteristics, attitudes toward science, and country-level indicators in 37 countries/regions (random intercepts across countries/regions, aligned scores from weighted country-specific CFA; Fig. S1, Path A1)**

| | **Block 1: Demographic characteristics** | | | | | | **Block 2: Ideological views** | | | | | |
|---|---|---|---|---|---|---|---|---|---|---|---|---|
| *Predictors* | *Beta* | *SE* | *CI* | *t* | *p* | *df* | *Beta* | *SE* | *CI* | *t* | *p* | *df* |
| Intercept | -0.334 | 0.066 | -0.467 – -0.200 | -5.074 | **<0.001** | 35.939 | -0.269 | 0.069 | -0.408 – -0.129 | -3.906 | **<0.001** | 34.844 |
| Sex (male) | -0.026 | 0.006 | -0.037 – -0.014 | -4.463 | **<0.001** | 45672.089 | -0.020 | 0.007 | -0.033 – -0.007 | -2.985 | **0.003** | 34199.462 |
| Age | 0.012 | 0.006 | 0.001 – 0.023 | 2.051 | **0.040** | 45672.310 | -0.005 | 0.007 | -0.018 – 0.007 | -0.824 | 0.410 | 34198.066 |
| Education (tertiary) | 0.074 | 0.006 | 0.063 – 0.086 | 12.603 | **<0.001** | 45671.358 | 0.065 | 0.007 | 0.052 – 0.079 | 9.737 | **<0.001** | 34196.929 |
| Income | 0.061 | 0.006 | 0.049 – 0.072 | 10.313 | **<0.001** | 45671.488 | 0.048 | 0.007 | 0.035 – 0.062 | 7.099 | **<0.001** | 34197.257 |
| Residence place (urban) | 0.053 | 0.006 | 0.042 – 0.065 | 9.143 | **<0.001** | 45671.258 | 0.054 | 0.007 | 0.041 – 0.067 | 8.081 | **<0.001** | 34196.784 |
| Political orientation (right) | | | | | | | 0.010 | 0.008 | -0.006 – 0.025 | 1.229 | 0.219 | 34196.732 |
| Political orientation (conservative) | | | | | | | -0.065 | 0.008 | -0.080 – -0.049 | -8.201 | **<0.001** | 34197.389 |
| Religiosity | | | | | | | 0.115 | 0.007 | 0.102 – 0.129 | 16.949 | **<0.001** | 34197.363 |
| Social dominance orientation | | | | | | | -0.154 | 0.007 | -0.167 – -0.140 | -22.570 | **<0.001** | 34197.135 |
| Science-related populist attitudes | | | | | | | | | | | | |
| Perceived benefit of science | | | | | | | | | | | | |
| Willingness to be vulnerable to science | | | | | | | | | | | | |
| Trust in the scientific method | | | | | | | | | | | | |
| GDP per capita | | | | | | | | | | | | |
| Govt expenditure on education (% of GDP) | | | | | | | | | | | | |
| Gini index | | | | | | | | | | | | |
| Science literacy (PISA) | | | | | | | | | | | | |
| Academic freedom | | | | | | | | | | | | |
| Degree of populism in politics | | | | | | | | | | | | |
| **Random Effects** | | | | | | | | | | | | |
| $\sigma^2$ | 1.51 | | | | | | 1.43 | | | | | |
| $\tau_{00}$ | 0.16 COUNTRY_NAME | | | | | | 0.17 COUNTRY_NAME | | | | | |
| ICC | 0.09 | | | | | | 0.10 | | | | | |
| N | 37 COUNTRY_NAME | | | | | | 36 COUNTRY_NAME | | | | | |
| Observations | 45713 | | | | | | 34241 | | | | | |
| Marginal $R^2$ / Conditional $R^2$ | 0.009 / 0.103 | | | | | | 0.031 / 0.132 | | | | | |
| AIC | 148854.945 | | | | | | 109698.994 | | | | | |

| | Block 3: Attitudes to science | | | | | | Block 4: Country indicators | | | | | |
|---|---|---|---|---|---|---|---|---|---|---|---|---|
| *Predictors* | *Beta* | *SE* | *CI* | *t* | *p* | *df* | *Beta* | *SE* | *CI* | *t* | *p* | *df* |
| Intercept | -0.333 | 0.070 | -0.475 – -0.191 | -4.761 | **<0.001** | 34.931 | -0.284 | 0.084 | -0.457 – -0.112 | -3.397 | **0.002** | 25.040 |
| Sex (male) | -0.047 | 0.005 | -0.057 – -0.037 | -8.823 | **<0.001** | 33635.237 | -0.042 | 0.005 | -0.053 – -0.032 | -7.867 | **<0.001** | 32048.739 |
| Age | -0.007 | 0.005 | -0.017 – 0.004 | -1.242 | 0.214 | 33634.376 | -0.004 | 0.005 | -0.014 – 0.007 | -0.665 | 0.506 | 32048.417 |
| Education (tertiary) | -0.021 | 0.005 | -0.032 – -0.011 | -3.927 | **<0.001** | 33633.912 | -0.019 | 0.005 | -0.030 – -0.008 | -3.482 | **<0.001** | 32048.112 |
| Income | 0.004 | 0.005 | -0.006 – 0.015 | 0.783 | 0.433 | 33633.825 | 0.002 | 0.006 | -0.009 – 0.013 | 0.305 | 0.760 | 32047.902 |
| Residence place (urban) | 0.023 | 0.005 | 0.012 – 0.034 | 4.274 | **<0.001** | 33633.501 | 0.024 | 0.005 | 0.013 – 0.035 | 4.403 | **<0.001** | 32047.598 |
| Political orientation (right) | 0.014 | 0.006 | 0.001 – 0.026 | 2.183 | **0.029** | 33633.497 | 0.017 | 0.006 | 0.004 – 0.030 | 2.609 | **0.009** | 32047.493 |
| Political orientation (conservative) | -0.022 | 0.006 | -0.034 – -0.010 | -3.456 | **0.001** | 33634.036 | -0.025 | 0.006 | -0.038 – -0.013 | -3.936 | **<0.001** | 32048.002 |
| Religiosity | 0.097 | 0.006 | 0.086 – 0.108 | 17.496 | **<0.001** | 33633.858 | 0.094 | 0.006 | 0.083 – 0.105 | 16.743 | **<0.001** | 32047.754 |
| Social dominance orientation | -0.011 | 0.006 | -0.022 – -0.000 | -1.968 | **0.049** | 33633.613 | -0.018 | 0.006 | -0.029 – -0.007 | -3.141 | **0.002** | 32047.593 |
| Science-related populist attitudes | -0.002 | 0.006 | -0.013 – 0.009 | -0.398 | 0.691 | 33634.472 | 0.001 | 0.006 | -0.010 – 0.013 | 0.247 | 0.805 | 32048.141 |
| Perceived benefit of science | 0.278 | 0.006 | 0.266 – 0.291 | 43.465 | **<0.001** | 33633.548 | 0.273 | 0.006 | 0.260 – 0.286 | 42.092 | **<0.001** | 32047.692 |
| Willingness to be vulnerable to science | 0.380 | 0.006 | 0.367 – 0.392 | 58.677 | **<0.001** | 33633.751 | 0.384 | 0.007 | 0.371 – 0.397 | 58.645 | **<0.001** | 32047.862 |
| Trust in the scientific method | 0.345 | 0.006 | 0.332 – 0.357 | 53.542 | **<0.001** | 33633.571 | 0.345 | 0.007 | 0.332 – 0.358 | 52.904 | **<0.001** | 32047.610 |
| GDP per capita | | | | | | | 0.053 | 0.087 | -0.126 – 0.233 | 0.610 | 0.547 | 25.322 |
| Govt expenditure on education (% of GDP) | | | | | | | -0.023 | 0.074 | -0.176 – 0.130 | -0.315 | 0.755 | 25.199 |
| Gini index | | | | | | | 0.294 | 0.084 | 0.122 – 0.466 | 3.523 | **0.002** | 25.193 |
| Science literacy (PISA) | | | | | | | 0.141 | 0.096 | -0.056 – 0.339 | 1.474 | 0.153 | 25.459 |
| Academic freedom | | | | | | | -0.036 | 0.094 | -0.229 – 0.157 | -0.384 | 0.704 | 25.394 |
| Degree of populism in politics | | | | | | | -0.039 | 0.072 | -0.187 – 0.110 | -0.532 | 0.599 | 25.502 |
| **Random Effects** | | | | | | | | | | | | |
| $\sigma^2$ | 0.91 | | | | | | 0.88 | | | | | |
| $\tau_{00}$ | 0.17 COUNTRY_NAME | | | | | | 0.12 COUNTRY_NAME | | | | | |
| ICC | 0.16 | | | | | | 0.12 | | | | | |
| N | 36 COUNTRY_NAME | | | | | | 32 COUNTRY_NAME | | | | | |
| Observations | 33682 | | | | | | 32092 | | | | | |
| Marginal $R^2$ / Conditional $R^2$ | 0.350 / 0.455 | | | | | | 0.385 / 0.456 | | | | | |
| AIC | 92629.67 | | | | | | 87392.006 | | | | | |

*Note.* Significant testing based on two-sided t tests. AIC = Akaike information criterion, ICC = Intraclass Correlation Coefficient. $\sigma^2$ = within country (residual) variance. $\tau_{00}$ = between-country variance (variation between individual intercepts and average intercept).

**Table S19. Weighted linear multilevel regression testing the association of perceived scientist trustworthiness with demographic characteristics, attitudes toward science, and country-level indicators in 24 countries/regions (random intercepts across countries/regions, aligned scores from country-specific weighted CFA; Fig. S1, Path A1)**

| | **Block 1: Demographic characteristics** | | | | | | **Block 2: Ideological views** | | | | | |
|---|---|---|---|---|---|---|---|---|---|---|---|---|
| *Predictors* | *Beta* | *SE* | *CI* | *t* | *p* | *df* | *Beta* | *SE* | *CI* | *t* | *p* | *df* |
| Intercept | -0.267 | 0.073 | -0.418 – -0.117 | -3.678 | **0.001** | 21.702 | -0.211 | 0.075 | -0.367 – -0.056 | -2.831 | **0.010** | 21.087 |
| Sex (male) | -0.021 | 0.006 | -0.032 – -0.009 | -3.586 | **<0.001** | 34073.511 | -0.003 | 0.006 | -0.015 – 0.010 | -0.442 | 0.659 | 26462.340 |
| Age | 0.039 | 0.005 | 0.029 – 0.050 | 7.321 | **<0.001** | 34081.773 | 0.017 | 0.006 | 0.006 – 0.029 | 2.911 | **0.004** | 26468.418 |
| Education (tertiary) | 0.073 | 0.006 | 0.061 – 0.085 | 12.063 | **<0.001** | 34085.584 | 0.063 | 0.007 | 0.050 – 0.076 | 9.591 | **<0.001** | 26478.905 |
| Income | 0.070 | 0.006 | 0.058 – 0.081 | 12.005 | **<0.001** | 34075.134 | 0.051 | 0.007 | 0.038 – 0.064 | 7.764 | **<0.001** | 26462.307 |
| Residence place (urban) | 0.058 | 0.006 | 0.047 – 0.069 | 10.263 | **<0.001** | 34073.609 | 0.061 | 0.006 | 0.048 – 0.073 | 9.713 | **<0.001** | 26461.666 |
| Political orientation (right) | | | | | | | 0.002 | 0.008 | -0.013 – 0.017 | 0.283 | 0.777 | 26462.090 |
| Political orientation (conservative) | | | | | | | -0.075 | 0.008 | -0.089 – -0.060 | -9.812 | **<0.001** | 26465.579 |
| Religiosity | | | | | | | 0.083 | 0.007 | 0.070 – 0.096 | 12.759 | **<0.001** | 26463.599 |
| Social dominance orientation | | | | | | | -0.176 | 0.007 | -0.189 – -0.163 | -26.789 | **<0.001** | 26462.214 |
| Science-related populist attitudes | | | | | | | | | | | | |
| Perceived benefit of science | | | | | | | | | | | | |
| Willingness to be vulnerable to science | | | | | | | | | | | | |
| Trust in the scientific method | | | | | | | | | | | | |
| GDP per capita | | | | | | | | | | | | |
| Govt expenditure on education (% of GDP) | | | | | | | | | | | | |
| Gini index | | | | | | | | | | | | |
| Science literacy (PISA) | | | | | | | | | | | | |
| Academic freedom | | | | | | | | | | | | |
| Degree of populism in politics | | | | | | | | | | | | |
| **Random Effects** | | | | | | | | | | | | |
| $\sigma^2$ | 0.77 | | | | | | 0.71 | | | | | |
| $\tau_{00}$ | 0.12 COUNTRY_NAME | | | | | | 0.13 COUNTRY_NAME | | | | | |
| ICC | 0.14 | | | | | | 0.16 | | | | | |
| N | 24 COUNTRY_NAME | | | | | | 24 COUNTRY_NAME | | | | | |
| Observations | 34103 | | | | | | 26495 | | | | | |
| Marginal $R^2$ / Conditional $R^2$ | 0.019 / 0.156 | | | | | | 0.063 / 0.208 | | | | | |
| AIC | 109578.446 | | | | | | 83267.868 | | | | | |

| Predictors | Block 3: Attitudes to science | | | | | | Block 4: Country indicators | | | | | |
|---|---|---|---|---|---|---|---|---|---|---|---|---|
| *Predictors* | *Beta* | *SE* | *CI* | *t* | *p* | *df* | *Beta* | *SE* | *CI* | *t* | *p* | *df* |
| Intercept | -0.288 | 0.075 | -0.444 – -0.133 | -3.840 | **0.001** | 21.908 | -0.361 | 0.067 | -0.504 – -0.217 | -5.361 | **<0.001** | 15.135 |
| Sex (male) | -0.035 | 0.005 | -0.044 – -0.025 | -7.129 | **<0.001** | 26445.304 | -0.034 | 0.005 | -0.043 – -0.024 | -6.765 | **<0.001** | 25364.463 |
| Age | 0.008 | 0.005 | -0.000 – 0.017 | 1.852 | 0.064 | 26448.640 | 0.008 | 0.005 | -0.001 – 0.017 | 1.646 | 0.100 | 25365.973 |
| Education (tertiary) | -0.018 | 0.005 | -0.028 – -0.008 | -3.555 | **<0.001** | 26467.744 | -0.017 | 0.005 | -0.027 – -0.007 | -3.285 | **0.001** | 25377.963 |
| Income | 0.010 | 0.005 | 0.000 – 0.020 | 1.996 | **0.046** | 26445.141 | 0.009 | 0.005 | -0.001 – 0.019 | 1.681 | 0.093 | 25363.896 |
| Residence place (urban) | 0.025 | 0.005 | 0.016 – 0.034 | 5.239 | **<0.001** | 26444.801 | 0.025 | 0.005 | 0.015 – 0.034 | 5.099 | **<0.001** | 25363.648 |
| Political orientation (right) | 0.014 | 0.006 | 0.003 – 0.026 | 2.416 | **0.016** | 26444.967 | 0.014 | 0.006 | 0.002 – 0.026 | 2.347 | **0.019** | 25363.650 |
| Political orientation (conservative) | -0.029 | 0.006 | -0.041 – -0.018 | -5.048 | **<0.001** | 26447.393 | -0.028 | 0.006 | -0.040 – -0.017 | -4.770 | **<0.001** | 25364.369 |
| Religiosity | 0.073 | 0.005 | 0.063 – 0.083 | 14.512 | **<0.001** | 26445.626 | 0.072 | 0.005 | 0.062 – 0.082 | 14.049 | **<0.001** | 25363.782 |
| Social dominance orientation | -0.038 | 0.005 | -0.048 – -0.028 | -7.412 | **<0.001** | 26444.907 | -0.040 | 0.005 | -0.051 – -0.030 | -7.641 | **<0.001** | 25363.222 |
| Science-related populist attitudes | -0.022 | 0.005 | -0.032 – -0.012 | -4.218 | **<0.001** | 26446.076 | -0.022 | 0.005 | -0.032 – -0.012 | -4.157 | **<0.001** | 25364.116 |
| Perceived benefit of science | 0.231 | 0.006 | 0.219 – 0.242 | 39.640 | **<0.001** | 26445.802 | 0.228 | 0.006 | 0.217 – 0.240 | 38.459 | **<0.001** | 25363.818 |
| Willingness to be vulnerable to science | 0.327 | 0.006 | 0.315 – 0.338 | 55.527 | **<0.001** | 26444.481 | 0.327 | 0.006 | 0.316 – 0.339 | 54.539 | **<0.001** | 25363.502 |
| Trust in the scientific method | 0.307 | 0.006 | 0.295 – 0.318 | 52.002 | **<0.001** | 26444.409 | 0.307 | 0.006 | 0.295 – 0.319 | 50.952 | **<0.001** | 25363.562 |
| GDP per capita | | | | | | | 0.084 | 0.063 | -0.049 – 0.218 | 1.341 | 0.200 | 15.201 |
| Govt expenditure on education (% of GDP) | | | | | | | 0.109 | 0.065 | -0.030 – 0.248 | 1.668 | 0.116 | 15.076 |
| Gini index | | | | | | | 0.228 | 0.073 | 0.072 – 0.384 | 3.108 | **0.007** | 15.062 |
| Science literacy (PISA) | | | | | | | 0.117 | 0.114 | -0.126 – 0.361 | 1.028 | 0.320 | 15.051 |
| Academic freedom | | | | | | | -0.079 | 0.076 | -0.240 – 0.082 | -1.043 | 0.313 | 15.257 |
| Degree of populism in politics | | | | | | | -0.065 | 0.074 | -0.223 – 0.093 | -0.879 | 0.393 | 14.926 |
| **Random Effects** | | | | | | | | | | | | |
| $\sigma^2$ | 0.42 | | | | | | 0.43 | | | | | |
| $\tau_{00}$ | 0.13 COUNTRY_NAME | | | | | | 0.05 COUNTRY_NAME | | | | | |
| ICC | 0.24 | | | | | | 0.11 | | | | | |
| N | 24 COUNTRY_NAME | | | | | | 22 COUNTRY_NAME | | | | | |
| Observations | 26482 | | | | | | 25398 | | | | | |
| Marginal $R^2$ / Conditional $R^2$ | 0.452 / 0.585 | | | | | | 0.504 / 0.560 | | | | | |
| AIC | 69058.807 | | | | | | 64409.567 | | | | | |

*Note.* Significant testing based on two-sided t tests. AIC = Akaike information criterion, ICC = Intraclass Correlation Coefficient. $\sigma^2$ = within country (residual) variance. $\tau_{00}$ = between-country variance (variation between individual intercepts and average intercept).

**Table S20. Unweighted linear multilevel regression testing the association of perceived scientist trustworthiness with demographic characteristics, attitudes toward science, and country-level indicators in 24 countries/regions (random intercepts across countries/regions, aligned scores from weighted country-specific CFA; Fig. S1, Path A1)**

| | **Block 1: Demographic characteristics** | | | | | | **Block 2: Ideological views** | | | | | |
|---|---|---|---|---|---|---|---|---|---|---|---|---|
| *Predictors* | *Beta* | *SE* | *CI* | *t* | *p* | *df* | *Beta* | *SE* | *CI* | *t* | *p* | *df* |
| Intercept | -0.303 | 0.063 | -0.434 – -0.172 | -4.779 | **<0.001** | 22.953 | -0.247 | 0.063 | -0.378 – -0.116 | -3.898 | **0.001** | 22.904 |
| Sex (male) | -0.017 | 0.006 | -0.029 – -0.006 | -3.097 | **0.002** | 34075.137 | -0.005 | 0.006 | -0.017 – 0.008 | -0.743 | 0.458 | 26465.228 |
| Age | 0.023 | 0.006 | 0.012 – 0.034 | 4.130 | **<0.001** | 34074.792 | 0.006 | 0.006 | -0.006 – 0.018 | 0.965 | 0.335 | 26463.119 |
| Education (tertiary) | 0.074 | 0.006 | 0.062 – 0.085 | 12.883 | **<0.001** | 34074.233 | 0.063 | 0.006 | 0.050 – 0.075 | 9.807 | **<0.001** | 26462.725 |
| Income | 0.057 | 0.006 | 0.046 – 0.069 | 10.002 | **<0.001** | 34074.083 | 0.043 | 0.006 | 0.030 – 0.055 | 6.588 | **<0.001** | 26462.578 |
| Residence place (urban) | 0.050 | 0.006 | 0.039 – 0.061 | 8.855 | **<0.001** | 34074.206 | 0.050 | 0.006 | 0.037 – 0.062 | 7.856 | **<0.001** | 26462.569 |
| Political orientation (right) | | | | | | | 0.003 | 0.008 | -0.011 – 0.018 | 0.457 | 0.647 | 26462.439 |
| Political orientation (conservative) | | | | | | | -0.070 | 0.008 | -0.085 – -0.055 | -9.271 | **<0.001** | 26462.600 |
| Religiosity | | | | | | | 0.095 | 0.006 | 0.082 – 0.107 | 14.616 | **<0.001** | 26463.144 |
| Social dominance orientation | | | | | | | -0.150 | 0.007 | -0.163 – -0.137 | -23.039 | **<0.001** | 26462.861 |
| Science-related populist attitudes | | | | | | | | | | | | |
| Perceived benefit of science | | | | | | | | | | | | |
| Willingness to be vulnerable to science | | | | | | | | | | | | |
| Trust in the scientific method | | | | | | | | | | | | |
| GDP per capita | | | | | | | | | | | | |
| Govt expenditure on education (% of GDP) | | | | | | | | | | | | |
| Gini index | | | | | | | | | | | | |
| Science literacy (PISA) | | | | | | | | | | | | |
| Academic freedom | | | | | | | | | | | | |
| Degree of populism in politics | | | | | | | | | | | | |
| **Random Effects** | | | | | | | | | | | | |
| $\sigma^2$ | 1.07 | | | | | | 1 | | | | | |
| $\tau_{00}$ | 0.10 COUNTRY_NAME | | | | | | 0.09 COUNTRY_NAME | | | | | |
| ICC | 0.08 | | | | | | 0.09 | | | | | |
| N | 24 COUNTRY_NAME | | | | | | 24 COUNTRY_NAME | | | | | |
| Observations | 34103 | | | | | | 26495 | | | | | |
| Marginal $R^2$ / Conditional $R^2$ | 0.012 / 0.093 | | | | | | 0.039 / 0.122 | | | | | |
| AIC | 99132.911 | | | | | | 75472.877 | | | | | |

| | Block 3: Attitudes to science | | | | | | Block 4: Country indicators | | | | | |
|---|---|---|---|---|---|---|---|---|---|---|---|---|
| *Predictors* | *Beta* | *SE* | *CI* | *t* | *p* | *df* | *Beta* | *SE* | *CI* | *t* | *p* | *df* |
| Intercept | -0.303 | 0.064 | -0.436 – -0.170 | -4.712 | **<0.001** | 22.960 | -0.359 | 0.064 | -0.496 – -0.222 | -5.583 | **<0.001** | 14.970 |
| Sex (male) | -0.035 | 0.005 | -0.045 – -0.026 | -7.162 | **<0.001** | 26446.952 | -0.030 | 0.005 | -0.040 – -0.021 | -6.098 | **<0.001** | 25364.810 |
| Age | 0.003 | 0.005 | -0.006 – 0.013 | 0.660 | 0.509 | 26445.676 | 0.005 | 0.005 | -0.004 – 0.015 | 1.062 | 0.288 | 25363.786 |
| Education (tertiary) | -0.017 | 0.005 | -0.027 – -0.007 | -3.398 | **0.001** | 26445.426 | -0.014 | 0.005 | -0.024 – -0.004 | -2.815 | **0.005** | 25363.531 |
| Income | 0.003 | 0.005 | -0.007 – 0.013 | 0.583 | 0.560 | 26445.362 | 0.001 | 0.005 | -0.009 – 0.011 | 0.192 | 0.848 | 25363.488 |
| Residence place (urban) | 0.019 | 0.005 | 0.010 – 0.029 | 3.916 | **<0.001** | 26445.364 | 0.019 | 0.005 | 0.010 – 0.029 | 3.879 | **<0.001** | 25363.322 |
| Political orientation (right) | 0.011 | 0.006 | -0.001 – 0.022 | 1.819 | 0.069 | 26445.322 | 0.014 | 0.006 | 0.002 – 0.025 | 2.266 | **0.023** | 25363.237 |
| Political orientation (conservative) | -0.028 | 0.006 | -0.040 – -0.017 | -4.786 | **<0.001** | 26445.392 | -0.031 | 0.006 | -0.042 – -0.019 | -5.134 | **<0.001** | 25363.620 |
| Religiosity | 0.078 | 0.005 | 0.068 – 0.088 | 15.221 | **<0.001** | 26445.632 | 0.075 | 0.005 | 0.065 – 0.085 | 14.599 | **<0.001** | 25363.356 |
| Social dominance orientation | -0.025 | 0.005 | -0.036 – -0.015 | -4.876 | **<0.001** | 26445.385 | -0.032 | 0.005 | -0.042 – -0.021 | -6.002 | **<0.001** | 25363.396 |
| Science-related populist attitudes | -0.003 | 0.005 | -0.014 – 0.007 | -0.646 | 0.518 | 26445.812 | -0.002 | 0.005 | -0.012 – 0.009 | -0.310 | 0.757 | 25363.723 |
| Perceived benefit of science | 0.236 | 0.006 | 0.224 – 0.248 | 39.481 | **<0.001** | 26445.430 | 0.231 | 0.006 | 0.219 – 0.242 | 38.137 | **<0.001** | 25363.633 |
| Willingness to be vulnerable to science | 0.333 | 0.006 | 0.321 – 0.344 | 55.504 | **<0.001** | 26445.229 | 0.335 | 0.006 | 0.323 – 0.347 | 55.328 | **<0.001** | 25363.360 |
| Trust in the scientific method | 0.293 | 0.006 | 0.281 – 0.305 | 48.884 | **<0.001** | 26445.349 | 0.292 | 0.006 | 0.281 – 0.304 | 48.295 | **<0.001** | 25363.371 |
| GDP per capita | | | | | | | 0.085 | 0.060 | -0.043 – 0.213 | 1.418 | 0.176 | 15.229 |
| Govt expenditure on education (% of GDP) | | | | | | | 0.106 | 0.062 | -0.027 – 0.239 | 1.700 | 0.110 | 15.101 |
| Gini index | | | | | | | 0.225 | 0.070 | 0.075 – 0.375 | 3.207 | **0.006** | 14.985 |
| Science literacy (PISA) | | | | | | | 0.122 | 0.109 | -0.111 – 0.354 | 1.113 | 0.283 | 14.980 |
| Academic freedom | | | | | | | -0.081 | 0.073 | -0.236 – 0.073 | -1.120 | 0.280 | 15.250 |
| Degree of populism in politics | | | | | | | -0.057 | 0.071 | -0.208 – 0.095 | -0.798 | 0.437 | 14.900 |
| **Random Effects** | | | | | | | | | | | | |
| $\sigma^2$ | 0.61 | | | | | | 0.59 | | | | | |
| $\tau_{00}$ | 0.10 COUNTRY_NAME | | | | | | 0.05 COUNTRY_NAME | | | | | |
| ICC | 0.14 | | | | | | 0.08 | | | | | |
| N | 24 COUNTRY_NAME | | | | | | 22 COUNTRY_NAME | | | | | |
| Observations | 26482 | | | | | | 25398 | | | | | |
| Marginal $R^2$ / Conditional $R^2$ | 0.385 / 0.471 | | | | | | 0.425 / 0.469 | | | | | |
| AIC | 62181.813 | | | | | | 59048.592 | | | | | |

*Note.* Significant testing based on two-sided t tests. AIC = Akaike information criterion, ICC = Intraclass Correlation Coefficient. $\sigma^2$ = within country (residual) variance. $\tau_{00}$ = between-country variance (variation between individual intercepts and average intercept).

**Table S21. Weighted linear multilevel regression testing the association of perceived scientist trustworthiness with demographic characteristics, attitudes toward science, and country-level indicators in 56 countries/regions (random intercepts across countries/regions, aligned scores from unweighted country-specific CFA; Fig. S1, Path A2)**

| | Block 1: Demographic characteristics | | | | | | Block 2: Ideological views | | | | | |
|---|---|---|---|---|---|---|---|---|---|---|---|---|
| *Predictors* | *Beta* | *SE* | *CI* | *t* | *p* | *df* | *Beta* | *SE* | *CI* | *t* | *p* | *df* |
| Intercept | -0.250 | 0.061 | -0.373 – -0.127 | -4.085 | **<0.001** | 53.145 | -0.187 | 0.062 | -0.311 – -0.063 | -3.023 | **0.004** | 52.343 |
| Sex (male) | -0.038 | 0.005 | -0.048 – -0.028 | -7.295 | **<0.001** | 57544.401 | -0.022 | 0.006 | -0.033 – -0.010 | -3.669 | **<0.001** | 42858.882 |
| Age | 0.041 | 0.005 | 0.031 – 0.050 | 8.487 | **<0.001** | 57568.476 | 0.016 | 0.005 | 0.005 – 0.027 | 2.942 | **0.003** | 42882.379 |
| Education (tertiary) | 0.069 | 0.005 | 0.058 – 0.080 | 12.603 | **<0.001** | 57241.212 | 0.060 | 0.006 | 0.048 – 0.072 | 9.887 | **<0.001** | 42625.048 |
| Income | 0.070 | 0.005 | 0.060 – 0.080 | 13.434 | **<0.001** | 57548.928 | 0.049 | 0.006 | 0.038 – 0.061 | 8.210 | **<0.001** | 42862.850 |
| Residence place (urban) | 0.055 | 0.005 | 0.045 – 0.065 | 10.808 | **<0.001** | 57540.733 | 0.057 | 0.006 | 0.046 – 0.068 | 9.855 | **<0.001** | 42856.888 |
| Political orientation (right) | | | | | | | 0.009 | 0.007 | -0.004 – 0.023 | 1.326 | 0.185 | 42859.949 |
| Political orientation (conservative) | | | | | | | -0.066 | 0.007 | -0.080 – -0.052 | -9.401 | **<0.001** | 42863.981 |
| Religiosity | | | | | | | 0.098 | 0.006 | 0.087 – 0.110 | 16.305 | **<0.001** | 42859.183 |
| Social dominance orientation | | | | | | | -0.190 | 0.006 | -0.202 – -0.178 | -31.255 | **<0.001** | 42864.193 |
| Science-related populist attitudes | | | | | | | | | | | | |
| Perceived benefit of science | | | | | | | | | | | | |
| Willingness to be vulnerable to science | | | | | | | | | | | | |
| Trust in the scientific method | | | | | | | | | | | | |
| GDP per capita | | | | | | | | | | | | |
| Govt expenditure on education (% of GDP) | | | | | | | | | | | | |
| Gini index | | | | | | | | | | | | |
| Science literacy (PISA) | | | | | | | | | | | | |
| Academic freedom | | | | | | | | | | | | |
| Degree of populism in politics | | | | | | | | | | | | |
| **Random Effects** | | | | | | | | | | | | |
| $\sigma^2$ | 0.91 | | | | | | 0.86 | | | | | |
| $\tau_{00}$ | 0.20 COUNTRY_NAME | | | | | | 0.20 COUNTRY_NAME | | | | | |
| ICC | 0.18 | | | | | | 0.19 | | | | | |
| N | 56 COUNTRY_NAME | | | | | | 55 COUNTRY_NAME | | | | | |
| Observations | 57601 | | | | | | 42916 | | | | | |
| Marginal $R^2$ / Conditional $R^2$ | 0.016 / 0.195 | | | | | | 0.054 / 0.232 | | | | | |
| AIC | 221861.078 | | | | | | 161899.418 | | | | | |

| | Block 3: Attitudes to science | | | | | | Block 4: Country indicators | | | | | |
|---|---|---|---|---|---|---|---|---|---|---|---|---|
| *Predictors* | *Beta* | *SE* | *CI* | *t* | *p* | *df* | *Beta* | *SE* | *CI* | *t* | *p* | *df* |
| Intercept | -0.298 | 0.061 | -0.421 – -0.175 | -4.850 | **<0.001** | 53.100 | -0.317 | 0.056 | -0.431 – -0.204 | -5.663 | **<0.001** | 37.295 |
| Sex (male) | -0.046 | 0.005 | -0.055 – -0.037 | -9.763 | **<0.001** | 42289.544 | -0.044 | 0.005 | -0.054 – -0.034 | -8.968 | **<0.001** | 37956.180 |
| Age | 0.001 | 0.004 | -0.008 – 0.009 | 0.227 | 0.820 | 42306.260 | 0.002 | 0.005 | -0.007 – 0.011 | 0.490 | 0.624 | 37965.347 |
| Education (tertiary) | -0.026 | 0.005 | -0.036 – -0.016 | -5.342 | **<0.001** | 42263.749 | -0.026 | 0.005 | -0.036 – -0.016 | -5.094 | **<0.001** | 37987.905 |
| Income | 0.008 | 0.005 | -0.001 – 0.018 | 1.712 | 0.087 | 42292.592 | 0.009 | 0.005 | -0.001 – 0.019 | 1.808 | 0.071 | 37955.068 |
| Residence place (urban) | 0.024 | 0.005 | 0.015 – 0.033 | 5.133 | **<0.001** | 42288.358 | 0.026 | 0.005 | 0.017 – 0.036 | 5.429 | **<0.001** | 37954.298 |
| Political orientation (right) | 0.015 | 0.006 | 0.004 – 0.026 | 2.602 | **0.009** | 42290.348 | 0.016 | 0.006 | 0.004 – 0.028 | 2.665 | **0.008** | 37955.129 |
| Political orientation (conservative) | -0.017 | 0.006 | -0.028 – -0.006 | -3.044 | **0.002** | 42293.639 | -0.021 | 0.006 | -0.032 – -0.009 | -3.464 | **0.001** | 37955.012 |
| Religiosity | 0.095 | 0.005 | 0.086 – 0.105 | 19.550 | **<0.001** | 42289.667 | 0.096 | 0.005 | 0.086 – 0.106 | 18.752 | **<0.001** | 37954.713 |
| Social dominance orientation | -0.034 | 0.005 | -0.044 – -0.024 | -6.824 | **<0.001** | 42292.791 | -0.036 | 0.005 | -0.047 – -0.026 | -6.986 | **<0.001** | 37955.133 |
| Science-related populist attitudes | -0.030 | 0.005 | -0.040 – -0.020 | -6.066 | **<0.001** | 42288.581 | -0.029 | 0.005 | -0.039 – -0.019 | -5.651 | **<0.001** | 37955.155 |
| Perceived benefit of science | 0.280 | 0.006 | 0.269 – 0.291 | 50.718 | **<0.001** | 42292.907 | 0.276 | 0.006 | 0.265 – 0.287 | 47.702 | **<0.001** | 37956.120 |
| Willingness to be vulnerable to science | 0.347 | 0.006 | 0.336 – 0.359 | 61.552 | **<0.001** | 42289.229 | 0.352 | 0.006 | 0.341 – 0.364 | 59.540 | **<0.001** | 37956.199 |
| Trust in the scientific method | 0.355 | 0.006 | 0.344 – 0.366 | 63.004 | **<0.001** | 42290.654 | 0.354 | 0.006 | 0.343 – 0.366 | 59.957 | **<0.001** | 37955.348 |
| GDP per capita | | | | | | | 0.089 | 0.075 | -0.062 – 0.241 | 1.195 | 0.240 | 36.094 |
| Govt expenditure on education (% of GDP) | | | | | | | 0.055 | 0.062 | -0.071 – 0.181 | 0.883 | 0.383 | 36.555 |
| Gini index | | | | | | | 0.249 | 0.070 | 0.107 – 0.392 | 3.550 | **0.001** | 36.999 |
| Science literacy (PISA) | | | | | | | 0.035 | 0.063 | -0.094 – 0.163 | 0.545 | 0.589 | 39.351 |
| Academic freedom | | | | | | | -0.063 | 0.066 | -0.196 – 0.070 | -0.954 | 0.346 | 37.594 |
| Degree of populism in politics | | | | | | | -0.041 | 0.055 | -0.152 – 0.070 | -0.748 | 0.459 | 37.640 |
| **Random Effects** | | | | | | | | | | | | |
| $\sigma^2$ | 0.54 | | | | | | 0.57 | | | | | |
| $\tau_{00}$ | 0.20 COUNTRY_NAME | | | | | | 0.11 COUNTRY_NAME | | | | | |
| ICC | 0.27 | | | | | | 0.16 | | | | | |
| N | 55 COUNTRY_NAME | | | | | | 43 COUNTRY_NAME | | | | | |
| Observations | 42353 | | | | | | 38008 | | | | | |
| Marginal $R^2$ / Conditional $R^2$ | 0.433 / 0.587 | | | | | | 0.470 / 0.556 | | | | | |
| AIC | 140236.492 | | | | | | 117319.599 | | | | | |

*Note.* Significant testing based on two-sided t tests. AIC = Akaike information criterion, ICC = Intraclass Correlation Coefficient. $\sigma^2$ = within country (residual) variance. $\tau_{00}$ = between-country variance (variation between individual intercepts and average intercept).

**Table S22. Unweighted linear multilevel regression testing the association of perceived scientist trustworthiness with demographic characteristics, attitudes toward science, and country-level indicators in 56 countries/regions (random intercepts across countries/regions, aligned scores from unweighted country-specific CFA; Fig. S1, Path A2)**

| | **Block 1: Demographic characteristics** | | | | | | **Block 2: Ideological views** | | | | | |
|---|---|---|---|---|---|---|---|---|---|---|---|---|
| *Predictors* | *Beta* | *SE* | *CI* | *t* | *p* | *df* | *Beta* | *SE* | *CI* | *t* | *p* | *df* |
| Intercept | -0.313 | 0.056 | -0.425 – -0.201 | -5.587 | **<0.001** | 54.934 | -0.243 | 0.058 | -0.358 – -0.127 | -4.215 | **<0.001** | 53.931 |
| Sex (male) | -0.033 | 0.005 | -0.043 – -0.023 | -6.338 | **<0.001** | 57541.092 | -0.026 | 0.006 | -0.038 – -0.015 | -4.462 | **<0.001** | 42856.856 |
| Age | 0.016 | 0.005 | 0.006 – 0.026 | 3.134 | **0.002** | 57541.605 | -0.001 | 0.006 | -0.012 – 0.010 | -0.184 | 0.854 | 42854.985 |
| Education (tertiary) | 0.066 | 0.005 | 0.056 – 0.076 | 12.537 | **<0.001** | 57540.454 | 0.058 | 0.006 | 0.046 – 0.069 | 9.550 | **<0.001** | 42853.414 |
| Income | 0.056 | 0.005 | 0.046 – 0.066 | 10.635 | **<0.001** | 57540.933 | 0.043 | 0.006 | 0.031 – 0.055 | 7.108 | **<0.001** | 42853.683 |
| Residence place (urban) | 0.043 | 0.005 | 0.033 – 0.054 | 8.317 | **<0.001** | 57540.386 | 0.046 | 0.006 | 0.034 – 0.057 | 7.611 | **<0.001** | 42853.154 |
| Political orientation (right) | | | | | | | 0.021 | 0.007 | 0.007 – 0.035 | 3.028 | **0.002** | 42853.245 |
| Political orientation (conservative) | | | | | | | -0.063 | 0.007 | -0.076 – -0.049 | -8.978 | **<0.001** | 42854.732 |
| Religiosity | | | | | | | 0.121 | 0.006 | 0.109 – 0.133 | 19.931 | **<0.001** | 42854.644 |
| Social dominance orientation | | | | | | | -0.148 | 0.006 | -0.160 – -0.136 | -24.442 | **<0.001** | 42853.862 |
| Science-related populist attitudes | | | | | | | | | | | | |
| Perceived benefit of science | | | | | | | | | | | | |
| Willingness to be vulnerable to science | | | | | | | | | | | | |
| Trust in the scientific method | | | | | | | | | | | | |
| GDP per capita | | | | | | | | | | | | |
| Govt expenditure on education (% of GDP) | | | | | | | | | | | | |
| Gini index | | | | | | | | | | | | |
| Science literacy (PISA) | | | | | | | | | | | | |
| Academic freedom | | | | | | | | | | | | |
| Degree of populism in politics | | | | | | | | | | | | |
| **Random Effects** | | | | | | | | | | | | |
| $\sigma^2$ | 1.52 | | | | | | 1.43 | | | | | |
| $\tau_{00}$ | 0.17 COUNTRY_NAME | | | | | | 0.18 COUNTRY_NAME | | | | | |
| ICC | 0.10 | | | | | | 0.11 | | | | | |
| N | 56 COUNTRY_NAME | | | | | | 55 COUNTRY_NAME | | | | | |
| Observations | 57601 | | | | | | 42916 | | | | | |
| Marginal $R^2$ / Conditional $R^2$ | 0.007 / 0.109 | | | | | | 0.028 / 0.136 | | | | | |
| AIC | 187714.197 | | | | | | 137580.835 | | | | | |

| | Block 3: Attitudes to science | | | | | | Block 4: Country indicators | | | | | |
|---|---|---|---|---|---|---|---|---|---|---|---|---|
| *Predictors* | *Beta* | *SE* | *CI* | *t* | *p* | *df* | *Beta* | *SE* | *CI* | *t* | *p* | *df* |
| Intercept | -0.302 | 0.059 | -0.420 – -0.185 | -5.154 | **<0.001** | 53.985 | -0.329 | 0.056 | -0.443 – -0.216 | -5.885 | **<0.001** | 36.053 |
| Sex (male) | -0.053 | 0.005 | -0.062 – -0.043 | -10.894 | **<0.001** | 42288.218 | -0.047 | 0.005 | -0.056 – -0.037 | -9.311 | **<0.001** | 37954.629 |
| Age | -0.010 | 0.005 | -0.020 – -0.001 | -2.119 | **0.034** | 42287.041 | -0.006 | 0.005 | -0.016 – 0.004 | -1.239 | 0.215 | 37953.796 |
| Education (tertiary) | -0.019 | 0.005 | -0.029 – -0.009 | -3.841 | **<0.001** | 42286.234 | -0.021 | 0.005 | -0.031 – -0.011 | -4.165 | **<0.001** | 37953.375 |
| Income | 0.006 | 0.005 | -0.004 – 0.015 | 1.103 | 0.270 | 42286.126 | 0.004 | 0.005 | -0.006 – 0.014 | 0.711 | 0.477 | 37953.310 |
| Residence place (urban) | 0.017 | 0.005 | 0.008 – 0.027 | 3.547 | **<0.001** | 42285.774 | 0.019 | 0.005 | 0.009 – 0.029 | 3.781 | **<0.001** | 37952.731 |
| Political orientation (right) | 0.015 | 0.006 | 0.004 – 0.027 | 2.679 | **0.007** | 42285.885 | 0.015 | 0.006 | 0.003 – 0.027 | 2.464 | **0.014** | 37952.701 |
| Political orientation (conservative) | -0.018 | 0.006 | -0.030 – -0.007 | -3.155 | **0.002** | 42286.927 | -0.021 | 0.006 | -0.033 – -0.009 | -3.525 | **<0.001** | 37953.536 |
| Religiosity | 0.102 | 0.005 | 0.092 – 0.112 | 20.266 | **<0.001** | 42286.801 | 0.100 | 0.005 | 0.090 – 0.111 | 19.291 | **<0.001** | 37953.586 |
| Social dominance orientation | -0.012 | 0.005 | -0.022 – -0.002 | -2.280 | **0.023** | 42286.180 | -0.021 | 0.005 | -0.031 – -0.010 | -3.888 | **<0.001** | 37952.921 |
| Science-related populist attitudes | -0.004 | 0.005 | -0.014 – 0.006 | -0.733 | 0.464 | 42286.972 | -0.003 | 0.005 | -0.013 – 0.007 | -0.602 | 0.547 | 37953.766 |
| Perceived benefit of science | 0.283 | 0.006 | 0.272 – 0.294 | 49.184 | **<0.001** | 42286.076 | 0.278 | 0.006 | 0.267 – 0.290 | 46.829 | **<0.001** | 37952.864 |
| Willingness to be vulnerable to science | 0.344 | 0.006 | 0.333 – 0.356 | 58.626 | **<0.001** | 42285.975 | 0.358 | 0.006 | 0.346 – 0.370 | 59.119 | **<0.001** | 37953.108 |
| Trust in the scientific method | 0.339 | 0.006 | 0.328 – 0.351 | 58.270 | **<0.001** | 42285.923 | 0.339 | 0.006 | 0.327 – 0.351 | 56.321 | **<0.001** | 37952.698 |
| GDP per capita | | | | | | | 0.100 | 0.075 | -0.052 – 0.253 | 1.331 | 0.192 | 36.250 |
| Govt expenditure on education (% of GDP) | | | | | | | 0.031 | 0.063 | -0.096 – 0.158 | 0.492 | 0.626 | 36.228 |
| Gini index | | | | | | | 0.246 | 0.070 | 0.103 – 0.389 | 3.490 | **0.001** | 36.293 |
| Science literacy (PISA) | | | | | | | 0.047 | 0.063 | -0.080 – 0.174 | 0.750 | 0.458 | 36.488 |
| Academic freedom | | | | | | | -0.044 | 0.066 | -0.177 – 0.089 | -0.668 | 0.508 | 36.474 |
| Degree of populism in politics | | | | | | | -0.024 | 0.055 | -0.135 – 0.087 | -0.438 | 0.664 | 36.581 |
| **Random Effects** | | | | | | | | | | | | |
| $\sigma^2$ | 0.95 | | | | | | 0.90 | | | | | |
| $\tau_{00}$ | 0.19 COUNTRY_NAME | | | | | | 0.11 COUNTRY_NAME | | | | | |
| ICC | 0.16 | | | | | | 0.11 | | | | | |
| N | 55 COUNTRY_NAME | | | | | | 43 COUNTRY_NAME | | | | | |
| Observations | 42353 | | | | | | 38008 | | | | | |
| Marginal $R^2$ / Conditional $R^2$ | 0.318 / 0.429 | | | | | | 0.363 / 0.434 | | | | | |
| AIC | 118634.047 | | | | | | 104416.093 | | | | | |

*Note.* Significant testing based on two-sided t tests. AIC = Akaike information criterion, ICC = Intraclass Correlation Coefficient. $\sigma^2$ = within country (residual) variance. $\tau_{00}$ = between-country variance (variation between individual intercepts and average intercept).

**Table S23. Weighted linear multilevel regression testing the association of perceived scientist trustworthiness with demographic characteristics, attitudes toward science, and country-level indicators in 34 countries/regions (random intercepts across countries/regions, aligned scores from unweighted country-specific CFA; Fig. S1, Path A2)**

| | **Block 1: Demographic characteristics** | | | | | | **Block 2: Ideological views** | | | | | |
|---|---|---|---|---|---|---|---|---|---|---|---|---|
| *Predictors* | *Beta* | *SE* | *CI* | *t* | *p* | *df* | *Beta* | *SE* | *CI* | *t* | *p* | *df* |
| Intercept | -0.231 | 0.070 | -0.373 – -0.089 | -3.312 | **0.002** | 30.618 | -0.160 | 0.068 | -0.299 – -0.021 | -2.356 | **0.025** | 29.569 |
| Sex (male) | -0.033 | 0.006 | -0.044 – -0.021 | -5.614 | **<0.001** | 39784.239 | -0.013 | 0.007 | -0.025 – 0.000 | -1.921 | 0.055 | 30288.385 |
| Age | 0.038 | 0.005 | 0.028 – 0.049 | 7.084 | **<0.001** | 39800.175 | 0.017 | 0.006 | 0.005 – 0.029 | 2.840 | **0.005** | 30302.254 |
| Education (tertiary) | 0.074 | 0.006 | 0.062 – 0.086 | 11.861 | **<0.001** | 39285.138 | 0.061 | 0.007 | 0.047 – 0.074 | 8.898 | **<0.001** | 29825.298 |
| Income | 0.079 | 0.006 | 0.068 – 0.091 | 13.421 | **<0.001** | 39788.284 | 0.060 | 0.007 | 0.047 – 0.073 | 8.882 | **<0.001** | 30291.817 |
| Residence place (urban) | 0.059 | 0.006 | 0.048 – 0.070 | 10.320 | **<0.001** | 39781.798 | 0.059 | 0.006 | 0.047 – 0.072 | 9.263 | **<0.001** | 30285.835 |
| Political orientation (right) | | | | | | | 0.009 | 0.008 | -0.007 – 0.024 | 1.069 | 0.285 | 30287.996 |
| Political orientation (conservative) | | | | | | | -0.069 | 0.008 | -0.085 – -0.054 | -8.723 | **<0.001** | 30290.593 |
| Religiosity | | | | | | | 0.101 | 0.007 | 0.088 – 0.114 | 15.009 | **<0.001** | 30288.212 |
| Social dominance orientation | | | | | | | -0.201 | 0.007 | -0.214 – -0.187 | -29.601 | **<0.001** | 30288.446 |
| Science-related populist attitudes | | | | | | | | | | | | |
| Perceived benefit of science | | | | | | | | | | | | |
| Willingness to be vulnerable to science | | | | | | | | | | | | |
| Trust in the scientific method | | | | | | | | | | | | |
| GDP per capita | | | | | | | | | | | | |
| Govt expenditure on education (% of GDP) | | | | | | | | | | | | |
| Gini index | | | | | | | | | | | | |
| Science literacy (PISA) | | | | | | | | | | | | |
| Academic freedom | | | | | | | | | | | | |
| Degree of populism in politics | | | | | | | | | | | | |
| **Random Effects** | | | | | | | | | | | | |
| $\sigma^2$ | 0.81 | | | | | | 0.76 | | | | | |
| $\tau_{00}$ | 0.16 COUNTRY_NAME | | | | | | 0.14 COUNTRY_NAME | | | | | |
| ICC | 0.16 | | | | | | 0.16 | | | | | |
| N | 34 COUNTRY_NAME | | | | | | 33 COUNTRY_NAME | | | | | |
| Observations | 39821 | | | | | | 30325 | | | | | |
| Marginal $R^2$ / Conditional $R^2$ | 0.020 / 0.180 | | | | | | 0.070 / 0.217 | | | | | |
| AIC | 149147.009 | | | | | | 110402.631 | | | | | |

| | Block 3: Attitudes to science | | | | | | Block 4: Country indicators | | | | | |
|---|---|---|---|---|---|---|---|---|---|---|---|---|
| *Predictors* | *Beta* | *SE* | *CI* | *t* | *p* | *df* | *Beta* | *SE* | *CI* | *t* | *p* | *df* |
| Intercept | -0.274 | 0.069 | -0.414 – -0.134 | -3.980 | **<0.001** | 30.713 | -0.349 | 0.062 | -0.479 – -0.218 | -5.586 | **<0.001** | 19.545 |
| Sex (male) | -0.043 | 0.005 | -0.053 – -0.033 | -8.581 | **<0.001** | 30267.282 | -0.043 | 0.005 | -0.053 – -0.033 | -8.212 | **<0.001** | 27831.643 |
| Age | 0.013 | 0.005 | 0.004 – 0.022 | 2.833 | **0.005** | 30276.348 | 0.013 | 0.005 | 0.003 – 0.022 | 2.602 | **0.009** | 27834.683 |
| Education (tertiary) | -0.025 | 0.005 | -0.036 – -0.015 | -4.805 | **<0.001** | 30173.747 | -0.024 | 0.005 | -0.035 – -0.014 | -4.485 | **<0.001** | 27844.101 |
| Income | 0.011 | 0.005 | 0.001 – 0.022 | 2.217 | **0.027** | 30269.812 | 0.011 | 0.005 | -0.000 – 0.021 | 1.954 | 0.051 | 27830.486 |
| Residence place (urban) | 0.024 | 0.005 | 0.014 – 0.034 | 4.905 | **<0.001** | 30265.920 | 0.025 | 0.005 | 0.015 – 0.035 | 4.921 | **<0.001** | 27830.108 |
| Political orientation (right) | 0.019 | 0.006 | 0.006 – 0.031 | 3.008 | **0.003** | 30267.098 | 0.019 | 0.006 | 0.006 – 0.032 | 2.948 | **0.003** | 27831.105 |
| Political orientation (conservative) | -0.022 | 0.006 | -0.034 – -0.010 | -3.676 | **<0.001** | 30269.127 | -0.022 | 0.006 | -0.035 – -0.010 | -3.516 | **<0.001** | 27830.804 |
| Religiosity | 0.083 | 0.005 | 0.073 – 0.094 | 15.982 | **<0.001** | 30266.820 | 0.082 | 0.005 | 0.072 – 0.093 | 15.190 | **<0.001** | 27830.384 |
| Social dominance orientation | -0.047 | 0.005 | -0.057 – -0.036 | -8.744 | **<0.001** | 30266.487 | -0.049 | 0.006 | -0.060 – -0.038 | -8.834 | **<0.001** | 27830.610 |
| Science-related populist attitudes | -0.027 | 0.005 | -0.037 – -0.016 | -5.038 | **<0.001** | 30266.092 | -0.027 | 0.006 | -0.038 – -0.016 | -4.800 | **<0.001** | 27830.215 |
| Perceived benefit of science | 0.257 | 0.006 | 0.245 – 0.269 | 42.768 | **<0.001** | 30270.703 | 0.252 | 0.006 | 0.240 – 0.264 | 40.329 | **<0.001** | 27830.628 |
| Willingness to be vulnerable to science | 0.356 | 0.006 | 0.344 – 0.368 | 58.582 | **<0.001** | 30266.856 | 0.358 | 0.006 | 0.346 – 0.371 | 56.588 | **<0.001** | 27831.386 |
| Trust in the scientific method | 0.338 | 0.006 | 0.326 – 0.350 | 55.494 | **<0.001** | 30267.679 | 0.339 | 0.006 | 0.326 – 0.351 | 53.336 | **<0.001** | 27831.202 |
| GDP per capita | | | | | | | 0.107 | 0.075 | -0.049 – 0.264 | 1.441 | 0.166 | 19.171 |
| Govt expenditure on education (% of GDP) | | | | | | | 0.100 | 0.076 | -0.058 – 0.259 | 1.328 | 0.200 | 19.013 |
| Gini index | | | | | | | 0.138 | 0.100 | -0.072 – 0.349 | 1.379 | 0.184 | 19.069 |
| Science literacy (PISA) | | | | | | | -0.047 | 0.072 | -0.198 – 0.103 | -0.655 | 0.520 | 20.951 |
| Academic freedom | | | | | | | -0.144 | 0.074 | -0.299 – 0.011 | -1.946 | 0.066 | 19.461 |
| Degree of populism in politics | | | | | | | -0.034 | 0.079 | -0.198 – 0.131 | -0.427 | 0.674 | 19.279 |
| **Random Effects** | | | | | | | | | | | | |
| $\sigma^2$ | 0.45 | | | | | | 0.47 | | | | | |
| $\tau_{00}$ | 0.15 COUNTRY_NAME | | | | | | 0.08 COUNTRY_NAME | | | | | |
| ICC | 0.25 | | | | | | 0.15 | | | | | |
| N | 33 COUNTRY_NAME | | | | | | 26 COUNTRY_NAME | | | | | |
| Observations | 30310 | | | | | | 27868 | | | | | |
| Marginal $R^2$ / Conditional $R^2$ | 0.478 / 0.610 | | | | | | 0.511 / 0.585 | | | | | |
| AIC | 94108.885 | | | | | | 81572.805 | | | | | |

*Note.* Significant testing based on two-sided t tests. AIC = Akaike information criterion, ICC = Intraclass Correlation Coefficient. $\sigma^2$ = within country (residual) variance. $\tau_{00}$ = between-country variance (variation between individual intercepts and average intercept).

**Table S24. Unweighted linear multilevel regression testing the association of perceived scientist trustworthiness with demographic characteristics, attitudes toward science, and country-level indicators in 34 countries/regions (random intercepts across countries/regions, aligned scores from unweighted country-specific CFA; Fig. S1, Path A2)**

| | **Block 1: Demographic characteristics** | | | | | | **Block 2: Ideological views** | | | | | |
|---|---|---|---|---|---|---|---|---|---|---|---|---|
| *Predictors* | *Beta* | *SE* | *CI* | *t* | *p* | *df* | *Beta* | *SE* | *CI* | *t* | *p* | *df* |
| Intercept | -0.293 | 0.062 | -0.419 – -0.167 | -4.725 | **<0.001** | 32.857 | -0.215 | 0.062 | -0.341 – -0.088 | -3.465 | **0.002** | 31.853 |
| Sex (male) | -0.030 | 0.006 | -0.041 – -0.018 | -5.121 | **<0.001** | 39783.092 | -0.023 | 0.007 | -0.036 – -0.010 | -3.507 | **<0.001** | 30286.073 |
| Age | 0.017 | 0.006 | 0.005 – 0.028 | 2.855 | **0.004** | 39782.688 | 0.002 | 0.006 | -0.011 – 0.014 | 0.242 | 0.809 | 30285.008 |
| Education (tertiary) | 0.067 | 0.006 | 0.055 – 0.079 | 11.283 | **<0.001** | 39782.196 | 0.056 | 0.007 | 0.043 – 0.070 | 8.416 | **<0.001** | 30283.847 |
| Income | 0.062 | 0.006 | 0.050 – 0.073 | 10.389 | **<0.001** | 39782.443 | 0.049 | 0.007 | 0.036 – 0.063 | 7.249 | **<0.001** | 30283.803 |
| Residence place (urban) | 0.046 | 0.006 | 0.035 – 0.058 | 7.821 | **<0.001** | 39782.222 | 0.047 | 0.007 | 0.034 – 0.060 | 7.032 | **<0.001** | 30283.570 |
| Political orientation (right) | | | | | | | 0.010 | 0.008 | -0.005 – 0.025 | 1.284 | 0.199 | 30283.831 |
| Political orientation (conservative) | | | | | | | -0.060 | 0.008 | -0.075 – -0.044 | -7.577 | **<0.001** | 30284.791 |
| Religiosity | | | | | | | 0.120 | 0.007 | 0.107 – 0.133 | 17.698 | **<0.001** | 30285.404 |
| Social dominance orientation | | | | | | | -0.154 | 0.007 | -0.167 – -0.141 | -22.800 | **<0.001** | 30284.747 |
| Science-related populist attitudes | | | | | | | | | | | | |
| Perceived benefit of science | | | | | | | | | | | | |
| Willingness to be vulnerable to science | | | | | | | | | | | | |
| Trust in the scientific method | | | | | | | | | | | | |
| GDP per capita | | | | | | | | | | | | |
| Govt expenditure on education (% of GDP) | | | | | | | | | | | | |
| Gini index | | | | | | | | | | | | |
| Science literacy (PISA) | | | | | | | | | | | | |
| Academic freedom | | | | | | | | | | | | |
| Degree of populism in politics | | | | | | | | | | | | |
| **Random Effects** | | | | | | | | | | | | |
| $\sigma^2$ | 1.33 | | | | | | 1.25 | | | | | |
| $\tau_{00}$ | 0.13 COUNTRY_NAME | | | | | | 0.12 COUNTRY_NAME | | | | | |
| ICC | 0.09 | | | | | | 0.09 | | | | | |
| N | 34 COUNTRY_NAME | | | | | | 33 COUNTRY_NAME | | | | | |
| Observations | 39821 | | | | | | 30325 | | | | | |
| Marginal $R^2$ / Conditional $R^2$ | 0.009 / 0.096 | | | | | | 0.034 / 0.122 | | | | | |
| AIC | 124653.29 | | | | | | 93011.595 | | | | | |

| | Block 3: Attitudes to science | | | | | | Block 4: Country indicators | | | | | |
|---|---|---|---|---|---|---|---|---|---|---|---|---|
| *Predictors* | *Beta* | *SE* | *CI* | *t* | *p* | *df* | *Beta* | *SE* | *CI* | *t* | *p* | *df* |
| Intercept | -0.276 | 0.064 | -0.406 – -0.145 | -4.289 | **<0.001** | 31.920 | -0.351 | 0.060 | -0.478 – -0.225 | -5.809 | **<0.001** | 19.062 |
| Sex (male) | -0.050 | 0.005 | -0.060 – -0.040 | -9.593 | **<0.001** | 30265.768 | -0.046 | 0.005 | -0.056 – -0.035 | -8.597 | **<0.001** | 27830.153 |
| Age | 0.002 | 0.005 | -0.008 – 0.012 | 0.462 | 0.644 | 30265.224 | 0.006 | 0.005 | -0.004 – 0.016 | 1.135 | 0.256 | 27830.141 |
| Education (tertiary) | -0.022 | 0.005 | -0.032 – -0.011 | -4.041 | **<0.001** | 30264.454 | -0.021 | 0.005 | -0.032 – -0.010 | -3.876 | **<0.001** | 27829.570 |
| Income | 0.006 | 0.005 | -0.005 – 0.016 | 1.084 | 0.278 | 30264.478 | 0.003 | 0.006 | -0.008 – 0.014 | 0.562 | 0.574 | 27829.613 |
| Residence place (urban) | 0.015 | 0.005 | 0.005 – 0.025 | 2.821 | **0.005** | 30264.346 | 0.016 | 0.005 | 0.005 – 0.026 | 2.887 | **0.004** | 27829.275 |
| Political orientation (right) | 0.013 | 0.006 | 0.001 – 0.025 | 2.061 | **0.039** | 30264.546 | 0.017 | 0.006 | 0.005 – 0.030 | 2.660 | **0.008** | 27829.511 |
| Political orientation (conservative) | -0.018 | 0.006 | -0.031 – -0.006 | -2.928 | **0.003** | 30265.069 | -0.020 | 0.006 | -0.033 – -0.008 | -3.140 | **0.002** | 27829.784 |
| Religiosity | 0.091 | 0.005 | 0.080 – 0.101 | 16.738 | **<0.001** | 30265.433 | 0.086 | 0.006 | 0.075 – 0.097 | 15.447 | **<0.001** | 27830.050 |
| Social dominance orientation | -0.019 | 0.006 | -0.030 – -0.009 | -3.518 | **<0.001** | 30264.958 | -0.030 | 0.006 | -0.041 – -0.019 | -5.349 | **<0.001** | 27829.628 |
| Science-related populist attitudes | -0.008 | 0.005 | -0.019 – 0.003 | -1.500 | 0.134 | 30264.802 | -0.004 | 0.006 | -0.015 – 0.007 | -0.745 | 0.456 | 27829.525 |
| Perceived benefit of science | 0.262 | 0.006 | 0.250 – 0.274 | 41.750 | **<0.001** | 30264.733 | 0.256 | 0.006 | 0.244 – 0.269 | 39.727 | **<0.001** | 27829.685 |
| Willingness to be vulnerable to science | 0.356 | 0.006 | 0.344 – 0.369 | 56.132 | **<0.001** | 30264.448 | 0.364 | 0.007 | 0.352 – 0.377 | 55.986 | **<0.001** | 27829.596 |
| Trust in the scientific method | 0.320 | 0.006 | 0.307 – 0.332 | 50.521 | **<0.001** | 30264.466 | 0.320 | 0.006 | 0.307 – 0.333 | 49.223 | **<0.001** | 27829.487 |
| GDP per capita | | | | | | | 0.111 | 0.073 | -0.041 – 0.263 | 1.528 | 0.143 | 19.296 |
| Govt expenditure on education (% of GDP) | | | | | | | 0.100 | 0.074 | -0.055 – 0.254 | 1.351 | 0.192 | 19.133 |
| Gini index | | | | | | | 0.135 | 0.098 | -0.070 – 0.340 | 1.382 | 0.183 | 19.113 |
| Science literacy (PISA) | | | | | | | -0.053 | 0.069 | -0.197 – 0.091 | -0.774 | 0.448 | 19.199 |
| Academic freedom | | | | | | | -0.138 | 0.072 | -0.288 – 0.012 | -1.918 | 0.070 | 19.249 |
| Degree of populism in politics | | | | | | | -0.028 | 0.076 | -0.188 – 0.131 | -0.371 | 0.714 | 19.012 |
| **Random Effects** | | | | | | | | | | | | |
| $\sigma^2$ | 0.78 | | | | | | 0.75 | | | | | |
| $\tau_{00}$ | 0.13 COUNTRY_NAME | | | | | | 0.08 COUNTRY_NAME | | | | | |
| ICC | 0.15 | | | | | | 0.10 | | | | | |
| N | 33 COUNTRY_NAME | | | | | | 26 COUNTRY_NAME | | | | | |
| Observations | 30310 | | | | | | 27868 | | | | | |
| Marginal $R^2$ / Conditional $R^2$ | 0.361 / 0.455 | | | | | | 0.401 / 0.458 | | | | | |
| AIC | 78802.796 | | | | | | 71372.611 | | | | | |

*Note.* Significant testing based on two-sided t tests. AIC = Akaike information criterion, ICC = Intraclass Correlation Coefficient. $\sigma^2$ = within country (residual) variance. $\tau_{00}$ = between-country variance (variation between individual intercepts and average intercept).

**Table S25. Weighted linear multilevel regression testing the association of perceived scientist trustworthiness with demographic characteristics, attitudes to science, and country-level indicators in 68 Countries/regions (random intercepts across countries/regions, aligned scores from pooled weighted CFA; Fig. S1, Path B1)**

| | **Block 1: Demographic characteristics** | | | | | | **Block 2: Ideological views** | | | | | |
|---|---|---|---|---|---|---|---|---|---|---|---|---|
| *Predictors* | *Beta* | *SE* | *CI* | *t* | *p* | *df* | *Beta* | *SE* | *CI* | *t* | *p* | *df* |
| Intercept | 1.127 | 0.115 | 0.897 – 1.356 | 9.799 | **<0.001** | 65.686 | 1.276 | 0.117 | 1.042 – 1.511 | 10.866 | **<0.001** | 64.880 |
| Sex (male) | -0.072 | 0.010 | -0.093 – -0.052 | -6.976 | **<0.001** | 64364.074 | -0.050 | 0.012 | -0.074 – -0.027 | -4.188 | **<0.001** | 47587.699 |
| Age | 0.036 | 0.010 | 0.017 – 0.055 | 3.721 | **<0.001** | 64402.848 | -0.019 | 0.011 | -0.040 – 0.003 | -1.672 | 0.095 | 47624.554 |
| Education (tertiary) | 0.116 | 0.011 | 0.095 – 0.138 | 10.538 | **<0.001** | 63832.530 | 0.103 | 0.012 | 0.079 – 0.127 | 8.350 | **<0.001** | 47215.631 |
| Income | 0.105 | 0.010 | 0.084 – 0.125 | 10.021 | **<0.001** | 64371.496 | 0.075 | 0.012 | 0.051 – 0.099 | 6.098 | **<0.001** | 47591.255 |
| Residence place (urban) | 0.117 | 0.010 | 0.098 – 0.137 | 11.571 | **<0.001** | 64360.617 | 0.126 | 0.012 | 0.103 – 0.149 | 10.699 | **<0.001** | 47586.888 |
| Political orientation (right) | | | | | | | 0.042 | 0.014 | 0.014 – 0.070 | 2.933 | **0.003** | 47590.092 |
| Political orientation (conservative) | | | | | | | -0.132 | 0.014 | -0.160 – -0.104 | -9.280 | **<0.001** | 47592.852 |
| Religiosity | | | | | | | 0.235 | 0.012 | 0.211 – 0.260 | 19.094 | **<0.001** | 47589.008 |
| Social dominance orientation | | | | | | | -0.278 | 0.012 | -0.302 – -0.254 | -22.518 | **<0.001** | 47594.823 |
| Science-related populist attitudes | | | | | | | | | | | | |
| Perceived benefit of science | | | | | | | | | | | | |
| Willingness to be vulnerable to science | | | | | | | | | | | | |
| Trust in the scientific method | | | | | | | | | | | | |
| GDP per capita | | | | | | | | | | | | |
| Govt expenditure on education (% of GDP) | | | | | | | | | | | | |
| Gini index | | | | | | | | | | | | |
| Science literacy (PISA) | | | | | | | | | | | | |
| Academic freedom | | | | | | | | | | | | |
| Degree of populism in politics | | | | | | | | | | | | |
| **Random Effects** | | | | | | | | | | | | |
| $\sigma^2$ | 4.13 | | | | | | 3.97 | | | | | |
| $\tau_{00}$ | 0.86 COUNTRY_NAME | | | | | | 0.87 COUNTRY_NAME | | | | | |
| ICC | 0.17 | | | | | | 0.18 | | | | | |
| N | 68 COUNTRY_NAME | | | | | | 67 COUNTRY_NAME | | | | | |
| Observations | 64431 | | | | | | 47654 | | | | | |
| Marginal $R^2$ / Conditional $R^2$ | 0.010 / 0.181 | | | | | | 0.035 / 0.209 | | | | | |
| AIC | 346520.165 | | | | | | 253572.438 | | | | | |

| | Block 3: Attitudes to science | | | | | | Block 4: Country indicators | | | | | |
|---|---|---|---|---|---|---|---|---|---|---|---|---|
| *Predictors* | *Beta* | *SE* | *CI* | *t* | *p* | *df* | *Beta* | *SE* | *CI* | *t* | *p* | *df* |
| Intercept | 1.039 | 0.117 | 0.806 – 1.273 | 8.903 | **<0.001** | 64.305 | 0.928 | 0.118 | 0.690 – 1.166 | 7.848 | **<0.001** | 44.580 |
| Sex (male) | -0.098 | 0.010 | -0.118 – -0.078 | -9.665 | **<0.001** | 46626.256 | -0.094 | 0.011 | -0.115 – -0.073 | -8.820 | **<0.001** | 41562.081 |
| Age | -0.047 | 0.009 | -0.065 – -0.028 | -4.996 | **<0.001** | 46656.007 | -0.046 | 0.010 | -0.065 – -0.026 | -4.641 | **<0.001** | 41577.375 |
| Education (tertiary) | -0.058 | 0.010 | -0.078 – -0.037 | -5.516 | **<0.001** | 46497.021 | -0.057 | 0.011 | -0.079 – -0.035 | -5.194 | **<0.001** | 41599.418 |
| Income | 0.001 | 0.010 | -0.020 – 0.021 | 0.053 | 0.958 | 46629.039 | 0.001 | 0.011 | -0.021 – 0.022 | 0.055 | 0.956 | 41560.786 |
| Residence place (urban) | 0.061 | 0.010 | 0.041 – 0.080 | 6.117 | **<0.001** | 46625.767 | 0.068 | 0.010 | 0.047 – 0.088 | 6.486 | **<0.001** | 41560.135 |
| Political orientation (right) | 0.040 | 0.012 | 0.017 – 0.064 | 3.343 | **0.001** | 46627.874 | 0.041 | 0.013 | 0.016 – 0.066 | 3.248 | **0.001** | 41561.332 |
| Political orientation (conservative) | -0.042 | 0.012 | -0.065 – -0.018 | -3.460 | **0.001** | 46630.451 | -0.048 | 0.013 | -0.073 – -0.023 | -3.768 | **<0.001** | 41560.511 |
| Religiosity | 0.221 | 0.010 | 0.201 – 0.242 | 21.127 | **<0.001** | 46627.276 | 0.219 | 0.011 | 0.197 – 0.240 | 19.822 | **<0.001** | 41561.000 |
| Social dominance orientation | 0.018 | 0.011 | -0.003 – 0.039 | 1.718 | 0.086 | 46631.660 | 0.014 | 0.011 | -0.008 – 0.036 | 1.224 | 0.221 | 41562.738 |
| Science-related populist attitudes | 0.003 | 0.011 | -0.018 – 0.024 | 0.285 | 0.775 | 46625.770 | 0.008 | 0.011 | -0.013 – 0.030 | 0.755 | 0.450 | 41560.838 |
| Perceived benefit of science | 0.578 | 0.012 | 0.554 – 0.601 | 48.697 | **<0.001** | 46631.758 | 0.573 | 0.012 | 0.549 – 0.598 | 45.912 | **<0.001** | 41561.990 |
| Willingness to be vulnerable to science | 0.686 | 0.012 | 0.662 – 0.710 | 56.466 | **<0.001** | 46629.529 | 0.695 | 0.013 | 0.670 – 0.720 | 54.321 | **<0.001** | 41563.021 |
| Trust in the scientific method | 0.666 | 0.012 | 0.643 – 0.690 | 54.955 | **<0.001** | 46629.277 | 0.662 | 0.013 | 0.637 – 0.688 | 51.894 | **<0.001** | 41561.010 |
| GDP per capita | | | | | | | 0.092 | 0.161 | -0.232 – 0.417 | 0.573 | 0.569 | 44.037 |
| Govt expenditure on education (% of GDP) | | | | | | | 0.124 | 0.139 | -0.157 – 0.404 | 0.889 | 0.379 | 45.065 |
| Gini index | | | | | | | 0.350 | 0.127 | 0.093 – 0.606 | 2.743 | **0.009** | 44.800 |
| Science literacy (PISA) | | | | | | | 0.008 | 0.138 | -0.270 – 0.287 | 0.060 | 0.953 | 46.999 |
| Academic freedom | | | | | | | 0.023 | 0.136 | -0.251 – 0.297 | 0.167 | 0.868 | 44.991 |
| Degree of populism in politics | | | | | | | -0.041 | 0.123 | -0.288 – 0.206 | -0.336 | 0.738 | 45.395 |
| **Random Effects** | | | | | | | | | | | | |
| $\sigma^2$ | 2.75 | | | | | | 2.94 | | | | | |
| $\tau_{00}$ | 0.86 COUNTRY_NAME | | | | | | 0.61 COUNTRY_NAME | | | | | |
| ICC | 0.24 | | | | | | 0.17 | | | | | |
| N | 66 COUNTRY_NAME | | | | | | 51 COUNTRY_NAME | | | | | |
| Observations | 46698 | | | | | | 41621 | | | | | |
| Marginal $R^2$ / Conditional $R^2$ | 0.366 / 0.517 | | | | | | 0.381 / 0.487 | | | | | |
| AIC | 231699.516 | | | | | | 196549.075 | | | | | |

*Note.* Significant testing based on two-sided t tests. AIC = Akaike information criterion, ICC = Intraclass Correlation Coefficient. $\sigma^2$ = within country (residual) variance. $\tau_{00}$ = between-country variance (variation between individual intercepts and average intercept).

**Table S26. Unweighted linear multilevel regression testing the association of perceived scientist trustworthiness with demographic characteristics, attitudes to science, and country-level indicators in 68 Countries/regions (random intercepts across countries/regions, aligned scores from pooled weighted CFA; Fig. S1, Path B1)**

| | **Block 1: Demographic characteristics** | | | | | | **Block 2: Ideological views** | | | | | |
|---|---|---|---|---|---|---|---|---|---|---|---|---|
| *Predictors* | *Beta* | *SE* | *CI* | *t* | *p* | *df* | *Beta* | *SE* | *CI* | *t* | *p* | *df* |
| Intercept | 1.015 | 0.104 | 0.807 – 1.224 | 9.728 | **<0.001** | 67.102 | 1.155 | 0.108 | 0.938 – 1.371 | 10.642 | **<0.001** | 66.208 |
| Sex (male) | -0.049 | 0.010 | -0.069 – -0.028 | -4.658 | **<0.001** | 64359.440 | -0.047 | 0.012 | -0.071 – -0.023 | -3.888 | **<0.001** | 47583.956 |
| Age | 0.004 | 0.010 | -0.017 – 0.024 | 0.362 | 0.717 | 64360.294 | -0.036 | 0.012 | -0.060 – -0.013 | -3.017 | **0.003** | 47583.038 |
| Education (tertiary) | 0.113 | 0.011 | 0.092 – 0.134 | 10.568 | **<0.001** | 64358.954 | 0.101 | 0.012 | 0.077 – 0.125 | 8.140 | **<0.001** | 47580.733 |
| Income | 0.092 | 0.011 | 0.071 – 0.113 | 8.631 | **<0.001** | 64359.251 | 0.077 | 0.013 | 0.052 – 0.101 | 6.127 | **<0.001** | 47580.394 |
| Residence place (urban) | 0.091 | 0.011 | 0.070 – 0.112 | 8.610 | **<0.001** | 64358.735 | 0.102 | 0.012 | 0.077 – 0.126 | 8.244 | **<0.001** | 47580.279 |
| Political orientation (right) | | | | | | | 0.066 | 0.014 | 0.038 – 0.094 | 4.674 | **<0.001** | 47580.043 |
| Political orientation (conservative) | | | | | | | -0.123 | 0.014 | -0.151 – -0.096 | -8.675 | **<0.001** | 47581.671 |
| Religiosity | | | | | | | 0.265 | 0.012 | 0.241 – 0.290 | 21.239 | **<0.001** | 47581.428 |
| Social dominance orientation | | | | | | | -0.202 | 0.012 | -0.226 – -0.177 | -16.283 | **<0.001** | 47580.946 |
| Science-related populist attitudes | | | | | | | | | | | | |
| Perceived benefit of science | | | | | | | | | | | | |
| Willingness to be vulnerable to science | | | | | | | | | | | | |
| Trust in the scientific method | | | | | | | | | | | | |
| GDP per capita | | | | | | | | | | | | |
| Govt expenditure on education (% of GDP) | | | | | | | | | | | | |
| Gini index | | | | | | | | | | | | |
| Science literacy (PISA) | | | | | | | | | | | | |
| Academic freedom | | | | | | | | | | | | |
| Degree of populism in politics | | | | | | | | | | | | |
| **Random Effects** | | | | | | | | | | | | |
| $\sigma^2$ | 6.97 | | | | | | 6.71 | | | | | |
| $\tau_{00}$ | 0.73 COUNTRY_NAME | | | | | | 0.77 COUNTRY_NAME | | | | | |
| ICC | 0.09 | | | | | | 0.1 | | | | | |
| N | 68 COUNTRY_NAME | | | | | | 67 COUNTRY_NAME | | | | | |
| Observations | 64431 | | | | | | 47654 | | | | | |
| Marginal $R^2$ / Conditional $R^2$ | 0.005 / 0.099 | | | | | | 0.019 / 0.120 | | | | | |
| AIC | 308310.959 | | | | | | 226315.048 | | | | | |

| | Block 3: Attitudes to science | | | | | | Block 4: Country indicators | | | | | |
|---|---|---|---|---|---|---|---|---|---|---|---|---|
| *Predictors* | *Beta* | *SE* | *CI* | *t* | *p* | *df* | *Beta* | *SE* | *CI* | *t* | *p* | *df* |
| Intercept | 1.022 | 0.111 | 0.801 – 1.244 | 9.223 | **<0.001** | 65.189 | 0.904 | 0.119 | 0.664 – 1.143 | 7.603 | **<0.001** | 44.137 |
| Sex (male) | -0.097 | 0.010 | -0.117 – -0.076 | -9.249 | **<0.001** | 46623.325 | -0.085 | 0.011 | -0.107 – -0.064 | -7.937 | **<0.001** | 41559.937 |
| Age | -0.050 | 0.010 | -0.071 – -0.030 | -4.879 | **<0.001** | 46622.74 | -0.055 | 0.011 | -0.076 – -0.035 | -5.207 | **<0.001** | 41559.219 |
| Education (tertiary) | -0.043 | 0.011 | -0.064 – -0.022 | -3.979 | **<0.001** | 46621.371 | -0.046 | 0.011 | -0.068 – -0.025 | -4.200 | **<0.001** | 41558.656 |
| Income | 0.006 | 0.011 | -0.015 – 0.028 | 0.598 | 0.550 | 46620.757 | -0.001 | 0.011 | -0.023 – 0.021 | -0.088 | 0.930 | 41558.476 |
| Residence place (urban) | 0.047 | 0.011 | 0.027 – 0.068 | 4.454 | **<0.001** | 46620.564 | 0.052 | 0.011 | 0.031 – 0.073 | 4.785 | **<0.001** | 41558.238 |
| Political orientation (right) | 0.045 | 0.012 | 0.021 – 0.069 | 3.654 | **<0.001** | 46620.543 | 0.04 | 0.013 | 0.014 – 0.065 | 3.089 | **0.002** | 41557.909 |
| Political orientation (conservative) | -0.042 | 0.012 | -0.066 – -0.018 | -3.421 | **0.001** | 46621.832 | -0.046 | 0.013 | -0.071 – -0.021 | -3.566 | **<0.001** | 41558.614 |
| Religiosity | 0.225 | 0.011 | 0.203 – 0.246 | 20.641 | **<0.001** | 46621.502 | 0.220 | 0.011 | 0.198 – 0.242 | 19.644 | **<0.001** | 41558.702 |
| Social dominance orientation | 0.058 | 0.011 | 0.037 – 0.080 | 5.286 | **<0.001** | 46621.018 | 0.045 | 0.011 | 0.023 – 0.068 | 3.985 | **<0.001** | 41558.122 |
| Science-related populist attitudes | 0.050 | 0.011 | 0.029 – 0.072 | 4.592 | **<0.001** | 46621.71 | 0.059 | 0.011 | 0.037 – 0.081 | 5.245 | **<0.001** | 41558.797 |
| Perceived benefit of science | 0.580 | 0.012 | 0.556 – 0.604 | 47.016 | **<0.001** | 46621.098 | 0.572 | 0.013 | 0.547 – 0.597 | 44.860 | **<0.001** | 41558.303 |
| Willingness to be vulnerable to science | 0.680 | 0.013 | 0.655 – 0.705 | 53.721 | **<0.001** | 46620.805 | 0.714 | 0.013 | 0.688 – 0.739 | 54.693 | **<0.001** | 41558.439 |
| Trust in the scientific method | 0.638 | 0.013 | 0.613 – 0.663 | 50.848 | **<0.001** | 46620.408 | 0.635 | 0.013 | 0.610 – 0.661 | 49.057 | **<0.001** | 41557.919 |
| GDP per capita | | | | | | | 0.119 | 0.163 | -0.209 – 0.447 | 0.731 | 0.468 | 44.403 |
| Govt expenditure on education (% of GDP) | | | | | | | 0.064 | 0.139 | -0.217 – 0.345 | 0.458 | 0.649 | 44.394 |
| Gini index | | | | | | | 0.340 | 0.128 | 0.081 – 0.598 | 2.649 | **0.011** | 44.577 |
| Science literacy (PISA) | | | | | | | 0.036 | 0.138 | -0.241 – 0.313 | 0.262 | 0.795 | 44.607 |
| Academic freedom | | | | | | | 0.057 | 0.137 | -0.218 – 0.332 | 0.418 | 0.678 | 44.571 |
| Degree of populism in politics | | | | | | | 0.007 | 0.123 | -0.241 – 0.256 | 0.060 | 0.952 | 44.708 |
| **Random Effects** | | | | | | | | | | | | |
| $\sigma^2$ | 4.98 | | | | | | 4.6 | | | | | |
| $\tau_{00}$ | 0.80 COUNTRY_NAME | | | | | | 0.62 COUNTRY_NAME | | | | | |
| ICC | 0.14 | | | | | | 0.12 | | | | | |
| N | 66 COUNTRY_NAME | | | | | | 51 COUNTRY_NAME | | | | | |
| Observations | 46698 | | | | | | 41621 | | | | | |
| Marginal $R^2$ / Conditional $R^2$ | 0.258 / 0.362 | | | | | | 0.289 / 0.373 | | | | | |
| AIC | 207086.195 | | | | | | 182005.733 | | | | | |

*Note.* Significant testing based on two-sided t tests. AIC = Akaike information criterion, ICC = Intraclass Correlation Coefficient. $\sigma^2$ = within country (residual) variance. $\tau_{00}$ = between-country variance (variation between individual intercepts and average intercept).

**Table S27. Weighted linear multilevel regression testing the association of perceived scientist trustworthiness with demographic characteristics, attitudes to science, and country-level indicators in 68 Countries/regions (random intercepts across countries/regions, aligned scores from pooled unweighted CFA; Fig. S1, Path B2)**

| | **Block 1: Demographic characteristics** | | | | | | **Block 2: Ideological views** | | | | | |
|---|---|---|---|---|---|---|---|---|---|---|---|---|
| *Predictors* | *Beta* | *SE* | *CI* | *t* | *p* | *df* | *Beta* | *SE* | *CI* | *t* | *p* | *df* |
| Intercept | 0.820 | 0.061 | 0.698 – 0.943 | 13.424 | **<0.001** | 65.026 | 0.909 | 0.062 | 0.785 – 1.033 | 14.664 | **<0.001** | 63.978 |
| Sex (male) | -0.041 | 0.005 | -0.052 – -0.031 | -7.865 | **<0.001** | 64369.853 | -0.026 | 0.006 | -0.038 – -0.014 | -4.343 | **<0.001** | 47587.978 |
| Age | 0.049 | 0.005 | 0.040 – 0.059 | 10.019 | **<0.001** | 64406.886 | 0.024 | 0.006 | 0.013 – 0.035 | 4.341 | **<0.001** | 47623.451 |
| Education (tertiary) | 0.067 | 0.006 | 0.056 – 0.078 | 11.914 | **<0.001** | 63951.444 | 0.058 | 0.006 | 0.046 – 0.070 | 9.355 | **<0.001** | 47308.424 |
| Income | 0.072 | 0.005 | 0.062 – 0.083 | 13.628 | **<0.001** | 64376.880 | 0.049 | 0.006 | 0.037 – 0.061 | 7.976 | **<0.001** | 47591.442 |
| Residence place (urban) | 0.057 | 0.005 | 0.047 – 0.067 | 10.999 | **<0.001** | 64366.620 | 0.059 | 0.006 | 0.048 – 0.071 | 10.022 | **<0.001** | 47587.251 |
| Political orientation (right) | | | | | | | 0.018 | 0.007 | 0.004 – 0.032 | 2.483 | **0.013** | 47590.229 |
| Political orientation (conservative) | | | | | | | -0.069 | 0.007 | -0.083 – -0.055 | -9.684 | **<0.001** | 47592.788 |
| Religiosity | | | | | | | 0.104 | 0.006 | 0.092 – 0.116 | 16.856 | **<0.001** | 47589.208 |
| Social dominance orientation | | | | | | | -0.209 | 0.006 | -0.221 – -0.197 | -33.756 | **<0.001** | 47594.778 |
| Science-related populist attitudes | | | | | | | | | | | | |
| Perceived benefit of science | | | | | | | | | | | | |
| Willingness to be vulnerable to science | | | | | | | | | | | | |
| Trust in the scientific method | | | | | | | | | | | | |
| GDP per capita | | | | | | | | | | | | |
| Govt expenditure on education (% of GDP) | | | | | | | | | | | | |
| Gini index | | | | | | | | | | | | |
| Science literacy (PISA) | | | | | | | | | | | | |
| Academic freedom | | | | | | | | | | | | |
| Degree of populism in politics | | | | | | | | | | | | |
| **Random Effects** | | | | | | | | | | | | |
| $\sigma^2$ | 1.07 | | | | | | 1 | | | | | |
| $\tau_{00}$ | 0.24 COUNTRY_NAME | | | | | | 0.24 COUNTRY_NAME | | | | | |
| ICC | 0.19 | | | | | | 0.2 | | | | | |
| N | 68 COUNTRY_NAME | | | | | | 67 COUNTRY_NAME | | | | | |
| Observations | 64438 | | | | | | 47656 | | | | | |
| Marginal $R^2$ / Conditional $R^2$ | 0.014 / 0.198 | | | | | | 0.054 / 0.239 | | | | | |
| AIC | 259526.863 | | | | | | 187930.412 | | | | | |

| | Block 3: Attitudes to science | | | | | | Block 4: Country indicators | | | | | |
|---|---|---|---|---|---|---|---|---|---|---|---|---|
| *Predictors* | *Beta* | *SE* | *CI* | *t* | *p* | *df* | *Beta* | *SE* | *CI* | *t* | *p* | *df* |
| Intercept | 0.776 | 0.062 | 0.652 – 0.900 | 12.498 | **<0.001** | 63.329 | 0.748 | 0.061 | 0.625 – 0.871 | 12.256 | **<0.001** | 43.845 |
| Sex (male) | -0.052 | 0.005 | -0.062 – -0.043 | -10.834 | **<0.001** | 46625.951 | -0.051 | 0.005 | -0.061 – -0.041 | -9.954 | **<0.001** | 41561.864 |
| Age | 0.007 | 0.004 | -0.002 – 0.015 | 1.493 | 0.135 | 46652.292 | 0.008 | 0.005 | -0.001 – 0.017 | 1.693 | 0.090 | 41576.042 |
| Education (tertiary) | -0.032 | 0.005 | -0.042 – -0.022 | -6.325 | **<0.001** | 46587.786 | -0.031 | 0.005 | -0.042 – -0.021 | -5.992 | **<0.001** | 41601.958 |
| Income | 0.008 | 0.005 | -0.001 – 0.018 | 1.706 | 0.088 | 46628.462 | 0.009 | 0.005 | -0.001 – 0.019 | 1.697 | 0.090 | 41560.688 |
| Residence place (urban) | 0.024 | 0.005 | 0.014 – 0.033 | 4.990 | **<0.001** | 46625.568 | 0.027 | 0.005 | 0.017 – 0.036 | 5.314 | **<0.001** | 41560.108 |
| Political orientation (right) | 0.019 | 0.006 | 0.008 – 0.030 | 3.312 | **0.001** | 46627.322 | 0.019 | 0.006 | 0.007 – 0.031 | 3.158 | **0.002** | 41561.172 |
| Political orientation (conservative) | -0.016 | 0.006 | -0.027 – -0.005 | -2.800 | **0.005** | 46629.484 | -0.019 | 0.006 | -0.031 – -0.007 | -3.121 | **0.002** | 41560.426 |
| Religiosity | 0.102 | 0.005 | 0.092 – 0.112 | 20.307 | **<0.001** | 46626.821 | 0.102 | 0.005 | 0.092 – 0.112 | 19.325 | **<0.001** | 41560.894 |
| Social dominance orientation | -0.042 | 0.005 | -0.052 – -0.032 | -8.160 | **<0.001** | 46630.694 | -0.044 | 0.005 | -0.055 – -0.034 | -8.258 | **<0.001** | 41562.544 |
| Science-related populist attitudes | -0.033 | 0.005 | -0.042 – -0.023 | -6.403 | **<0.001** | 46625.535 | -0.031 | 0.005 | -0.042 – -0.021 | -5.830 | **<0.001** | 41560.735 |
| Perceived benefit of science | 0.310 | 0.006 | 0.298 – 0.321 | 54.491 | **<0.001** | 46630.681 | 0.307 | 0.006 | 0.295 – 0.319 | 51.427 | **<0.001** | 41561.744 |
| Willingness to be vulnerable to science | 0.368 | 0.006 | 0.357 – 0.379 | 63.231 | **<0.001** | 46628.898 | 0.373 | 0.006 | 0.361 – 0.385 | 61.012 | **<0.001** | 41562.771 |
| Trust in the scientific method | 0.377 | 0.006 | 0.366 – 0.389 | 64.994 | **<0.001** | 46628.617 | 0.376 | 0.006 | 0.364 – 0.388 | 61.622 | **<0.001** | 41560.902 |
| GDP per capita | | | | | | | 0.029 | 0.083 | -0.139 – 0.196 | 0.348 | 0.730 | 43.373 |
| Govt expenditure on education (% of GDP) | | | | | | | 0.090 | 0.072 | -0.055 – 0.235 | 1.251 | 0.217 | 44.318 |
| Gini index | | | | | | | 0.200 | 0.066 | 0.068 – 0.333 | 3.042 | **0.004** | 44.046 |
| Science literacy (PISA) | | | | | | | 0.040 | 0.071 | -0.104 – 0.183 | 0.558 | 0.579 | 45.926 |
| Academic freedom | | | | | | | 0.049 | 0.070 | -0.093 – 0.190 | 0.696 | 0.490 | 44.193 |
| Degree of populism in politics | | | | | | | -0.045 | 0.063 | -0.173 – 0.082 | -0.717 | 0.477 | 44.527 |
| **Random Effects** | | | | | | | | | | | | |
| $\sigma^2$ | 0.63 | | | | | | 0.67 | | | | | |
| $\tau_{00}$ | 0.25 COUNTRY_NAME | | | | | | 0.16 COUNTRY_NAME | | | | | |
| ICC | 0.28 | | | | | | 0.19 | | | | | |
| N | 66 COUNTRY_NAME | | | | | | 51 COUNTRY_NAME | | | | | |
| Observations | 46700 | | | | | | 41622 | | | | | |
| Marginal $R^2$ / Conditional $R^2$ | 0.425 / 0.586 | | | | | | 0.451 / 0.558 | | | | | |
| AIC | 162995.218 | | | | | | 135191.765 | | | | | |

*Note.* Significant testing based on two-sided t tests. AIC = Akaike information criterion, ICC = Intraclass Correlation Coefficient. $\sigma^2$ = within country (residual) variance. $\tau_{00}$ = between-country variance (variation between individual intercepts and average intercept).

**Table S28. Unweighted linear multilevel regression testing the association of perceived scientist trustworthiness with demographic characteristics, attitudes to science, and country-level indicators in 68 Countries/regions (random intercepts across countries/regions, aligned scores from pooled unweighted CFA; Fig. S1, Path B2)**

| | **Block 1: Demographic characteristics** | | | | | | **Block 2: Ideological views** | | | | | |
|---|---|---|---|---|---|---|---|---|---|---|---|---|
| *Predictors* | *Beta* | *SE* | *CI* | *t* | *p* | *df* | *Beta* | *SE* | *CI* | *t* | *p* | *df* |
| Intercept | 0.762 | 0.056 | 0.651 – 0.873 | 13.704 | **<0.001** | 67.000 | 0.840 | 0.057 | 0.726 – 0.954 | 14.687 | **<0.001** | 66.018 |
| Sex (male) | -0.037 | 0.005 | -0.048 – -0.027 | -7.083 | **<0.001** | 64366.206 | -0.031 | 0.006 | -0.043 – -0.020 | -5.205 | **<0.001** | 47585.176 |
| Age | 0.026 | 0.005 | 0.016 – 0.037 | 5.004 | **<0.001** | 64366.965 | 0.008 | 0.006 | -0.004 – 0.020 | 1.363 | 0.173 | 47584.350 |
| Education (tertiary) | 0.065 | 0.005 | 0.054 – 0.076 | 12.081 | **<0.001** | 64365.762 | 0.055 | 0.006 | 0.043 – 0.067 | 8.887 | **<0.001** | 47582.288 |
| Income | 0.061 | 0.005 | 0.050 – 0.071 | 11.228 | **<0.001** | 64366.031 | 0.046 | 0.006 | 0.034 – 0.058 | 7.352 | **<0.001** | 47581.981 |
| Residence place (urban) | 0.046 | 0.005 | 0.036 – 0.057 | 8.684 | **<0.001** | 64365.573 | 0.050 | 0.006 | 0.038 – 0.062 | 8.158 | **<0.001** | 47581.879 |
| Political orientation (right) | | | | | | | 0.027 | 0.007 | 0.013 – 0.041 | 3.838 | **<0.001** | 47581.674 |
| Political orientation (conservative) | | | | | | | -0.066 | 0.007 | -0.080 – -0.052 | -9.292 | **<0.001** | 47583.128 |
| Religiosity | | | | | | | 0.125 | 0.006 | 0.113 – 0.137 | 20.046 | **<0.001** | 47582.939 |
| Social dominance orientation | | | | | | | -0.166 | 0.006 | -0.178 – -0.153 | -26.813 | **<0.001** | 47582.477 |
| Science-related populist attitudes | | | | | | | | | | | | |
| Perceived benefit of science | | | | | | | | | | | | |
| Willingness to be vulnerable to science | | | | | | | | | | | | |
| Trust in the scientific method | | | | | | | | | | | | |
| GDP per capita | | | | | | | | | | | | |
| Govt expenditure on education (% of GDP) | | | | | | | | | | | | |
| Gini index | | | | | | | | | | | | |
| Science literacy (PISA) | | | | | | | | | | | | |
| Academic freedom | | | | | | | | | | | | |
| Degree of populism in politics | | | | | | | | | | | | |
| **Random Effects** | | | | | | | | | | | | |
| $\sigma^2$ | 1.77 | | | | | | 1.67 | | | | | |
| $\tau_{00}$ | 0.21 COUNTRY_NAME | | | | | | 0.21 COUNTRY_NAME | | | | | |
| ICC | 0.1 | | | | | | 0.11 | | | | | |
| N | 68 COUNTRY_NAME | | | | | | 67 COUNTRY_NAME | | | | | |
| Observations | 64438 | | | | | | 47656 | | | | | |
| Marginal $R^2$ / Conditional $R^2$ | 0.007 / 0.111 | | | | | | 0.028 / 0.139 | | | | | |
| AIC | 220135.445 | | | | | | 160018.572 | | | | | |

| *Predictors* | Block 3: Attitudes to science | | | | | | Block 4: Country indicators | | | | | |
|---|---|---|---|---|---|---|---|---|---|---|---|---|
| | *Beta* | *SE* | *CI* | *t* | *p* | *df* | *Beta* | *SE* | *CI* | *t* | *p* | *df* |
| Intercept | 0.768 | 0.058 | 0.651 – 0.884 | 13.174 | **<0.001** | 65.058 | 0.738 | 0.060 | 0.617 – 0.860 | 12.271 | **<0.001** | 44.054 |
| Sex (male) | -0.059 | 0.005 | -0.069 – -0.049 | -11.752 | **<0.001** | 46624.504 | -0.052 | 0.005 | -0.062 – -0.042 | -10.120 | **<0.001** | 41560.569 |
| Age | -0.002 | 0.005 | -0.011 – 0.008 | -0.348 | 0.728 | 46624.014 | 0.000 | 0.005 | -0.010 – 0.010 | -0.056 | 0.955 | 41559.914 |
| Education (tertiary) | -0.025 | 0.005 | -0.035 – -0.014 | -4.766 | **<0.001** | 46622.881 | -0.026 | 0.005 | -0.036 – -0.016 | -4.961 | **<0.001** | 41559.416 |
| Income | 0.008 | 0.005 | -0.002 – 0.018 | 1.500 | 0.134 | 46622.365 | 0.003 | 0.005 | -0.007 – 0.014 | 0.626 | 0.531 | 41559.252 |
| Residence place (urban) | 0.021 | 0.005 | 0.011 – 0.031 | 4.021 | **<0.001** | 46622.205 | 0.019 | 0.005 | 0.009 – 0.029 | 3.693 | **<0.001** | 41559.043 |
| Political orientation (right) | 0.017 | 0.006 | 0.006 – 0.029 | 2.969 | **0.003** | 46622.192 | 0.016 | 0.006 | 0.004 – 0.028 | 2.615 | **0.009** | 41558.747 |
| Political orientation (conservative) | -0.018 | 0.006 | -0.030 – -0.006 | -3.049 | **0.002** | 46623.262 | -0.019 | 0.006 | -0.031 – -0.007 | -3.142 | **0.002** | 41559.383 |
| Religiosity | 0.108 | 0.005 | 0.098 – 0.118 | 20.634 | **<0.001** | 46623.006 | 0.106 | 0.005 | 0.095 – 0.116 | 19.647 | **<0.001** | 41559.474 |
| Social dominance orientation | -0.018 | 0.005 | -0.029 – -0.008 | -3.439 | **0.001** | 46622.584 | -0.027 | 0.005 | -0.037 – -0.016 | -4.905 | **<0.001** | 41558.937 |
| Science-related populist attitudes | -0.007 | 0.005 | -0.017 – 0.004 | -1.248 | 0.212 | 46623.169 | -0.006 | 0.005 | -0.017 – 0.004 | -1.125 | 0.260 | 41559.559 |
| Perceived benefit of science | 0.311 | 0.006 | 0.299 – 0.322 | 52.547 | **<0.001** | 46622.657 | 0.308 | 0.006 | 0.296 – 0.320 | 50.449 | **<0.001** | 41559.099 |
| Willingness to be vulnerable to science | 0.361 | 0.006 | 0.349 – 0.373 | 59.521 | **<0.001** | 46622.409 | 0.379 | 0.006 | 0.367 – 0.391 | 60.662 | **<0.001** | 41559.221 |
| Trust in the scientific method | 0.360 | 0.006 | 0.348 – 0.371 | 59.785 | **<0.001** | 46622.068 | 0.359 | 0.006 | 0.347 – 0.371 | 57.849 | **<0.001** | 41558.754 |
| GDP per capita | | | | | | | 0.039 | 0.082 | -0.127 – 0.205 | 0.471 | 0.640 | 44.292 |
| Govt expenditure on education (% of GDP) | | | | | | | 0.059 | 0.071 | -0.084 – 0.201 | 0.831 | 0.410 | 44.284 |
| Gini index | | | | | | | 0.192 | 0.065 | 0.061 – 0.322 | 2.955 | **0.005** | 44.447 |
| Science literacy (PISA) | | | | | | | 0.050 | 0.070 | -0.090 – 0.190 | 0.721 | 0.474 | 44.474 |
| Academic freedom | | | | | | | 0.065 | 0.069 | -0.074 – 0.204 | 0.940 | 0.352 | 44.443 |
| Degree of populism in politics | | | | | | | -0.027 | 0.062 | -0.153 – 0.098 | -0.435 | 0.666 | 44.563 |
| **Random Effects** | | | | | | | | | | | | |
| $\sigma^2$ | 1.13 | | | | | | 1.06 | | | | | |
| $\tau_{00}$ | 0.22 COUNTRY_NAME | | | | | | 0.16 COUNTRY_NAME | | | | | |
| ICC | 0.16 | | | | | | 0.13 | | | | | |
| N | 66 COUNTRY_NAME | | | | | | 51 COUNTRY_NAME | | | | | |
| Observations | 46700 | | | | | | 41622 | | | | | |
| Marginal $R^2$ / Conditional $R^2$ | 0.309 / 0.423 | | | | | | 0.350 / 0.435 | | | | | |
| AIC | 138481.609 | | | | | | 120798.544 | | | | | |

*Note.* Significant testing based on two-sided t tests. AIC = Akaike information criterion, ICC = Intraclass Correlation Coefficient. $\sigma^2$ = within country (residual) variance. $\tau_{00}$ = between-country variance (variation between individual intercepts and average intercept).

## Comparison

The analyses described above examined variations in country/region rankings and regression results across different weighting schemes, CFA specifications, and measurement alignment approaches. Preliminary findings indicated changes in country/region rankings when aligned scores were used under the analytical specification corresponding to the original study (Table S15). However, the aligned scores included only two dimensions retained after alignment, whereas the original scores covered all four dimensions. Thus, the observed differences could reflect changes in construct coverage as well as adjustments for measurement non-invariance. Furthermore, differences in statistical significance between the original and reanalyzed regressions do not establish differences between their coefficients without a direct statistical comparison that accounts for the dependence between estimates derived from the same data.

To address these concerns, we conducted additional analyses comparing three scoring approaches: (1) the original four-factor score, (2) a six-item unaligned score from the two retained factors, and (3) the aligned two-factor score. Respondents, countries/regions, weighting schemes, and other analytical specifications were held constant within each comparison. The following section presents these analyses in detail.

**Rank-order Stability**

Rank-order stability across the three scoring approaches was evaluated using Spearman's rank correlation (1961) and Kendall's $\tau$ (1938). To characterize country-specific rank displacement, we calculated the absolute difference in rank positions between each pair of scoring approaches, consistent with the general logic of rank-distance measures such as Spearman's footrule (Diaconis & Graham, 1977). These differences were summarized using the median, interquartile range (IQR), and range, and the number and percentage of countries whose exact rank position changed were also reported.

The results showed that overall rank-order agreement was high, but substantial country-level rank changes were still observed (Table 1). The original 12-item ranking and the six-item unaligned benevolence–openness ranking showed strong agreement (Spearman's $\rho$ = .971; Kendall's $\tau$ = .863), as did the original and aligned rankings (Spearman's $\rho$ = .971; Kendall's $\tau$ = .863). Nevertheless, 62 of 68 countries (91.2%) changed their exact rank position in both comparisons, with a median absolute displacement of 3 positions (IQR = 2–5 for the original 12-item versus six-item unaligned comparison; IQR = 1.75–5 for the original versus aligned comparison; range = 0–15 for both). Agreement between the six-item unaligned benevolence–openness ranking and the aligned ranking was even higher (Spearman's $\rho$ = .995; Kendall's $\tau$ = .963), yet 36 countries (52.9%) still changed rank, with a median absolute displacement of 1 position (IQR = 0–1; range = 0–11). These findings suggest that, although the overall country rankings remained highly consistent across scoring approaches, some countries experienced changes in their relative positions. These changes may still be substantively relevant, as even modest shifts in individual country rankings can affect the interpretation of cross-national differences, particularly when rankings are used to inform substantive conclusions or policy-related comparisons.

## Comparison of Regression Coefficients Across Scoring Methods

To examine whether different scoring methods yielded different associations between perceived trustworthiness of scientists and the predictors, we adopted a pairwise joint multilevel linear mixed-effects modeling approach, drawing on the cross-equation coefficient comparison framework of seemingly unrelated regression (SUE; Zellner, 1962). Cluster-robust variance estimation and Wald tests were used to directly assess differences in regression coefficients across scoring methods.

Three trustworthiness scores were considered: (1) the Raw4 score, calculated as the mean of available responses to 12 items covering four dimensions—competence, integrity, benevolence, and openness; (2) the Raw2 score, calculated as the mean of available responses to six items covering benevolence and openness; and (3) the Aligned score, calculated as the mean of the available measurement-aligned factor scores for benevolence and openness. Given differences in measurement content and scoring scales, all three scores were separately standardized to Z-scores ($M$ = 0, $SD$ = 1) within the common analytic sample for each regression block to facilitate comparisons of regression coefficients. Pearson correlations were also calculated to describe the linear relationships among the three scores before standardization.

Following the four sequentially adjusted regression models (Blocks 1–4) in the original analyses (Cologna et al., 2025), we conducted pairwise comparisons of the Original and Raw2 scores, the Original and Aligned scores, and the Raw2 and Aligned scores, resulting in 12 joint multilevel models. The four blocks sequentially included demographic characteristics, ideological views, science-related populist attitudes and other attitudes toward science, and country-level indicators. Within each block, all comparisons were restricted to a common analytic sample with valid observations on all three trustworthiness scores, the corresponding predictors, country identifiers, and the weighting variable. To ensure the comparability of regression coefficients across scoring methods, a common analytic sample was used within each regression block, retaining only respondents with valid observations on all three trustworthiness scores and the corresponding predictors. To assess the additional sample loss introduced by this restriction, we compared the sample sizes of the independently estimated models with those of the common-sample models. The results indicated that the additional exclusions were minimal across all four blocks, with a maximum exclusion rate of only 0.042%. In Block 4, which included all predictors, only eight additional respondents (0.019%) were excluded for both the Raw4 and Raw2 scores, while no additional exclusions occurred for the Aligned2 score. Given the negligible additional sample loss attributable to the common-sample restriction, no additional inverse probability weighting (IPW; Cole & Hernán, 2008) adjustment was applied specifically to this restriction.

For each pairwise comparison, the two standardized outcome scores were restructured into long format and jointly modeled, allowing separate intercepts and regression coefficients for each scoring method. To account for the hierarchical structure of respondents nested within countries, scoring-method-specific random intercepts were specified at the country level, with separate random-intercept and residual variances across scoring methods. The models retained the weighting variable used in the original analyses ("WEIGHT_MLVLM") and were estimated using restricted maximum likelihood (REML) (Harville, 1977; Patterson & Thompson, 1971). To account for within-country dependence and improve the robustness of statistical inference, we calculated bias-reduced cluster-robust variance estimates (CR2), with countries serving as the clustering units (Bell & Mccaffrey, 2002; Pustejovsky & Tipton, 2018). Differences in regression coefficients across scoring methods were then evaluated using Wald tests with the Hotelling's $T^2$ approximation (HTZ), which provides a small-sample adjustment for hypothesis testing based on cluster-robust variance estimates (Tipton & Pustejovsky, 2015).

Specifically, an omnibus Wald test was first conducted for each pairwise comparison to evaluate the joint null hypothesis that all corresponding predictor coefficients were equal across the two scoring methods within a given block. Individual Wald tests were then performed to assess differences in the corresponding regression coefficients for each predictor. For these comparisons, we reported the coefficient difference ($\Delta\beta$), CR2-robust standard error (SE), 95% confidence interval (CI), and HTZ-adjusted p-value. The coefficient difference was defined as the coefficient estimated for the first scoring method minus that estimated for the second. Confidence intervals were calculated using CR2-robust standard errors and the approximate degrees of freedom obtained from the corresponding HTZ tests.

To control the false discovery rate associated with multiple comparisons, p-values were adjusted using the Benjamini–Hochberg (BH) procedure (Benjamini & Hochberg, 1995). For individual coefficient comparisons, BH adjustments were applied both within each scoring-method pair and across all three pairwise comparisons within each regression block. For omnibus Wald tests, BH adjustments were applied across the three scoring-method comparisons within each block.

All analyses were conducted in R. Joint multilevel linear mixed-effects models were estimated using the `nlme` package (Pinheiro et al., 1999), while CR2 cluster-robust variance estimates and HTZ-adjusted Wald tests were implemented using the `clubSandwich` package (Pustejovsky, 2016).

**Regression Results of Different Score Methods**

Joint Wald tests revealed significant overall differences between the original four-dimensional score and the measurement-aligned two-dimensional score across all four regression blocks (all BH-adjusted *p*s < .001; Table S30). Predictor-specific comparisons indicated that these differences were concentrated in associations involving age, social dominance orientation, science-related populist attitudes, and several attitudes toward science (Table S29). To further distinguish the potential contributions of measurement content and alignment, we compared the original 12-item score with a six-item composite retaining only benevolence and openness. Significant overall differences were found between these two scores across all four regression blocks (all BH-adjusted *p*s < .001; Table S30). However, no significant differences were found between the original six-item composite and the measurement-aligned composite of the same two dimensions (BH-adjusted *p* = .096–.937; Table S30).

These findings suggest that different measurement specifications of scientist trustworthiness may yield different estimates of substantive associations. Specifically, significant differences were primarily observed between the original four-dimensional score and the two-dimensional scores, whereas no significant overall differences were found between the raw and aligned scores based on the same two dimensions. Nevertheless, the competence and integrity dimensions did not meet the criteria for successful measurement alignment, leaving their cross-country comparability uncertain. Therefore, rankings and substantive analyses based on the original composite score should be interpreted with caution. These findings highlight the importance of assessing measurement invariance in cross-country/region research, as evidence of measurement comparability may inform the choice of measurement specifications and the interpretation of substantive findings.

**Table S29. Differences in Regression Coefficients Across Trustworthiness-of-Scientists Scoring Methods**

| | Comparison | $\beta_1$ | $\beta_2$ | $\Delta\beta$ | 95% CI | Adjusted $p$ |
|---|---|---|---|---|---|---|
| Block 1 | | | | | | |
| Sex (male) | | | | | | |
| | Raw4 vs. Raw2 | -0.027 | -0.026 | -0.001 | [-0.007, 0.004] | 0.697 |
| | Raw4 vs. Aligned | -0.027 | -0.026 | -0.001 | [-0.007, 0.005] | 0.715 |
| | Raw2 vs. Aligned | -0.026 | -0.026 | 0.000 | [-0.001, 0.001] | 0.732 |
| Age | | | | | | |
| | Raw4 vs. Raw2 | 0.042 | 0.014 | 0.028 | [0.020, 0.035] | **<0.001** |
| | Raw4 vs. Aligned | 0.042 | 0.013 | 0.029 | [0.021, 0.036] | **<0.001** |
| | Raw2 vs. Aligned | 0.014 | 0.013 | 0.001 | [-0.000, 0.002] | 0.109 |
| Education (tertiary) | | | | | | |
| | Raw4 vs. Raw2 | 0.050 | 0.041 | 0.010 | [0.003, 0.017] | **0.023** |
| | Raw4 vs. Aligned | 0.050 | 0.042 | 0.008 | [0.001, 0.015] | 0.050 |
| | Raw2 vs. Aligned | 0.041 | 0.042 | -0.001 | [-0.002, -0.000] | 0.013 |
| Income | | | | | | |
| | Raw4 vs. Raw2 | 0.051 | 0.039 | 0.012 | [0.007, 0.017] | **<0.001** |
| | Raw4 vs. Aligned | 0.051 | 0.038 | 0.013 | [0.008, 0.018] | **<0.001** |
| | Raw2 vs. Aligned | 0.039 | 0.038 | 0.001 | [0.000, 0.002] | 0.090 |
| Residence place (urban) | | | | | | |
| | Raw4 vs. Raw2 | 0.041 | 0.043 | -0.002 | [-0.005, 0.002] | 0.546 |
| | Raw4 vs. Aligned | 0.041 | 0.042 | -0.001 | [-0.005, 0.003] | 0.669 |
| | Raw2 vs. Aligned | 0.043 | 0.042 | 0.000 | [-0.000, 0.001] | 0.533 |
| Block 2 | | | | | | |
| Sex (male) | | | | | | |

| | Comparison | $\beta_1$ | $\beta_2$ | $\Delta\beta$ | 95% CI | Adjusted $p$ |
|---|---|---|---|---|---|---|
| | Raw4 vs. Raw2 | -0.017 | -0.018 | 0.001 | [-0.004, 0.006] | 0.808 |
| | Raw4 vs. Aligned | -0.017 | -0.018 | 0.001 | [-0.004, 0.006] | 0.808 |
| | Raw2 vs. Aligned | -0.018 | -0.018 | 0.000 | [-0.001, 0.001] | 0.910 |
| Age | | | | | | |
| | Raw4 vs. Raw2 | 0.023 | -0.006 | 0.029 | [0.021, 0.037] | **<0.001** |
| | Raw4 vs. Aligned | 0.023 | -0.007 | 0.030 | [0.022, 0.038] | **<0.001** |
| | Raw2 vs. Aligned | -0.006 | -0.007 | 0.001 | [-0.000, 0.002] | 0.253 |
| Education (tertiary) | | | | | | |
| | Raw4 vs. Raw2 | 0.045 | 0.036 | 0.009 | [0.002, 0.015] | **0.039** |
| | Raw4 vs. Aligned | 0.045 | 0.037 | 0.008 | [0.001, 0.014] | 0.067 |
| | Raw2 vs. Aligned | 0.036 | 0.037 | -0.001 | [-0.002, -0.000] | 0.087 |
| Income | | | | | | |
| | Raw4 vs. Raw2 | 0.036 | 0.028 | 0.008 | [0.003, 0.014] | **0.030** |
| | Raw4 vs. Aligned | 0.036 | 0.027 | 0.009 | [0.004, 0.015] | **0.014** |
| | Raw2 vs. Aligned | 0.028 | 0.027 | 0.001 | [-0.000, 0.002] | 0.276 |
| Residence place (urban) | | | | | | |
| | Raw4 vs. Raw2 | 0.044 | 0.046 | -0.002 | [-0.006, 0.002] | 0.527 |
| | Raw4 vs. Aligned | 0.044 | 0.046 | -0.002 | [-0.006, 0.002] | 0.520 |
| | Raw2 vs. Aligned | 0.046 | 0.046 | 0.000 | [-0.001, 0.001] | 0.886 |
| Political orientation (right) | | | | | | |
| | Raw4 vs. Raw2 | 0.010 | 0.014 | -0.004 | [-0.014, 0.005] | 0.542 |
| | Raw4 vs. Aligned | 0.010 | 0.015 | -0.005 | [-0.014, 0.004] | 0.473 |
| | Raw2 vs. Aligned | 0.014 | 0.015 | -0.001 | [-0.002, 0.000] | 0.312 |
| Political orientation (conservative) | | | | | | |

| | Comparison | $\beta_1$ | $\beta_2$ | $\Delta\beta$ | 95% CI | Adjusted $p$ |
|---|---|---|---|---|---|---|
| | Raw4 vs. Raw2 | -0.050 | -0.048 | -0.002 | [-0.007, 0.003] | 0.542 |
| | Raw4 vs. Aligned | -0.050 | -0.048 | -0.002 | [-0.007, 0.003] | 0.594 |
| | Raw2 vs. Aligned | -0.048 | -0.048 | 0.000 | [-0.001, 0.002] | 0.884 |
| Religiosity | | | | | | |
| | Raw4 vs. Raw2 | 0.075 | 0.085 | -0.011 | [-0.019, -0.002] | **0.044** |
| | Raw4 vs. Aligned | 0.075 | 0.085 | -0.011 | [-0.019, -0.002] | **0.044** |
| | Raw2 vs. Aligned | 0.085 | 0.085 | 0.000 | [-0.001, 0.001] | 0.886 |
| Social dominance orientation | | | | | | |
| | Raw4 vs. Raw2 | -0.143 | -0.101 | -0.042 | [-0.053, -0.032] | **<0.001** |
| | Raw4 vs. Aligned | -0.143 | -0.101 | -0.042 | [-0.053, -0.032] | **<0.001** |
| | Raw2 vs. Aligned | -0.101 | -0.101 | 0.000 | [-0.001, 0.001] | 0.910 |
| Block 3 | | | | | | |
| Sex (male) | | | | | | |
| | Raw4 vs. Raw2 | -0.036 | -0.035 | -0.001 | [-0.006, 0.005] | 0.870 |
| | Raw4 vs. Aligned | -0.036 | -0.035 | -0.001 | [-0.006, 0.005] | 0.870 |
| | Raw2 vs. Aligned | -0.035 | -0.035 | 0.000 | [-0.001, 0.001] | 0.992 |
| Age | | | | | | |
| | Raw4 vs. Raw2 | 0.009 | -0.016 | 0.026 | [0.018, 0.033] | **<0.001** |
| | Raw4 vs. Aligned | 0.009 | -0.017 | 0.026 | [0.018, 0.034] | **<0.001** |
| | Raw2 vs. Aligned | -0.016 | -0.017 | 0.001 | [-0.000, 0.002] | 0.210 |
| Education (tertiary) | | | | | | |
| | Raw4 vs. Raw2 | -0.020 | -0.022 | 0.002 | [-0.003, 0.006] | 0.639 |
| | Raw4 vs. Aligned | -0.020 | -0.021 | 0.001 | [-0.004, 0.006] | 0.870 |
| | Raw2 vs. Aligned | -0.022 | -0.021 | -0.001 | [-0.002, 0.000] | 0.201 |

| | Comparison | $\beta_1$ | $\beta_2$ | $\Delta\beta$ | 95% CI | Adjusted *p* |
|---|---|---|---|---|---|---|
| Income | | | | | | |
| | Raw4 vs. Raw2 | 0.007 | 0.001 | 0.006 | [-0.000, 0.011] | 0.150 |
| | Raw4 vs. Aligned | 0.007 | 0.000 | 0.006 | [0.001, 0.012] | 0.089 |
| | Raw2 vs. Aligned | 0.001 | 0.000 | 0.001 | [-0.000, 0.002] | 0.228 |
| Residence place (urban) | | | | | | |
| | Raw4 vs. Raw2 | 0.018 | 0.022 | -0.004 | [-0.008, 0.000] | 0.154 |
| | Raw4 vs. Aligned | 0.018 | 0.022 | -0.004 | [-0.008, -0.000] | 0.122 |
| | Raw2 vs. Aligned | 0.022 | 0.022 | 0.000 | [-0.001, 0.001] | 0.761 |
| Political orientation (right) | | | | | | |
| | Raw4 vs. Raw2 | 0.012 | 0.014 | -0.002 | [-0.009, 0.005] | 0.717 |
| | Raw4 vs. Aligned | 0.012 | 0.015 | -0.003 | [-0.010, 0.004] | 0.610 |
| | Raw2 vs. Aligned | 0.014 | 0.015 | -0.001 | [-0.002, 0.000] | 0.251 |
| Political orientation (conservative) | | | | | | |
| | Raw4 vs. Raw2 | -0.011 | -0.015 | 0.004 | [-0.001, 0.009] | 0.228 |
| | Raw4 vs. Aligned | -0.011 | -0.015 | 0.004 | [-0.001, 0.009] | 0.228 |
| | Raw2 vs. Aligned | -0.015 | -0.015 | 0.000 | [-0.001, 0.002] | 0.870 |
| Religiosity | | | | | | |
| | Raw4 vs. Raw2 | 0.074 | 0.080 | -0.006 | [-0.013, 0.002] | 0.212 |
| | Raw4 vs. Aligned | 0.074 | 0.080 | -0.006 | [-0.014, 0.001] | 0.210 |
| | Raw2 vs. Aligned | 0.080 | 0.080 | -0.001 | [-0.002, 0.000] | 0.412 |
| Social dominance orientation | | | | | | |
| | Raw4 vs. Raw2 | -0.020 | 0.007 | -0.027 | [-0.037, -0.017] | **<0.001** |
| | Raw4 vs. Aligned | -0.020 | 0.007 | -0.027 | [-0.037, -0.017] | **<0.001** |
| | Raw2 vs. Aligned | 0.007 | 0.007 | 0.000 | [-0.001, 0.001] | 0.992 |

| | Comparison | $\beta_1$ | $\beta_2$ | $\Delta\beta$ | 95% CI | Adjusted $p$ |
|---|---|---|---|---|---|---|
| Science-related populist attitudes | | | | | | |
| | Raw4 vs. Raw2 | -0.032 | 0.002 | -0.034 | [-0.040, -0.028] | **<0.001** |
| | Raw4 vs. Aligned | -0.032 | 0.001 | -0.033 | [-0.040, -0.026] | **<0.001** |
| | Raw2 vs. Aligned | 0.002 | 0.001 | 0.001 | [-0.000, 0.002] | 0.201 |
| Perceived benefit of science | | | | | | |
| | Raw4 vs. Raw2 | 0.222 | 0.209 | 0.013 | [0.007, 0.018] | **<0.001** |
| | Raw4 vs. Aligned | 0.222 | 0.209 | 0.013 | [0.007, 0.019] | **<0.001** |
| | Raw2 vs. Aligned | 0.209 | 0.209 | 0.000 | [-0.001, 0.001] | 0.695 |
| Willingness to be vulnerable to science | | | | | | |
| | Raw4 vs. Raw2 | 0.268 | 0.249 | 0.019 | [0.010, 0.028] | **<0.001** |
| | Raw4 vs. Aligned | 0.268 | 0.248 | 0.019 | [0.010, 0.028] | **<0.001** |
| | Raw2 vs. Aligned | 0.249 | 0.248 | 0.000 | [-0.001, 0.002] | 0.761 |
| Trust in the scientific method | | | | | | |
| | Raw4 vs. Raw2 | 0.274 | 0.241 | 0.033 | [0.027, 0.039] | **<0.001** |
| | Raw4 vs. Aligned | 0.274 | 0.241 | 0.032 | [0.026, 0.039] | **<0.001** |
| | Raw2 vs. Aligned | 0.241 | 0.241 | -0.001 | [-0.002, 0.001] | 0.421 |
| Block 4 | | | | | | |
| Sex (male) | | | | | | |
| | Raw4 vs. Raw2 | -0.035 | -0.035 | 0.000 | [-0.006, 0.005] | 0.964 |
| | Raw4 vs. Aligned | -0.035 | -0.035 | 0.000 | [-0.006, 0.005] | 0.964 |
| | Raw2 vs. Aligned | -0.035 | -0.035 | 0.000 | [-0.001, 0.001] | 0.965 |
| Age | | | | | | |
| | Raw4 vs. Raw2 | 0.011 | -0.016 | 0.027 | [0.019, 0.035] | **<0.001** |
| | Raw4 vs. Aligned | 0.011 | -0.017 | 0.028 | [0.019, 0.036] | **<0.001** |

| | Comparison | $\beta_1$ | $\beta_2$ | $\Delta\beta$ | 95% CI | Adjusted $p$ |
|---|---|---|---|---|---|---|
| | Raw2 vs. Aligned | -0.016 | -0.017 | 0.001 | [-0.000, 0.002] | 0.360 |
| Education (tertiary) | | | | | | |
| | Raw4 vs. Raw2 | -0.021 | -0.022 | 0.001 | [-0.003, 0.006] | 0.746 |
| | Raw4 vs. Aligned | -0.021 | -0.021 | 0.000 | [-0.005, 0.005] | 0.964 |
| | Raw2 vs. Aligned | -0.022 | -0.021 | -0.001 | [-0.002, 0.000] | 0.274 |
| Income | | | | | | |
| | Raw4 vs. Raw2 | 0.007 | 0.001 | 0.006 | [0.000, 0.012] | 0.180 |
| | Raw4 vs. Aligned | 0.007 | 0.000 | 0.007 | [0.001, 0.013] | 0.103 |
| | Raw2 vs. Aligned | 0.001 | 0.000 | 0.001 | [-0.000, 0.002] | 0.292 |
| Residence place (urban) | | | | | | |
| | Raw4 vs. Raw2 | 0.021 | 0.025 | -0.004 | [-0.009, 0.000] | 0.192 |
| | Raw4 vs. Aligned | 0.021 | 0.025 | -0.004 | [-0.009, -0.000] | 0.180 |
| | Raw2 vs. Aligned | 0.025 | 0.025 | 0.000 | [-0.001, 0.001] | 0.968 |
| Political orientation (right) | | | | | | |
| | Raw4 vs. Raw2 | 0.012 | 0.015 | -0.003 | [-0.010, 0.005] | 0.671 |
| | Raw4 vs. Aligned | 0.012 | 0.015 | -0.003 | [-0.010, 0.004] | 0.613 |
| | Raw2 vs. Aligned | 0.015 | 0.015 | 0.000 | [-0.001, 0.001] | 0.613 |
| Political orientation (conservative) | | | | | | |
| | Raw4 vs. Raw2 | -0.013 | -0.018 | 0.004 | [-0.001, 0.010] | 0.292 |
| | Raw4 vs. Aligned | -0.013 | -0.018 | 0.004 | [-0.001, 0.010] | 0.292 |
| | Raw2 vs. Aligned | -0.018 | -0.018 | 0.000 | [-0.001, 0.002] | 0.965 |
| Religiosity | | | | | | |
| | Raw4 vs. Raw2 | 0.074 | 0.080 | -0.006 | [-0.013, 0.002] | 0.293 |
| | Raw4 vs. Aligned | 0.074 | 0.081 | -0.006 | [-0.014, 0.002] | 0.292 |

| | Comparison | $\beta_1$ | $\beta_2$ | $\Delta\beta$ | 95% CI | Adjusted *p* |
|---|---|---|---|---|---|---|
| | Raw2 vs. Aligned | 0.080 | 0.081 | -0.001 | [-0.002, 0.000] | 0.452 |
| Social dominance orientation | | | | | | |
| | Raw4 vs. Raw2 | -0.022 | 0.005 | -0.027 | [-0.037, -0.017] | **<0.001** |
| | Raw4 vs. Aligned | -0.022 | 0.005 | -0.027 | [-0.038, -0.017] | **<0.001** |
| | Raw2 vs. Aligned | 0.005 | 0.005 | 0.000 | [-0.001, 0.001] | 0.964 |
| Science-related populist attitudes | | | | | | |
| | Raw4 vs. Raw2 | -0.031 | 0.004 | -0.035 | [-0.042, -0.029] | **<0.001** |
| | Raw4 vs. Aligned | -0.031 | 0.003 | -0.035 | [-0.041, -0.028] | **<0.001** |
| | Raw2 vs. Aligned | 0.004 | 0.003 | 0.001 | [-0.000, 0.002] | 0.293 |
| Perceived benefit of science | | | | | | |
| | Raw4 vs. Raw2 | 0.224 | 0.212 | 0.013 | [0.007, 0.018] | **0.001** |
| | Raw4 vs. Aligned | 0.224 | 0.211 | 0.013 | [0.007, 0.019] | **0.001** |
| | Raw2 vs. Aligned | 0.212 | 0.211 | 0.000 | [-0.001, 0.001] | 0.964 |
| Willingness to be vulnerable to science | | | | | | |
| | Raw4 vs. Raw2 | 0.275 | 0.256 | 0.019 | [0.010, 0.028] | **0.003** |
| | Raw4 vs. Aligned | 0.275 | 0.256 | 0.019 | [0.010, 0.028] | **0.003** |
| | Raw2 vs. Aligned | 0.256 | 0.256 | 0.000 | [-0.001, 0.001] | 0.965 |
| Trust in the scientific method | | | | | | |
| | Raw4 vs. Raw2 | 0.277 | 0.244 | 0.033 | [0.027, 0.039] | **<0.001** |
| | Raw4 vs. Aligned | 0.277 | 0.244 | 0.032 | [0.026, 0.039] | **<0.001** |
| | Raw2 vs. Aligned | 0.244 | 0.244 | -0.001 | [-0.002, 0.001] | 0.474 |
| GDP per capita | | | | | | |
| | Raw4 vs. Raw2 | 0.036 | 0.033 | 0.002 | [-0.021, 0.026] | 0.964 |
| | Raw4 vs. Aligned | 0.036 | 0.034 | 0.002 | [-0.026, 0.030] | 0.965 |

| | Comparison | $\beta_1$ | $\beta_2$ | $\Delta\beta$ | 95% CI | Adjusted *p* |
|---|---|---|---|---|---|---|
| | Raw2 vs. Aligned | 0.033 | 0.034 | -0.001 | [-0.013, 0.011] | 0.964 |
| Govt expenditure on education (% of GDP) | | | | | | |
| | Raw4 vs. Raw2 | 0.066 | 0.053 | 0.013 | [-0.008, 0.033] | 0.393 |
| | Raw4 vs. Aligned | 0.066 | 0.046 | 0.020 | [-0.018, 0.058] | 0.475 |
| | Raw2 vs. Aligned | 0.053 | 0.046 | 0.007 | [-0.019, 0.033] | 0.747 |
| Gini index | | | | | | |
| | Raw4 vs. Raw2 | 0.140 | 0.138 | 0.003 | [-0.013, 0.018] | 0.950 |
| | Raw4 vs. Aligned | 0.140 | 0.129 | 0.011 | [-0.013, 0.036] | 0.534 |
| | Raw2 vs. Aligned | 0.138 | 0.129 | 0.009 | [-0.008, 0.025] | 0.474 |
| Science literacy (PISA) | | | | | | |
| | Raw4 vs. Raw2 | 0.025 | 0.014 | 0.011 | [-0.006, 0.029] | 0.371 |
| | Raw4 vs. Aligned | 0.025 | 0.003 | 0.022 | [-0.001, 0.046] | 0.196 |
| | Raw2 vs. Aligned | 0.014 | 0.003 | 0.011 | [-0.004, 0.026] | 0.293 |
| Academic freedom | | | | | | |
| | Raw4 vs. Raw2 | 0.020 | 0.014 | 0.006 | [-0.012, 0.024] | 0.671 |
| | Raw4 vs. Aligned | 0.020 | 0.008 | 0.012 | [-0.012, 0.035] | 0.516 |
| | Raw2 vs. Aligned | 0.014 | 0.008 | 0.006 | [-0.007, 0.018] | 0.572 |
| Degree of populism in politics | | | | | | |
| | Raw4 vs. Raw2 | -0.036 | -0.016 | -0.020 | [-0.043, 0.004] | 0.286 |
| | Raw4 vs. Aligned | -0.036 | -0.015 | -0.021 | [-0.048, 0.006] | 0.292 |
| | Raw2 vs. Aligned | -0.016 | -0.015 | -0.001 | [-0.011, 0.009] | 0.964 |

*Note.* “Raw4” refers to the mean score of 12 items covering four dimensions (competence, integrity, benevolence, and openness); “Raw2” refers to the mean score of six items covering benevolence and openness; and “Aligned” refers to the mean of the measurement-aligned factor scores for benevolence and openness. All outcome scores were standardized within each block’s common analytic sample. $\beta_1$ and $\beta_2$ represent the regression coefficients for the first and second

scoring methods, respectively, and $\Delta\beta = \beta_1 - \beta_2$. The 95% confidence intervals (CIs) refer to $\Delta\beta$ and were calculated using country-clustered CR2 robust standard errors and approximate degrees of freedom from the Hotelling's $T^2$ (HTZ) Wald tests. Adjusted *p*-values were obtained using the Benjamini–Hochberg (BH) procedure across all individual coefficient comparisons within each block. The CIs were not adjusted for multiple comparisons. Blocks 1–4 sequentially included demographic characteristics, ideological views, science-related populist attitudes and other attitudes toward science, and country-level indicators. Bold values indicate statistically significant coefficient differences (BH-adjusted $p < .05$).

**Table S30. Joint Wald Tests Comparing Regression Coefficients Across Trustworthiness-of-Scientists Scoring Methods**

| Comparison | *N* | Countries | *F* | $df_1$ | $df_2$ | BH-adjusted *p* |
|---|---|---|---|---|---|---|
| Block 1 | | | | | | |
| Raw4_vs_Raw2 | 64431 | 68 | 14.685 | 5 | 17.737 | **<0.001** |
| Raw4_vs_Aligned | 64431 | 68 | 15.483 | 5 | 17.738 | **<0.001** |
| Raw2_vs_Aligned | 64431 | 68 | 2.239 | 5 | 17.739 | 0.096 |
| Block 2 | | | | | | |
| Raw4_vs_Raw2 | 47654 | 67 | 20.822 | 9 | 10.222 | **<0.001** |
| Raw4_vs_Aligned | 47654 | 67 | 18.700 | 9 | 10.223 | **<0.001** |
| Raw2_vs_Aligned | 47654 | 67 | 0.725 | 9 | 10.223 | 0.680 |
| Block 3 | | | | | | |
| Raw4_vs_Raw2 | 46698 | 66 | 28.634 | 13 | 5.582 | **<0.001** |
| Raw4_vs_Aligned | 46698 | 66 | 27.270 | 13 | 5.582 | **<0.001** |
| Raw2_vs_Aligned | 46698 | 66 | 0.457 | 13 | 5.583 | 0.886 |
| Block 4 | | | | | | |
| Raw4_vs_Raw2 | 41621 | 51 | 29.282 | 19 | 6.480 | **<0.001** |
| Raw4_vs_Aligned | 41621 | 51 | 32.502 | 19 | 6.480 | **<0.001** |
| Raw2_vs_Aligned | 41621 | 51 | 0.415 | 19 | 6.479 | 0.937 |

*Note.* “Raw4” refers to the mean of available responses to 12 items covering four dimensions (competence, integrity, benevolence, and openness); “Raw2” refers to the mean of available responses to six items covering benevolence and openness; and “Aligned” refers to the mean of the available measurement-aligned factor scores for benevolence and openness. All outcome scores were standardized within each block's common analytic sample. Joint Wald tests assessed whether all corresponding predictor coefficients were equal across two scoring methods, using country-clustered CR2 robust variance estimates and the Hotelling's $T^2$ (HTZ) approximation. N denotes the number of respondents; Countries denotes the number of country-level clusters; *F* is the HTZ-adjusted Wald statistic; and $df_1$ and $df_2$ denote the numerator and approximate denominator degrees of freedom, respectively. *P*-values were adjusted using the Benjamini–Hochberg procedure across the three comparisons within each block. Blocks 1–4 sequentially included demographic characteristics, ideological views, science-related populist attitudes and other attitudes toward science, and country-level indicators.